\documentclass[11pt]{elsarticle}
\makeatletter
\def\ps@pprintTitle{%
 \let\@oddhead\@empty
 \let\@evenhead\@empty
 \def\@oddfoot{}%
 \let\@evenfoot\@oddfoot}
\makeatother
\biboptions{numbers,sort&compress}

\journal{Journal Name}

\newcommand{\aap}{Astron. Astrophys.}

\newcommand{\apjl}{Astrophys. J. Lett.}

\usepackage{helvet}

\usepackage[letterpaper, total={6in,9in}]{geometry}
\usepackage{multirow}
\usepackage{subcaption}
\usepackage{tabularx}
\usepackage{makecell}
\usepackage{amssymb}
\usepackage{fdsymbol}
\usepackage{graphicx}
\usepackage{hyperref}
\usepackage{enumerate}
\usepackage{xfrac}
\usepackage[table,xcdraw]{xcolor}
\usepackage{sectsty}
\usepackage[normalem]{ulem}
\usepackage{setspace}
\usepackage[verbose]{wrapfig}
\usepackage{lipsum}
\usepackage{ragged2e}
\usepackage{etoolbox} 
\usepackage{combelow}
\usepackage{lineno}
\usepackage[T1]{fontenc}
\usepackage{pdfpages}
\usepackage[compact]{titlesec}
\usepackage[utf8]{inputenc}
\usepackage{threeparttable}
\usepackage{ltablex}
\usepackage[all]{nowidow}

\definecolor{darkgreen}{rgb}{0,0.5,0}

\begin{document}
\pagenumbering{gobble} 
\begin{titlepage}
\includepdf{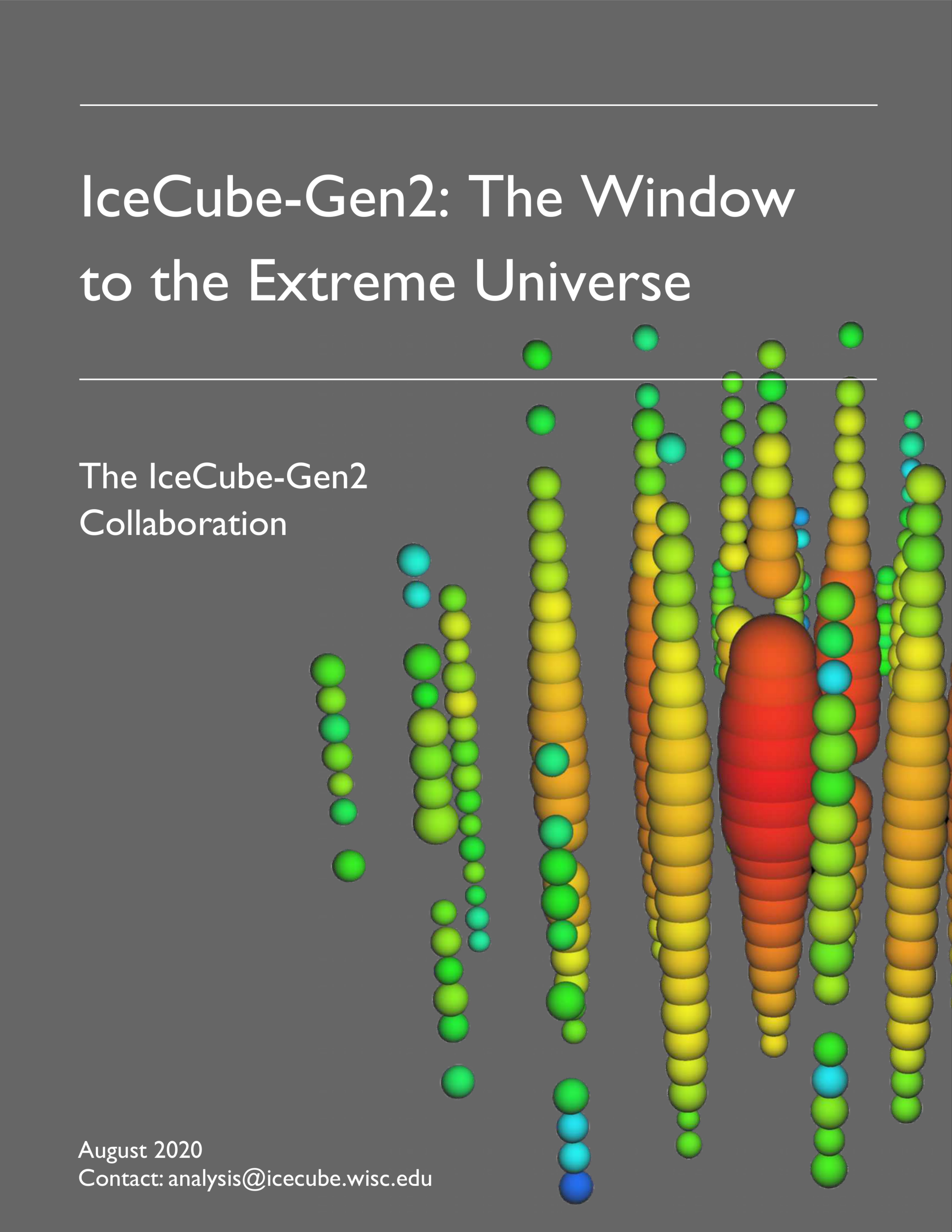}
\end{titlepage}
\begin{frontmatter}




\author[christchurch]{M. G. Aartsen}
\author[loyola]{R. Abbasi}
\author[zeuthen]{M. Ackermann}
\author[christchurch]{J. Adams}
\author[brusselslibre]{J. A. Aguilar}
\author[copenhagen]{M. Ahlers}
\author[stockholmokc]{M. Ahrens}
\author[geneva]{C. Alispach}
\author[ohioastro,ohio]{P. Allison}
\author[bartol]{N. M. Amin}

% ...
\address[brusselslibre]{Universit{\'e} Libre de Bruxelles, Science Faculty CP230, B-1050 Brussels, Belgium}
% ...
\address[loyola]{Department of Physics, Loyola University Chicago, Chicago, IL 60660, USA}
\address[christchurch]{Dept. of Physics and Astronomy, University of Canterbury, Private Bag 4800, Christchurch, New Zealand}
% ...
\address[ohioastro]{Dept. of Astronomy, Ohio State University, Columbus, OH 43210, USA}
\address[ohio]{Dept. of Physics and Center for Cosmology and Astro-Particle Physics, Ohio State University, Columbus, OH 43210, USA}
\address[copenhagen]{Niels Bohr Institute, University of Copenhagen, DK-2100 Copenhagen, Denmark}
% ...
\address[geneva]{D{\'e}partement de physique nucl{\'e}aire et corpusculaire, Universit{\'e} de Gen{\`e}ve, CH-1211 Gen{\`e}ve, Switzerland}
% ...
\address[bartol]{Bartol Research Institute and Dept. of Physics and Astronomy, University of Delaware, Newark, DE 19716, USA}
% ...
\address[stockholmokc]{Oskar Klein Centre and Dept. of Physics, Stockholm University, SE-10691 Stockholm, Sweden}
% ...
\address[zeuthen]{DESY, D-15738 Zeuthen, Germany \newpage}
\fntext[padova]{also at Universit{\`a} di Padova, I-35131 Padova, Italy}
\fntext[mephi]{also at National Research Nuclear University, Moscow Engineering Physics Institute (MEPhI), Moscow 115409, Russia}

\begin{abstract}
The observation of electromagnetic radiation from radio  to $\gamma$-ray wavelengths has provided a wealth of information about the universe. However, at PeV (10$^{15}$~eV) energies and above, most of the universe is impenetrable to photons. New messengers, namely cosmic neutrinos, are needed to explore the most extreme environments of the universe where black holes, neutron stars, and stellar explosions transform gravitational energy into non-thermal cosmic rays. These energetic particles have millions of times higher energies than those produced in the most powerful particle accelerators on Earth. As neutrinos can escape from regions otherwise opaque to radiation, they allow an unique view deep into exploding stars and the vicinity of the event horizons of black holes. 

The discovery of cosmic neutrinos with IceCube has opened this new window on the universe. IceCube has been successful in finding first evidence for cosmic particle acceleration in the jet of an active galactic nucleus. Yet, ultimately, its sensitivity is too limited to detect even the brightest neutrino sources with high significance, or to detect populations of less luminous sources. In this white paper, we present an overview of a next-generation instrument, IceCube-Gen2, which will sharpen our understanding of the processes and environments that govern the universe at the highest energies. IceCube-Gen2 is designed to: 

\begin{enumerate}
\item Resolve the high-energy neutrino sky from TeV to EeV energies
\item Investigate cosmic particle acceleration through multi-messenger observations
\item Reveal the sources and propagation of the highest energy particles in the universe
\item Probe fundamental physics with high-energy neutrinos
\end{enumerate}

IceCube-Gen2 will enhance the existing IceCube detector at the South Pole. It will increase the annual rate of observed cosmic neutrinos by a factor of ten compared to IceCube, and will be able to detect sources five times fainter than its predecessor. Furthermore, through the addition of a radio array, IceCube-Gen2 will extend the energy range by several orders of magnitude compared to IceCube. Construction will take 8 years and cost about \$350M. The goal is to have IceCube-Gen2 fully operational by 2033. 

IceCube-Gen2 will play an essential role in shaping the new era of multi-messenger astronomy, fundamentally advancing our knowledge of the high-energy universe. This challenging mission can be fully addressed only through the combination of the information from the neutrino, electromagnetic, and gravitational wave emission of high-energy sources, in concert with the new survey instruments across the electromagnetic spectrum and gravitational wave detectors which will be available in the coming years.
\end{abstract}



\end{frontmatter}

\newpage
\tableofcontents
\newpage


\pagenumbering{arabic}

\section{Introduction}
\label{S:1}

\begin{figure}[t]
\centering\includegraphics[width=1\linewidth]{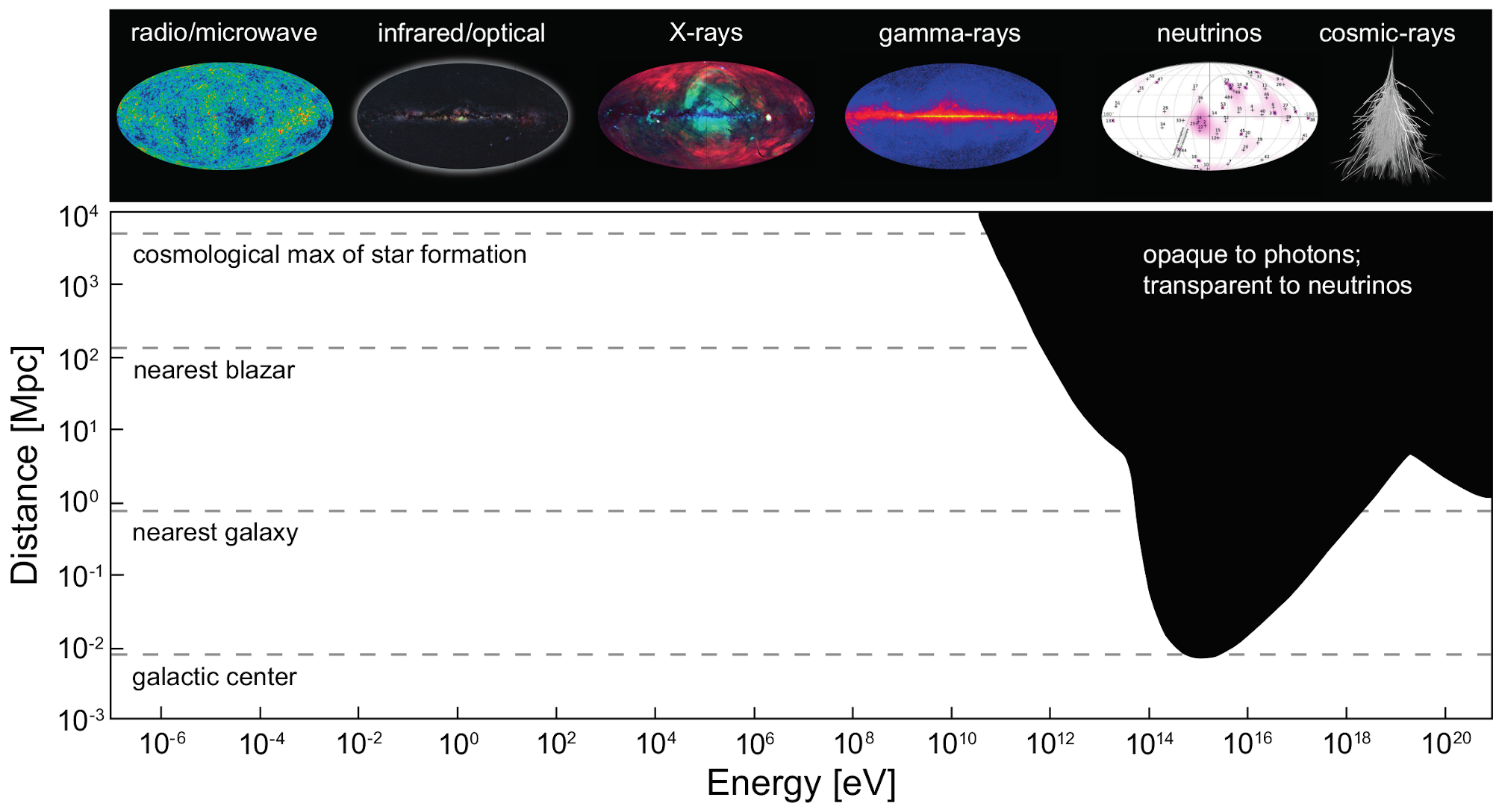}
\caption{Distance horizon at which the universe becomes non-transparent to electromagnetic radiation as a function of photon energy.}\label{fig:horizons}
\end{figure}

With the first detection of high-energy neutrinos of extraterrestrial origin in 2013~\citep{Aartsen:2013bka, Aartsen:2013jdh}, the IceCube Neutrino Observatory opened a new window to some of the most extreme regions of our universe. Neutrinos interact only weakly with matter and therefore escape energetic and dense astrophysical environments that are opaque to electromagnetic radiation. In addition, at PeV (10$^{15}$~eV) energies, extragalactic space becomes opaque to electromagnetic radiation due to the scattering of high-energy photons ($\gamma$-rays) on the cosmic microwave background and extragalactic background light~(EBL, e.g.,~\cite{Franceschini:2008tp}, see also Figure\ \ref{fig:horizons}). This leaves neutrinos as unique messengers to probe the most extreme particle accelerators in the cosmos - the sources of the ultra-high-energy cosmic rays (CR). There, CR with energies of more than 10$^{20}$~eV are produced, which is a factor of 10$^{7}$ times higher than the particle energy reached in the most powerful terrestrial particle accelerators. 

CR produce high-energy neutrinos through the interaction with ambient matter or radiation fields, either in the sources or during propagation in the interstellar and intergalactic medium. Unlike the charged CR, neutrinos are not deflected by magnetic fields on the way to the Earth, but point back to their source, thus resolving the long-standing question of CR origin(s). The power of this approach was recently demonstrated by IceCube and its multi-messenger partners, when a single high-energy neutrino was observed in coincidence with the flaring $\gamma$-ray blazar TXS~0506+056, identifying what appears to be the first, known extragalactic source of high-energy CR~\cite{IceCube:2018dnn, IceCube:2018cha}.   

IceCube instruments a gigaton of the very deep and clean South Pole ice. It has been taking data in full configuration since May 2011 with a duty cycle of about 99\%. 
With one cubic-kilometer instrumented volume, IceCube is more than an order of magnitude larger than previous and currently operating neutrino telescopes (Baikal Deep Under-water Neutrino Telescope~\cite{Belolaptikov:1997ry}, ANTARES~\cite{Collaboration:2011nsa}, AMANDA~\cite{Halzen:1998bp}). 
It has collected neutrino induced events with up to 10 PeV in energy, corresponding to the highest energy leptons ever observed and opening new scientific avenues not just for astronomy but also for probing physics beyond the Standard Model of particle physics (see, e.g., \cite{Aartsen:2017kpd}). In addition, its high uptime and low detector noise make it a valuable asset to search for and detect the MeV energy neutrinos from a Galactic supernova, thus providing a high-uptime alert system for what is expected to be a once-in-a-lifetime event. 

So far, the distribution of astrophysical neutrinos on the sky indicates an extragalactic origin. Given the limited statistics that IceCube collects at the highest energies, the identification of steady sources requires a very long integration time and the vast majority of flaring sources escape detection altogether. While the initial association of a cosmic neutrino with a blazar has been an essential first step, the sources of the bulk of the cosmic neutrino flux observed by IceCube remain to be resolved (see Section~\ref{sec:identify_sources} for a more detailed discussion of the origin of IceCube's neutrinos). The list of candidates is long; transients such as supernovae (SNe), neutron star mergers, or low luminosity Gamma Ray Bursts (GRBs) --- or steady sources such as Active Galactic Nuclei (AGN) or starburst galaxies --- are all very well motivated. And yet, with almost a decade of IceCube data having been analyzed, the need for new, larger instruments with improved sensitivity is becoming increasingly clear. 

\begin{figure}[t]
\centering\includegraphics[width=1\linewidth]{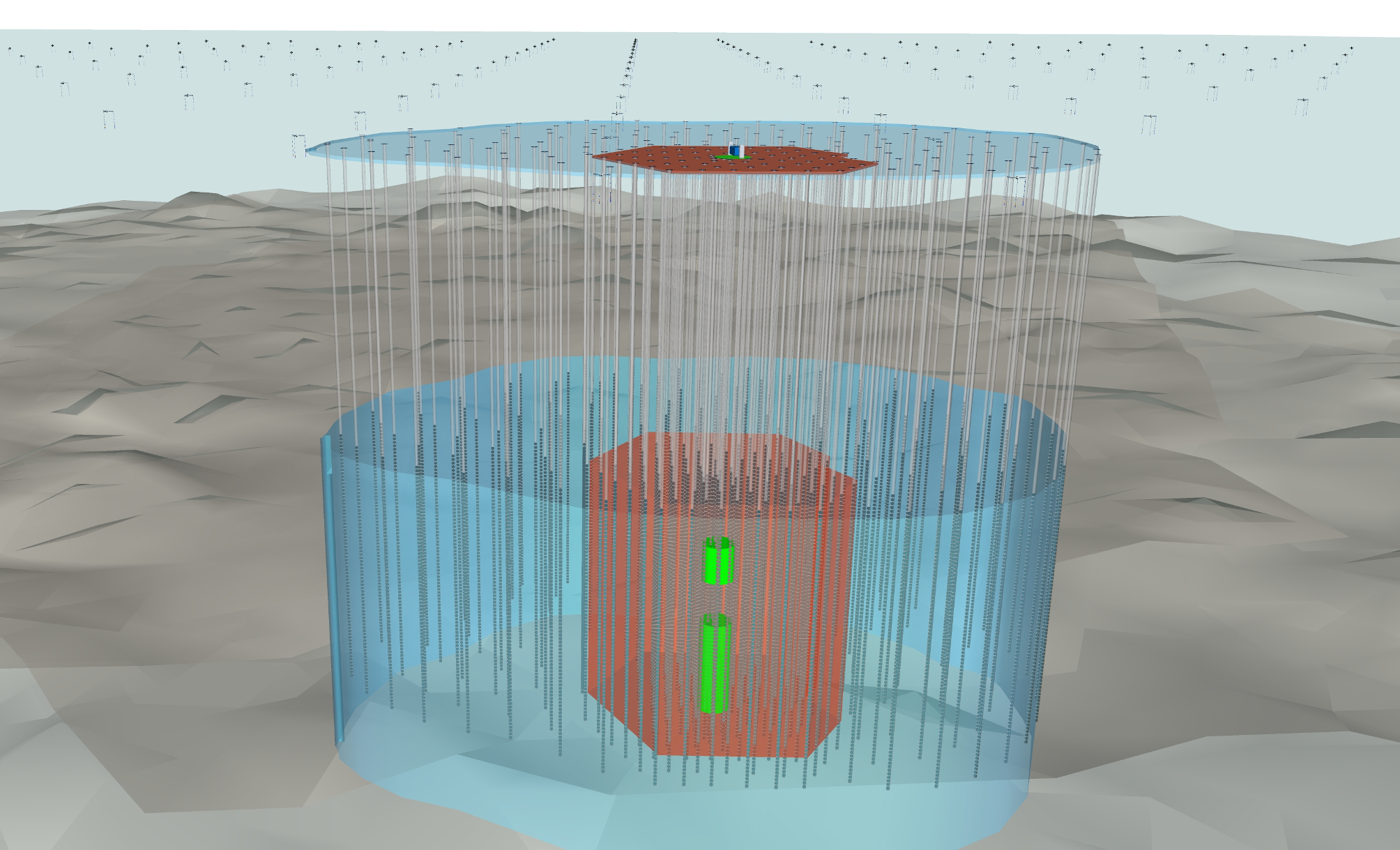}
\caption{Schematic drawing of the IceCube-Gen2 facility including the optical array (blue shaded region) that contains IceCube (red shaded region) and a densely instrumented core installed in the IceCube Upgrade (green shaded region). A surface array covers the footprint of the optical array. The stations of the giant radio array deployed at shallow depths and the surface extend all the way to the horizon in this perspective.}
\label{fig:icecube-gen2}
\end{figure}

With IceCube-Gen2 we propose a detector of sufficient volume to increase the neutrino collection rate by an order of magnitude. Meanwhile, the KM3NeT and GVD detectors under construction in the Mediterranean Sea and in Lake Baikal, respectively, target the size of one cubic-kilometer. They will complement IceCube-Gen2 in terms of sky coverage \cite{Adrian-Martinez:2016fdl, Avrorin:2014vca}, and will achieve astrophysical neutrino detection rates comparable to the present IceCube. Elaborate multi-messenger studies that combine information from other observatories, ranging from $\gamma$-rays to radio and also including gravitational waves, continue to provide opportunities for more associations of high-energy neutrinos with their sources.

IceCube-Gen2 will be a unique wide-band neutrino observatory (MeV--EeV)  (see Figure~\ref{fig:icecube-gen2} for a schematic overview) that employs two complementary detection technologies for neutrinos --- optical and radio, in combination with a surface detector array for CR air showers  --- to exploit the enormous scientific opportunities outlined in this document.

The IceCube-Gen2 facility will integrate the operating IceCube detector together with four new components: (1) an enlarged in-ice optical array, complemented by (2) a densely instrumented low-energy core, (3) the high-energy radio array, and (4) the surface CR detector array.  
Construction of the low-energy core has already started as part of the IceCube Upgrade project~\cite{Ishihara:2019aao} that is a (smaller) realization of the PINGU concept~\cite{Aartsen:2014oha}, with completion expected in 2023.
Hence, we focus  in this document on the science and instrumentation of the IceCube-Gen2 components for the detection of high-energy (TeV--EeV) neutrinos: the optical array, including its surface component, and the radio array.

After a brief review of IceCube and neutrino astronomy today (Section \ref{sec:icecube}), we lay out the science opportunities provided through the IceCube-Gen2 project (Section \ref{sec:Gen2Science}). Design considerations, performance studies, R\&D activities and logistic considerations are presented in Section \ref{sec:gen2_design}. The role of IceCube-Gen2 within the global landscape of observatories in astronomy and astroparticle physics is briefly discussed in Section \ref{sec:landscape} which  concludes this white paper.  A brief Glossary is also included at the end of this document.

\newpage
\section{IceCube and the discovery of high-energy cosmic neutrinos}
\label{sec:icecube}

IceCube was built between 2004 and 2010, financed by a Major Research Equipment and Facilities Construction (MREFC) grant from the U.S. National Science Foundation (NSF) with contributions from the funding agencies of several countries around the world. IceCube instruments one cubic kilometer of the deep glacial ice near the Amundsen-Scott South Pole Station in Antarctica. A total of 5160 digital optical modules (DOMs), each autonomously operating a 25 cm photomultiplier tube (PMT) in a glass pressure housing~\cite{Aartsen:2016nxy}, are currently deployed at depths between 1450\,m and 2450\,m along 86 cables (``strings'') connecting them to the surface. The glacial ice constitutes both the interaction medium and support structure for the IceCube array. Cherenkov radiation emitted by secondary charged particles, produced when a neutrino interacts in or near the active detector volume, carries the information on the neutrino's energy, direction, arrival time, and flavor. Digitized waveforms from each DOM provide the record of the event signature in IceCube, including the arrival time and number of the detected Cherenkov photons  (measured as  charge signals in the PMTs). 

\begin{figure}[tb]
\centering\includegraphics[width=\textwidth]{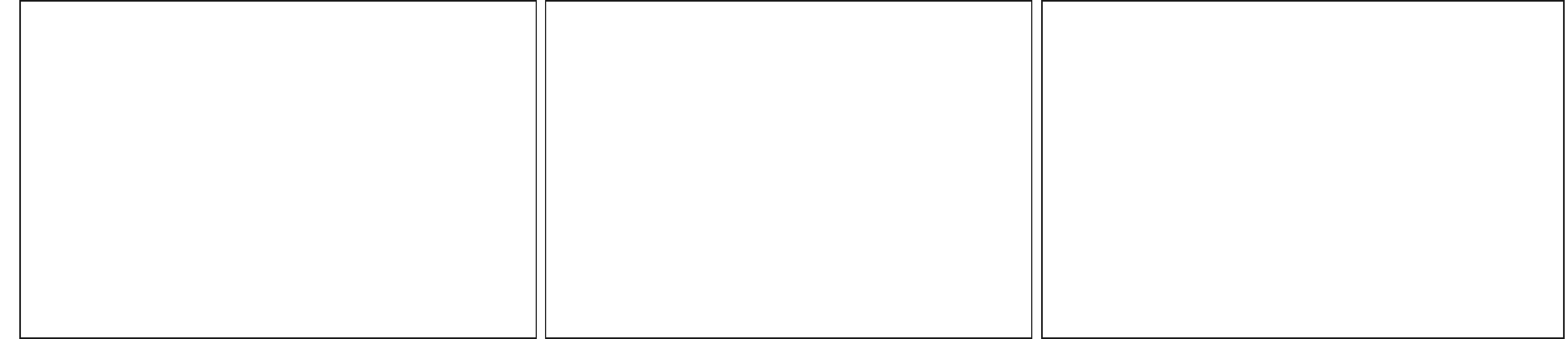}
\caption{Example event signatures observed by IceCube: a track-like event (left), a shower-like event (middle) and a simulated double-bang event (right). Each colored sphere marks a DOM that records Cherenkov light. The size of the spheres represents the amount of light that was observed. The colors indicate the relative time of the photons with respect to each other. Early photons are red, late photons are blue.}
\label{ic_event_top}
\end{figure}

IceCube records events at a rate of about 2.5~kHz, with the vast majority being muons from CR air showers. Only about one in a million events is a neutrino, most of them produced in the Earth's atmosphere, also from CR air showers. Yet, an unprecedentedly large sample of neutrinos is collected at this most remote place on Earth: $\sim$10$^5$~yr$^{-1}$, of which $\sim$30~yr$^{-1}$ are identified with high confidence as having astrophysical origin. The light deposition patterns from the recorded neutrino events fall into  three main event categories. Examples for each category are shown in Figure \ref{ic_event_top}: Track-like events from the charged-current interaction of muon neutrinos; cascade-like events from all neutrino flavors;  more complicated event signatures from very high-energy tau neutrinos, such as the so-called `double-bang' event shown in the figure, which are observed in rare cases.

\begin{figure}[hbt!]
\centering\includegraphics[width=1.0\linewidth]{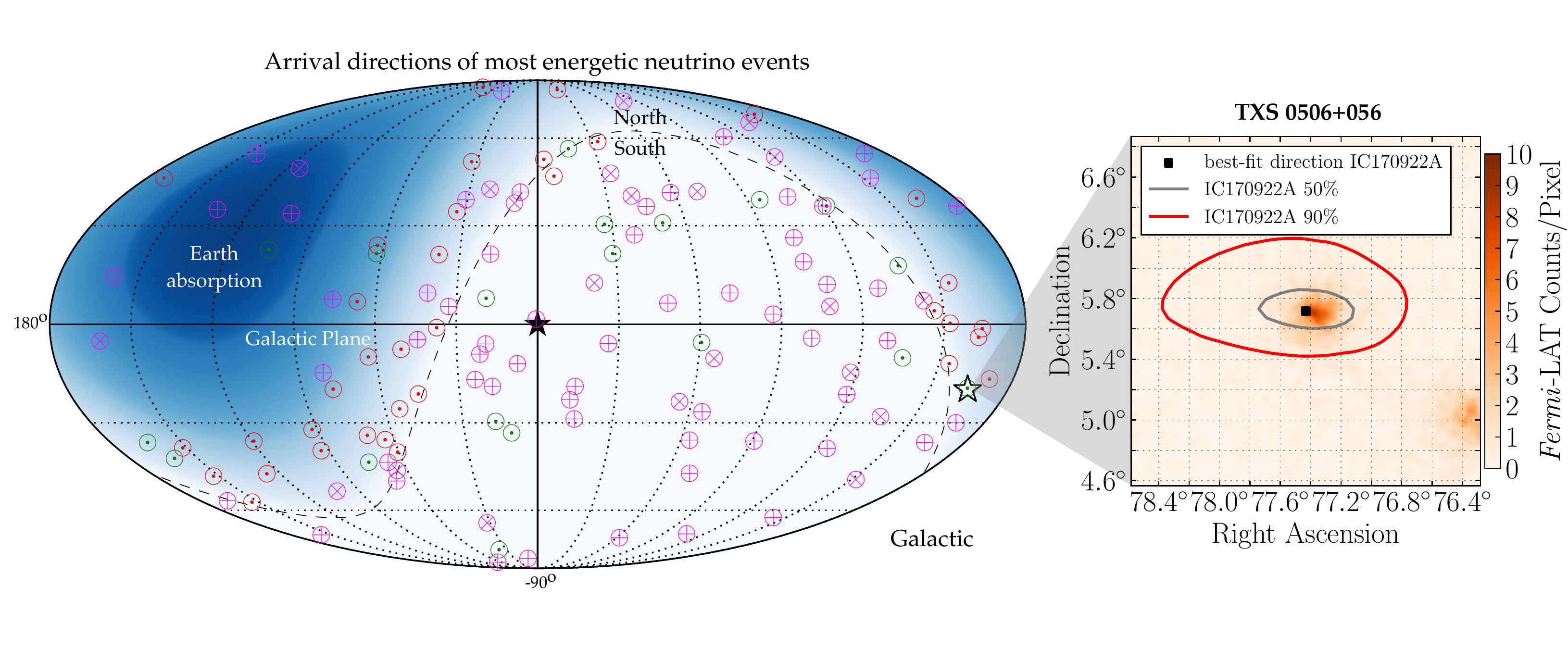}
\caption{A sky map of highly energetic neutrino events detected by IceCube. Shown are the best-fit directions for upgoing track events~\cite{Aartsen:2016xlq,Haack:2017dxi} collected in 8 years of IceCube operations (\textcolor{red}{$\odot$}), the high-energy starting events (HESE) (tracks \textcolor{magenta}{$\otimes$} and cascades \textcolor{magenta}{$\oplus$})~\cite{Aartsen:2014gkd,Kopper:2015vzf,Kopper:2017zzm} collected in 6 years, and additional track events published as public alerts (\textcolor{darkgreen}{$\odot$})~\cite{Smith:2012eu} since 2016. Note that the angular resolution for the different event categories varies from $\lesssim$1~deg for high-quality track events to $\gtrsim$10~deg for cascade-type events. The distribution of the events is consistent with isotropy once detector acceptance and neutrino Earth absorption are taken into account. The location of the first candidate neutrino source, the blazar TXS~0506+056, is marked with a star. Shown in the inset are the related \emph{Fermi Large Area Telescope}~(LAT) measurements of the region centered on TXS~0506+056 around the time that the high-energy neutrino IC-170922A was detected by IceCube (September 2017)~\cite{IceCube:2018dnn}.  The uncertainty on the reconstructed arrival direction of IC-170922A is shown for reference.}
\label{fig:skymap}
\end{figure}

IceCube has collected neutrino-induced events up to at least 10 PeV in energy, corresponding to the highest-energy neutrinos ever observed and opening new scientific avenues not just for astronomy but also for probing physics beyond the Standard Model of particle physics (see, e.g., \cite{Aartsen:2017kpd}). Evidence for astrophysical neutrinos comes from several independent detection channels, including: cascade-like events~\cite{Aartsen:2020aqd}, events that start inside the instrumented volume~\cite{Aartsen:2014gkd}; and through-going tracks~\cite{Aartsen:2015rwa}. IceCube has also observed two candidates for tau-neutrino events~\cite{Stachurska:2019srh} that are not expected to be produced in the atmosphere through conventional channels.

The significance for the cosmic origin of the observed neutrinos has collectively reached a level that puts it beyond any doubt. A decade of 
IceCube data taking has demonstrated the means to study the flavor composition of the cosmic neutrino flux via independent channels of tracks, cascades, the tau neutrino candidates, and one observed electron anti-neutrino candidate at the Glashow resonance of 6.3 PeV~\cite{Glashow:1960zz} to date~\cite{UHECRGlashowTalk,EPSGlashowTalk} (see Section \ref{Sec:Flavor}).
Clearly to exploit the full potential of all-flavor neutrino astronomy, much larger data samples are needed.     

\noindent
\subsection{Identifying the sources of high-energy neutrinos}
\label{sec:identify_sources}

One of the prime scientific goals of neutrino telescopes is the identification of the sources of high-energy neutrinos. However, the low statistics of such high-energy cosmic neutrinos, and the moderate angular resolution of $\sim$0.5$^{\circ}$ for track-like events from charged-current muon neutrino interactions and $\sim$10$^{\circ}$ for cascade-like events from all flavors of neutrinos, make identification of neutrino point sources challenging.
The distribution of astrophysical neutrinos to date in the sky is largely consistent with isotropy (see Figure~\ref{fig:skymap}), implying that a substantial fraction of IceCube's cosmic neutrinos are of extragalactic origin.

The most compelling evidence for a neutrino point source to date is the detection of one neutrino event (IC-170922A) in spatial and temporal coincidence with an enhanced $\gamma$-ray emission state of the blazar TXS~0506+056 \cite{IceCube:2018dnn}. Evidence for a period of enhanced neutrino emission from this source, in 2014/15, was revealed in a dedicated search in the IceCube archival data \cite{IceCube:2018cha}. The individual statistical significance of the blazar-neutrino association and the observed excess in the IceCube data alone are, respectively, of 3$\sigma$ and 3.5$\sigma$. 

Additional events of a similar nature are required to allow definitive conclusions about the production mechanism of neutrinos in blazars. At the same time, it is becoming increasingly clear that $\gamma$-ray blazars can not explain the majority of astrophysical neutrinos observed by IceCube: the number of observed coincidences is smaller than expected when compared to the total number of cosmic neutrino events~\cite{IceCube:2018dnn, Aartsen:2019gxs}. Further, a comparison of the full set of IceCube neutrinos with a catalog of $\gamma$-ray blazars does not produce evidence of a correlation and results in an upper bound of $\sim$30\% as the maximum contribution from these blazars to the diffuse astrophysical neutrino flux below 100 TeV~\cite{Glusenkamp:2015jca}. Accordingly, a blazar population responsible for the neutrinos would have to be appropriately dim in $\gamma$-rays (see, e.g., ~\cite{Halzen:2018iak, Neronov:2018wuo}). 

Another widely considered candidate source population of extragalactic neutrinos are $\gamma$-ray bursts (GRBs). Similar to blazars, the non-detection of neutrinos in spatial and temporal coincidence with GRBs over many years has placed a strict upper bound of 1\% for the maximum contribution from observed GRBs to the diffuse flux observed by IceCube~\cite{Aartsen:2014aqy}. 

Other source populations are anticipated to contribute to the observed cosmic neutrino flux. Starburst galaxies or galaxy clusters are candidates, but these source classes are usually considered transparent to $\gamma$-rays produced in the same hadronic interactions as the neutrinos. Indirect constraints on the multi-TeV neutrino flux from such sources exist from the observations of the total extragalactic $\gamma$-ray emission in the GeV band~\cite{Ackermann:2014usa} (see Section~\ref{sec:diffuse}), indicating that at least a fraction of the observed neutrinos needs to arise from source classes that are opaque to the $\gamma$-rays produced along with the neutrinos. An example class are Tidal Disruption Events (TDE). Recently, a potential association of a high-energy neutrino and a particular bright TDE was made \cite{Stein:2020xhk} consistent with model expectations~\cite{Winter:2020ptf,Murase:2020lnu}. 

IceCube has shown that the nature of the neutrino sky is complex, with the question of the sources and production mechanisms of the high-energy neutrino flux as yet largely unresolved and among the pressing unknowns in astronomy. IceCube-Gen2, through its larger size and improved technology (see Section \ref{sec:performance}), is designed to achieve a sensitivity 5 times that of IceCube, bringing within reach the goal of uncovering and disentangling the prospective populations of sources (\cite{Ackermann:2019ows}, see also Section \ref{sec:SourcePopulations}).

\subsection{The energy spectrum and flavor composition of cosmic neutrinos}
\label{sec:diffuse}

\begin{figure}[tb]
\centering\includegraphics[width=0.53\linewidth]{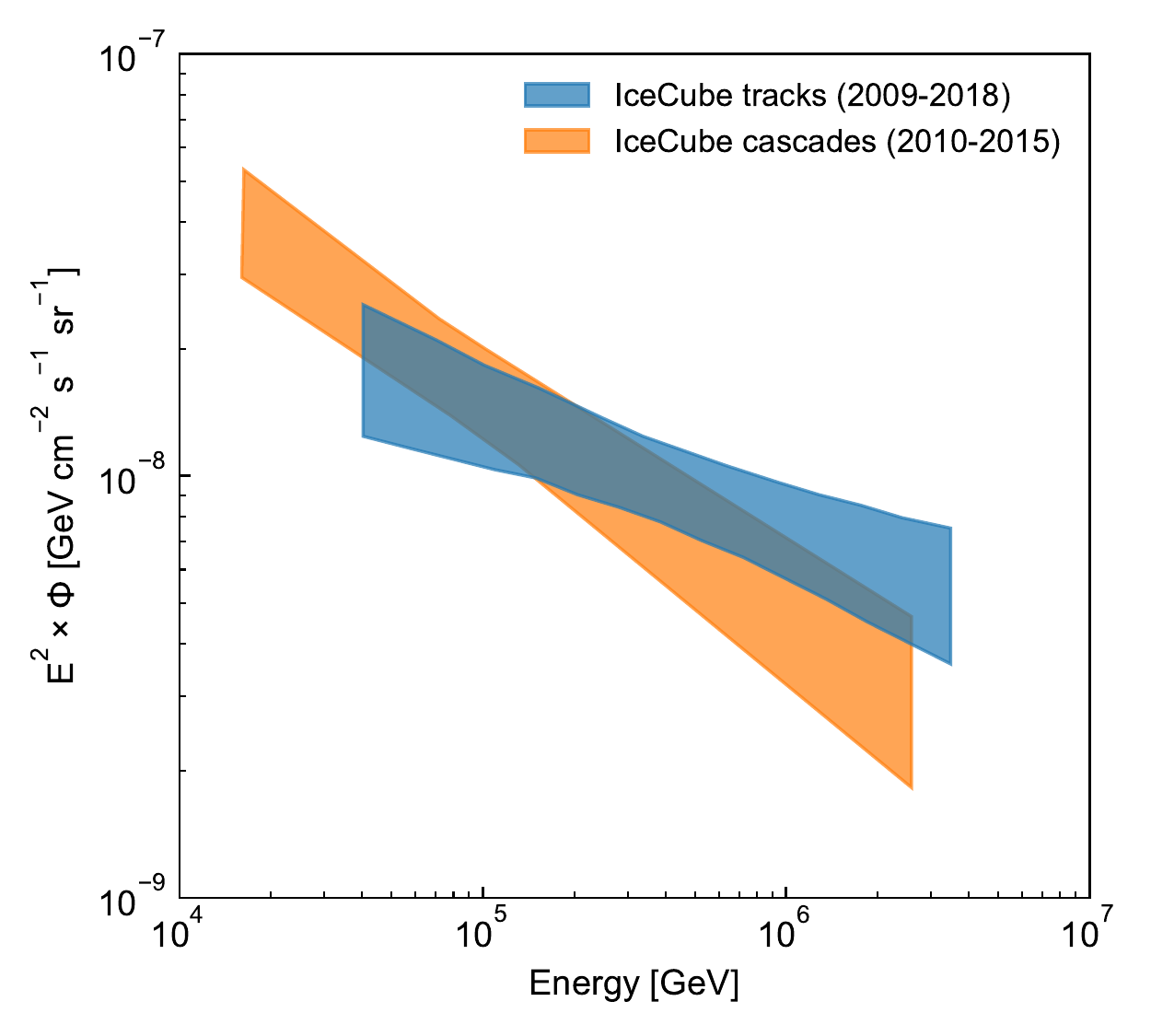}
\includegraphics[width=0.46\linewidth]{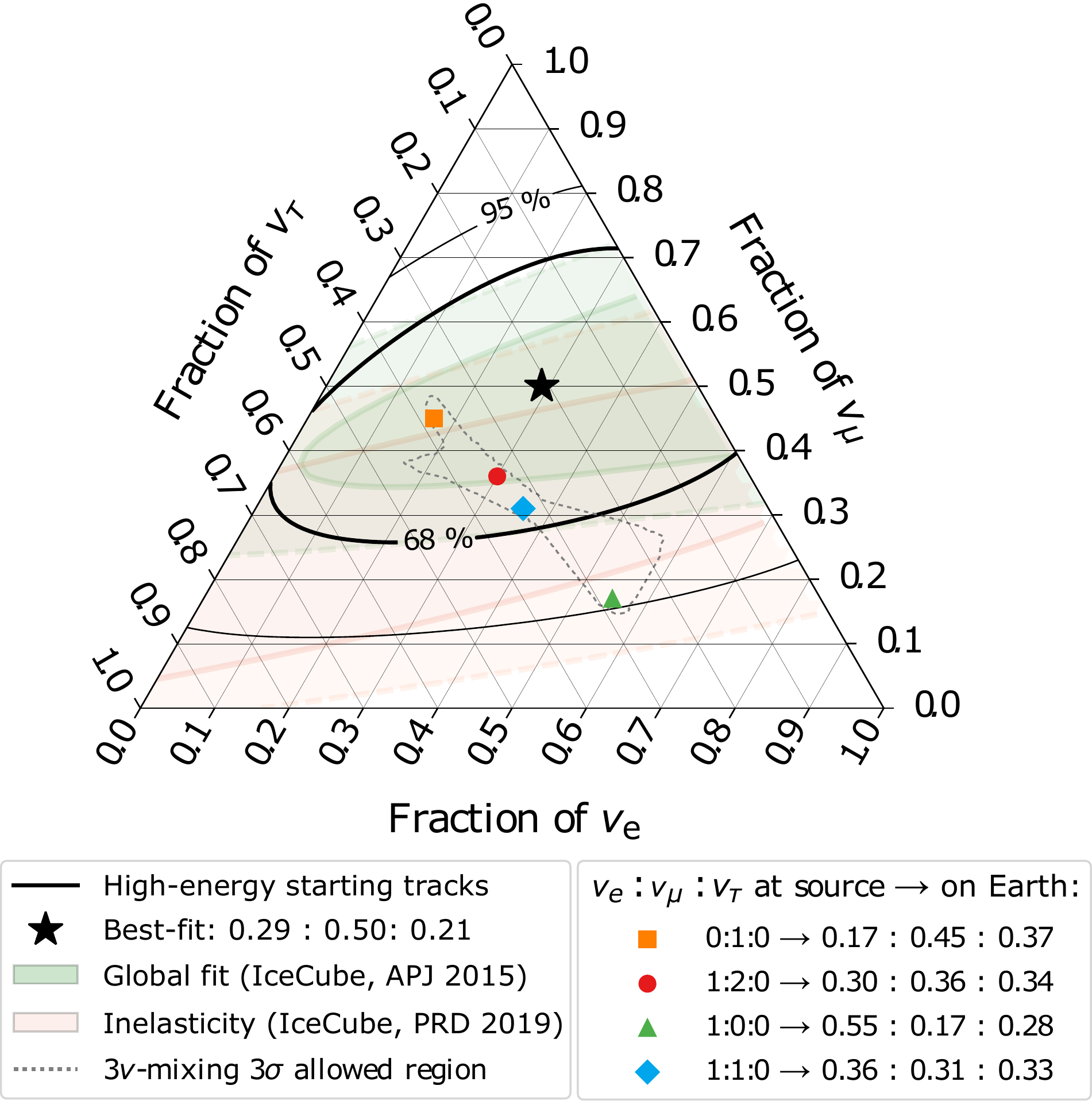}
\caption{{\it Left:} Spectrum of cosmic neutrinos measured in two independent detection channels assuming a power-law shape of the spectrum.  One measurement is based on neutrino-induced cascades collected over 6 years~\cite{Aartsen:2020aqd} (orange band), the other is based on the analysis of close to 10 years of through-going muons~\cite{Stettner:2019tok} (blue band). $\Phi$ corresponds to the per-flavor flux assuming a flavor composition of $\nu_{e}:\nu_{\mu}:\nu_{\tau}=1:1:1$. {\it Right:} Flavor constraints on the cosmic neutrino flux from various analyses of IceCube data. The preliminary constraints from an analysis identifying IceCube's first tau neutrino candidates~\cite{Stachurska:2019srh} is shown as black contours. Constraints from earlier measurements, a fit encompassing several IceCube datasets~\cite{Aartsen:2015knd}  and an analysis of the inelasticity distribution of IceCube high-energy events~\cite{Aartsen:2018vez} are shown as shaded regions. They are compared to different scenarios of neutrino production in astrophysical sources and the full range of possible flavor compositions assuming Standard Model flavor mixing (gray dotted region).}
\label{fig:icecube-spectrum}
\end{figure}

The spectrum and flavor composition of the diffuse cosmic neutrino flux, generated by all the sources that cannot be resolved, contain important information about the acceleration mechanisms, source environments, and population properties. A combined analysis of all available IceCube data in 2015 resulted in a spectrum consistent with an unbroken power law  with best-fit spectral index of -2.50~$\pm$~0.09 above 20~TeV~\cite{Aartsen:2015knd}. Newer measurements of the spectrum in individual detection channels are consistent with this early measurement.  An analysis of cascade-type events of all flavors collected by IceCube between 2010 and 2015 finds a spectral index -2.53~$\pm$~0.07 above 16~TeV~\cite{Aartsen:2020aqd}, while a preliminary analysis of high-energy muon tracks collected over almost 10 years finds the slightly harder spectral index of -2.28~$\pm$~0.09 above 40~TeV~\cite{Stettner:2019tok}  (see Figure \ref{fig:icecube-spectrum}). 
While no neutrinos have been observed by IceCube with inferred energies substantially above 10~PeV, searches for ultra-high-energy (UHE) neutrinos in this energy range have already placed significant constraints~\cite{Aartsen:2016ngq,Aartsen:2018vtx} on the composition of UHE cosmic rays, the redshift evolution of their sources (see also discussion below in this section), and generic astrophysical sources producing such UHE neutrinos (e.g.,~\cite{Murase:2014foa,Fang:2013vla}).

The flavor composition is only beginning to be meaningfully constrained by IceCube data. So far it is compatible with a standard astrophysical production scenario, the production of neutrinos in decays of pions and muons that have not been subject to significant previous energy loss. There, a flavor ratio of $\nu_{e}:\nu_{\mu}:\nu_{\tau} = 1:2:0$ at the source is expected. Neutrino oscillations change this into $\nu_{e}:\nu_{\mu}:\nu_{\tau} \approx 1:1:1$ at Earth \cite{Aartsen:2015knd}. 

\begin{figure}[t]\centering
\includegraphics[width=0.9\linewidth]{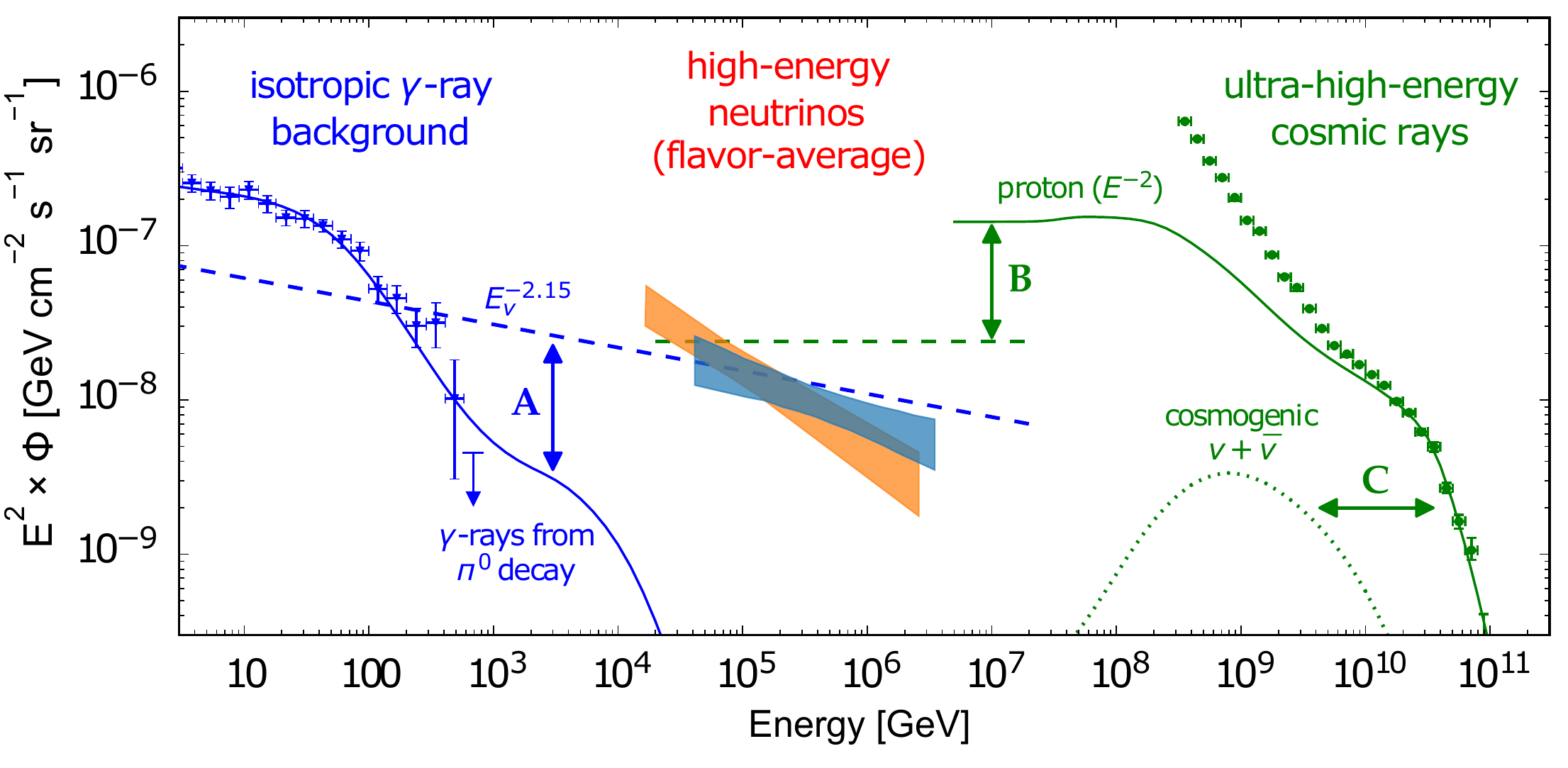}\\[-0.4cm]
\caption[]{The per-flavor flux ($\phi$) of neutrinos \cite{Aartsen:2020aqd,Stettner:2019tok} (orange and blue bands) compared to the flux of unresolved extragalactic $\gamma$-ray emission~\cite{Ackermann:2014usa} (blue data) and UHE cosmic rays~\cite{Valino:2015zdi} (green data). We highlight two upper limits on the neutrino flux (dashed lines) predicted by multi-messenger models~\cite{Murase:2013rfa,Ahlers:2018fkn}.}\label{fig:diffuse}
\end{figure}

Measurements of the isotropic neutrino flux ($\phi$) are shown in Figure~\ref{fig:diffuse}, along with the observed isotropic $\gamma$-ray background and the UHE cosmic-ray flux.  The correspondence among the energy densities (proportional to $E^2\phi$) observed in neutrinos, $\gamma$-rays, and UHE cosmic rays suggests a strong multi-messenger relationship. We highlight three areas: 

{\bf A) The multi-TeV range:} The simultaneous production of neutral and charged pions in CR interactions, which decay into $\gamma$-rays and neutrinos, respectively, suggests that the sources of high-energy neutrinos could also be strong 10~TeV~--~10~PeV $\gamma$-ray emitters. For extragalactic scenarios, this $\gamma$-ray emission is not directly observable because of the strong absorption on extragalactic background photons, resulting in $e^+e^-$ pair production. High-energy $\gamma$-rays initiate electromagnetic cascades of repeated inverse-Compton scattering and pair production that eventually contribute to the diffuse $\gamma$-ray flux below 100~GeV provided the source environment is transparent to such $\gamma$-rays. This leads to a theoretical constraint on the diffuse neutrino flux from $\gamma$-ray transparent sources~\cite{Berezinsky:1975zz,Mannheim:1998wp}.
The high flux of neutrinos below 100~TeV implied by the measurement of a spectral index significantly softer than -2 indicates that at least some neutrino sources are opaque to $\gamma$-rays \cite{Murase:2015xka,Bechtol:2015uqb}.  

{\bf B) The PeV universe:} Precision measurements of the neutrino flux can test the idea of cosmic particle unification, in which sub-TeV $\gamma$-rays, PeV neutrinos, and UHE cosmic rays can be explained simultaneously~\cite{Murase:2016gly,Fang:2017zjf,Kachelriess:2017tvs,Resconi:2017zkg}.
If the neutrino flux is related to the sources of UHE cosmic rays, then there is a theoretical upper limit (the dashed green line in Figure~\ref{fig:diffuse}) to the neutrino flux~(\cite{Ahlers:2018fkn}, also \cite{Waxman:1998yy,Bahcall:1999yr}). 
UHE cosmic-ray sources can be embedded in environments that act as ``CR reservoirs'' where magnetic fields trap CR with energies far below the highest CR energies. The trapped CR collide with gas and produce a flux of $\gamma$-rays and neutrinos at PeV energies. The measured IceCube flux is consistent with predictions of some of these models~\cite{Loeb:2006tw,Murase:2008yt,Kotera:2009ms}; see, however, \cite{Anchordoqui:2018qom}. The precise characterization of the spectrum and flavor composition beyond the energy range currently accessible by IceCube goes hand in hand with resolving the sources, as the combination of the two will provide novel avenues for  understanding the most extreme particle accelerators in the universe.

{\bf C) Ultra-high energies (UHE)}: The attenuation of UHE cosmic rays through resonant interactions with cosmic microwave background photons is the dominant mechanism for the production of UHE neutrinos during the propagation of the CR in the universe. This mechanism, first pointed out by Greisen, Zatsepin and Kuzmin (GZK), would cause a suppression of the UHE cosmic-ray proton flux beyond $5\times10^{10}$~GeV~\cite{Greisen:1966jv,Zatsepin:1966jv} and gives rise to a flux of UHE neutrinos~\cite{Beresinsky:1969qj} that is shown in Figure~\ref{fig:diffuse}, but has not yet been detected.
The observation of these \emph{cosmogenic} neutrinos in addition to the potential direct identification of astrophysical sources or transients producing neutrinos at $\sim$EeV energies, or a stringent upper limit on their flux, will provide information on the cosmological evolution of UHE cosmic-ray sources and restrict the models of acceleration, spectrum and composition of extragalactic CR  (e.g., \cite{Beresinsky:1969qj,Berezinsky:1975zz,Stecker:1978ah,Hill:1983xs,Yoshida:1993pt,Engel:2001hd,Anchordoqui:2007fi,Takami:2007pp,Ahlers:2009rf,Ahlers:2010fw,Kotera:2010yn,Yoshida:2012gf,Ahlers:2012rz,Aloisio:2015ega,Heinze:2015hhp,Romero-Wolf:2017xqe,AlvesBatista:2018zui,Moller:2018isk,vanVliet:2019nse,Heinze:2019jou}).      

To make significant progress on the above questions a future detector should provide several times higher neutrino statistics in the PeV range, flavor identification capabilities, and expand IceCube's energy range to provide sensitivity for neutrinos with 10$^{18}$~eV (1~EeV) at an energy flux level $E^{2}\Phi <$~10$^{-9}$ GeV~s$^{-1}$~cm$^{-2}$~sr$^{-1}$. Radio detection techniques are being developed that measure the radio pulses generated in the particle cascades induced by neutrino interactions in the ice to deliver a cost-effective way to explore the EeV energy range. The IceCube-Gen2 radio array is designed to provide the necessary sensitivity at EeV energies to study the cosmic neutrino flux and astrophysical neutrino sources directly connected to the highest energy CR. The design of the IceCube-Gen2 optical array to provide a 5 times higher sensitivity for detecting neutrino sources will increase the rate of detected neutrinos in the PeV range by a factor 10 compared to IceCube.

\subsection{Exploring fundamental physics with high-energy neutrinos}

\begin{figure}
    \centering
    \includegraphics[width=0.99\textwidth]{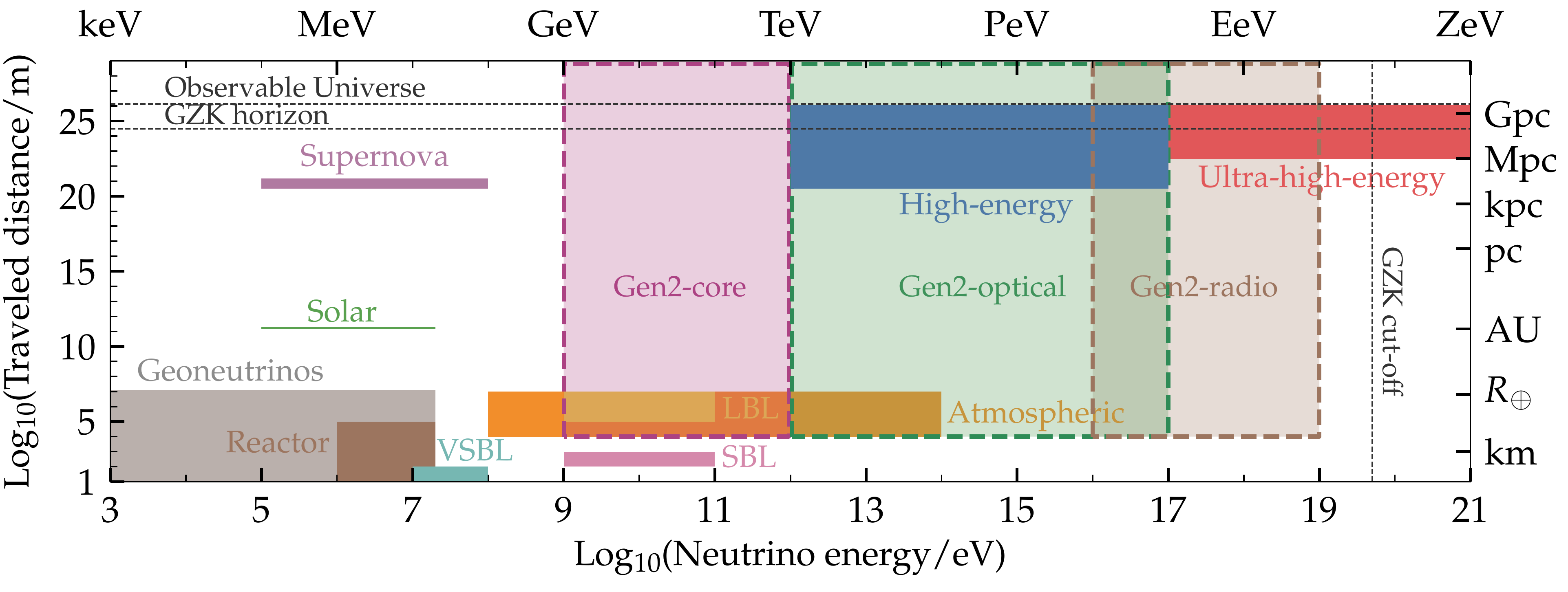}
    \caption{Range of travel distances and energies for neutrinos of different origin that are used for tests of fundamental physics. The IceCube-Gen2 observatory will cover a large range of energies and distances, observing both atmospheric and cosmic neutrinos.}
    \label{fig:nu_physics_overview}
\end{figure}

The flux of high-energy atmospheric and very high-energy cosmic neutrinos is invaluable not just for investigations of extreme, astronomical sources, but also for probing fundamental properties of the neutrino itself. Often these probes require the observation of neutrinos over specific travel distances (baselines) and in specific energy ranges (e.g., ~\cite{fundamentalWhitepaper,Arguelles:2019rbn}).

The neutrino interaction cross-section can be probed indirectly through absorption in the Earth (depending on the energy and angle dependent matter column), allowing us to test Standard Model predictions, and constrain hypothesized beyond Standard Model (BSM) physics, including new spatial dimensions and leptoquarks \cite{Aartsen:2017kpd}. The flavor composition of the cosmic neutrino flux, predicted to lie in a narrow range for various source models and standard neutrino oscillations, probes BSM physics and the structure of space-time itself through propagation effects over cosmic baselines. 

Below 100 TeV, the large statistics of atmospheric neutrino events observed by open water/ice detectors yields sensitivity to  anomalous oscillation signatures for which some of the strongest constraints originate from IceCube data. These include signatures due to additional sterile neutrinos~\cite{TheIceCube:2016oqi,Aartsen:2017bap,Aartsen:2020iky,Aartsen:2020fwb}, violation of Lorentz Invariance~\cite{Aartsen:2017ibm}, or previously unobserved neutrino production channels such as forward charm production in the atmosphere~\cite{Aartsen:2016xlq}. In addition, neutrinos are favorable messengers to search for signatures from the annihilation or decays of heavy dark matter with masses beyond about 100 TeV~\cite{Gondolo:1991rn,Murase:2012xs,Rott:2014kfa}. Finally, neutrino detectors are excellent instruments to search for  hypothesized exotic particles which would leave distinct traces in the detector, such as magnetic monopoles or supersymmetric charginos (e.g.,~\cite{Aartsen:2015exf,Albert:2017fud}).

The general requirements for future BSM measurements are similar to those for astrophysics: a much enlarged sample of cosmic neutrinos, and; an extension of the energy range beyond~10$^{18}$~eV. IceCube-Gen2 is ideally positioned to provide both of these elements through the combination of optical and radio detection methods (see also Section~\ref{sec:BSMPhysics}).
Figure~\ref{fig:nu_physics_overview} shows the unique coverage of both the high-energy and cosmological baseline domains that IceCube-Gen2 will provide.

\subsection{Summary of objectives and requirements for a next generation neutrino observatory}
\label{sec:objectives}

The broader science motivation for a next generation neutrino observatory follows from the recent observations (see above), and has been articulated by the astronomical community as part of the Astro2020 Decadal Survey. These contributions pursue a diverse set of research topics, focusing on neutrino astronomy~\cite{Ackermann:2019ows},  fundamental physics with cosmic neutrinos~\cite{fundamentalWhitepaper},  cosmic ray science~\cite{Astro2020_GCR_WhitePaper, 2019astro2020T.459F, 2019BAAS...51c..93S}, extragalactic sources ~\cite{2019BAAS...51c..92R,2019BAAS...51c.228S,2019BAAS...51c.396V},  Galactic sources~\cite{2019BAAS...51c.115C,2019BAAS...51c.194N}, multi-messenger studies with $\gamma$-rays~\cite{2019BAAS...51c.553V,2019BAAS...51c.267H,2019BAAS...51c.431O,2019astro2020T.427C} and multi-messenger studies with gravitational waves~\cite{2019BAAS...51c..38B, 2019BAAS...51c.239K, 2019BAAS...51c.247F}.

The scientific goals can been grouped according to the following topics:
\begin{enumerate}
  \item {\it Resolving the high-energy neutrino sky from TeV to EeV energies:} What are the sources of high-energy neutrinos detected by IceCube? The IceCube-Gen2 sensitivity should allow for identifying realistic candidate source populations . 

\item {\it Understanding cosmic particle acceleration through multi-messenger observations:} This involves studying particle acceleration and neutrino emission from a range of multi-messenger sources (e.g., AGN, GRBs, TDEs, SNe or kilonovae). Constraints on the physics within these sources can also come from measurements of the spectrum and flavor composition of the astrophysical neutrino flux.

\item {\it Revealing the sources and propagation of the highest-energy particles in the universe:} This includes studying Galactic and extragalactic cosmic ray sources and their neutrino emission, as well as the propagation of cosmic rays through the measurement of cosmogenic neutrinos. 

\item {\it Probing fundamental physics with high-energy neutrinos:} This entails the measurement of neutrino cross sections at energies far beyond the reach of particle accelerators, searching for new physics from neutrino flavor mixing over cosmic baselines, and searches for heavy dark matter particles.\\
\end{enumerate}

These goals have been translated into the following requirements for the IceCube-Gen2 detector: 

\begin{enumerate}
\item A neutrino point source sensitivity at least 5 times better than the current IceCube array (essential for  goals 1 and 2).
\item The reconstruction of individual high-energy neutrinos in near real-time with sub-degree resolution to enable follow-up observations and multi-messenger astronomy (essential for goals 1 and 2).
\item An order-of-magnitude higher collection rate than the current IceCube array for neutrinos of all flavors in the energy range 100 TeV to 10 PeV (essential for goals 2, 3 and 4). 
\item The expansion of the energy range beyond 10$^{18}$~eV with two orders of magnitude better sensitivity than what is currently available (essential for goals 3 and 4).
\item The enhancement of the sensitivity to neutrino flavors and improved ability for flavor identification (essential for goals 3 and 4).

\end{enumerate}

The baseline design for IceCube-Gen2 meets these requirements and is described in Section \ref{sec:gen2_design}. The scientific potential of the detector is discussed in Section \ref{sec:Gen2Science}.

\newpage

\section{IceCube-Gen2: Exploring the cosmic energy frontier}
\label{sec:Gen2Science}

IceCube-Gen2 is designed to observe the neutrino sky from TeV to EeV energies with a sensitivity to individual sources at least five times better than IceCube. It will collect at least ten times more neutrinos per year than IceCube and enable detailed studies of their distribution on the sky, energy spectrum, and flavor composition, as well as tests of new physics on cosmic baselines. In this section, we focus on the expected impact of the IceCube-Gen2 observatory in the young field of neutrino and multi-messenger astronomy. It is structured according to the four key science objectives defined in Section~\ref{sec:objectives}: resolving the high-energy sky (Section~\ref{sec:MappingSources}), understanding the cosmic particle acceleration (Section~\ref{sec:UnderstandingSources}), revealing the sources and propagation of cosmic rays (Section \ref{sec:UHECRSources}), and  probing fundamental physics with high-energy neutrinos (Section~\ref{sec:BSMPhysics}). The performance of IceCube-Gen2 shown in the various figures refers to the expected combined performance of the radio and optical array unless stated otherwise. 
A comprehensive overview of the science case for the study of fundamental neutrino properties with GeV neutrinos in a densely instrumented core was presented in \cite{Aartsen:2014oha,TheIceCube-Gen2:2016cap}. A description of the CR science using the surface instrumentation will follow in a separate publication.

\subsection{Resolving the high-energy sky from TeV to EeV energies}
\label{sec:MappingSources}
Neutrinos are the only messengers that can directly reveal the remote sites --- beyond our local universe --- where CR are accelerated to PeV and EeV energies. IceCube has been successful in finding first evidence for particle acceleration in the jet of an active galactic nucleus. However, ultimately, it is not sensitive enough to detect even the brightest  neutrino sources with high significance, or to detect populations of less luminous sources.

\subsubsection{Detection of persistent and transient sources}
\label{sec:Sensitivities}

The proposed IceCube-Gen2 observatory combines an 8~km$^3$ array for the detection of optical Cherenkov light with a 500~km$^2$ radio array for the detection of ultra-high-energy neutrinos. An angular resolution of 10 arcmin at PeV energies for the optical array and several degrees above 100 PeV energies for the radio array will ensure that individual neutrinos are well localized on the sky and can be correlated with potential counterparts in the electromagnetic spectrum. This will allow for sources to be distinguished from diffuse backgrounds. Details about the instrumentation and performance can be found in Section~\ref{sec:gen2_design}.

\begin{figure}[tb]
\centering \includegraphics[width=.99\textwidth]{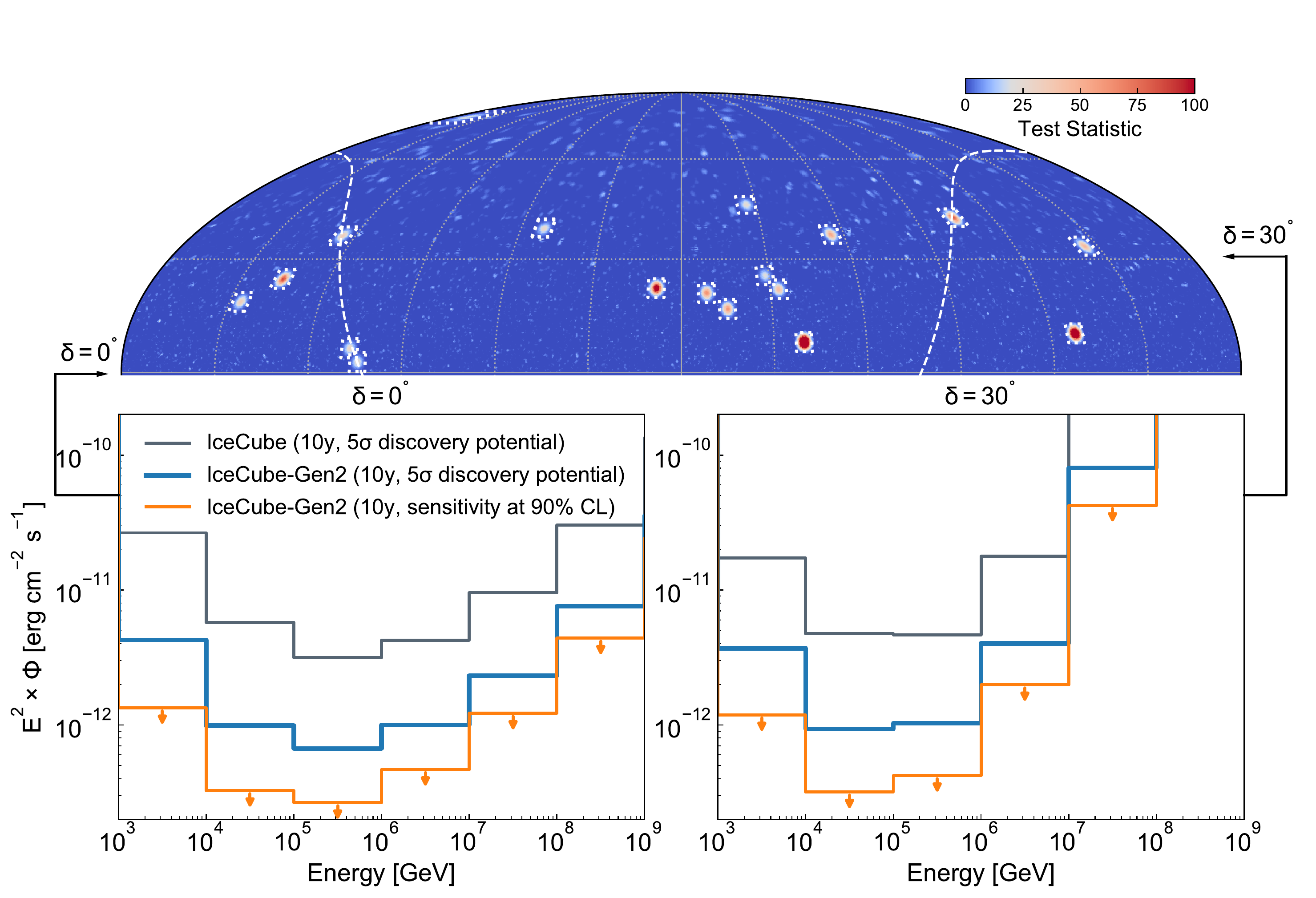}
\caption{Visualization of source detection capabilities expected for IceCube-Gen2. Source positions on the sky and intensities have been selected randomly from an intensity distribution expected for sources with a constant density in the local universe, and consistent with current IceCube neutrino flux constraints. Shown is the test statistic value determined in a mock-simulation of track-like events that can be obtained at the source position after 10 years of operation of IceCube-Gen2. For better visibility, the region around the sources (indicated by white dotted lines) has been magnified. The position of the Galactic plane is shown as a dashed curve. Below the map, differential sensitivities for the detection of point sources ($5\sigma$ discovery potential, and sensitivity at 90\% CL) are shown for two selected declinations, at the celestial horizon and at $\delta=30^{\circ}$. Absorption of neutrinos in the Earth limits the sensitivity in the PeV energies for higher declinations. The IceCube and IceCube-Gen2 sensitivities are calculated separately for each decade in energy, assuming a differential flux $dN/dE \propto E^{-2}$ in that decade only. Neutrino fluxes are shown as the per-flavor sum of neutrino plus anti-neutrino flux, assuming an equal flux in all flavors. The curves refer to the optical array only.}
\label{fig:gen2_sky}
\end{figure}

IceCube-Gen2 will allow the observation of sources at least five times fainter than sources observable with currently operating detectors. An impression of the neutrino sky that can be expected in the IceCube-Gen2 era is presented in Figure~\ref{fig:gen2_sky}. It shows a test statistic map obtained from the simulation of the arrival direction of tracks for a detector as sensitive as IceCube-Gen2 searching for point sources of neutrinos. The neutrino flux of the simulated sources has been chosen randomly from a model extragalactic source population that has a number density distribution expected of sources having an uniform density and luminosity in the local universe. The intensity of the model sources is consistent with current constraints from IceCube observations. Potential Galactic sources as discussed below in Section~\ref{sec:galacticSources} have been added. The differential sensitivity curves in Figure \ref{fig:gen2_sky} for two selected declinations allow a quantitative evaluation of the source detection potential of IceCube-Gen2. They refer to the sensitivity of the optical array only, as there are large uncertainties on the sensitivity of the radio array for long-duration observations of steady neutrino sources due to the unknown backgrounds at these energies, from, e.g., diffuse astrophysical and cosmogenic neutrinos.

 IceCube-Gen2 reaches its peak sensitivity in the region around the celestial equator.  Due to the huge atmospheric backgrounds and the increased absorption in the Earth at high neutrino energies, the sensitivity below 100 TeV is largest for events from the Northern Hemisphere, while above a few PeV, mainly the southern sky is observed.
Between 100~TeV and 1~PeV the Northern Hemisphere averaged 5$\sigma$ discovery potential for a neutrino energy flux is $1.3 \times 10^{-12}$~ergs cm$^{-2}$ s$^{-1}$ ---  similar to the energy flux level current generation high-energy and very-high-energy $\gamma$-ray telescopes can detect in the GeV to TeV range.

As $\gamma$-rays and neutrinos are produced by CR in the same interaction processes their energy fluxes are expected to be similar at production.  However, due to absorption of $\gamma$-rays in the sources and the intergalactic medium the photons are reprocessed to the GeV and TeV bands (or absorbed, in which case the neutrino energy flux could be even higher than the $\gamma$-ray energy flux). Consequently, IceCube-Gen2 will be able to measure or constrain CR acceleration processes for thousands of known $\gamma$-ray sources, as well as searching for cosmic accelerators opaque to high-energy electromagnetic radiation. 

Short, second-to-day-scale transients like GRBs, compact object mergers, or core-collapse supernovae (CCSN) explosions are different from persistent sources. Backgrounds from diffuse neutrinos, air showers, thermal and anthropogenic noise are usually negligible when searching for a short burst of neutrinos; therefore, the sensitivity scales differently with effective area, volume, and angular resolution than for persistent sources. 

\begin{figure}[tb]
\centering \includegraphics[width=.59\textwidth]{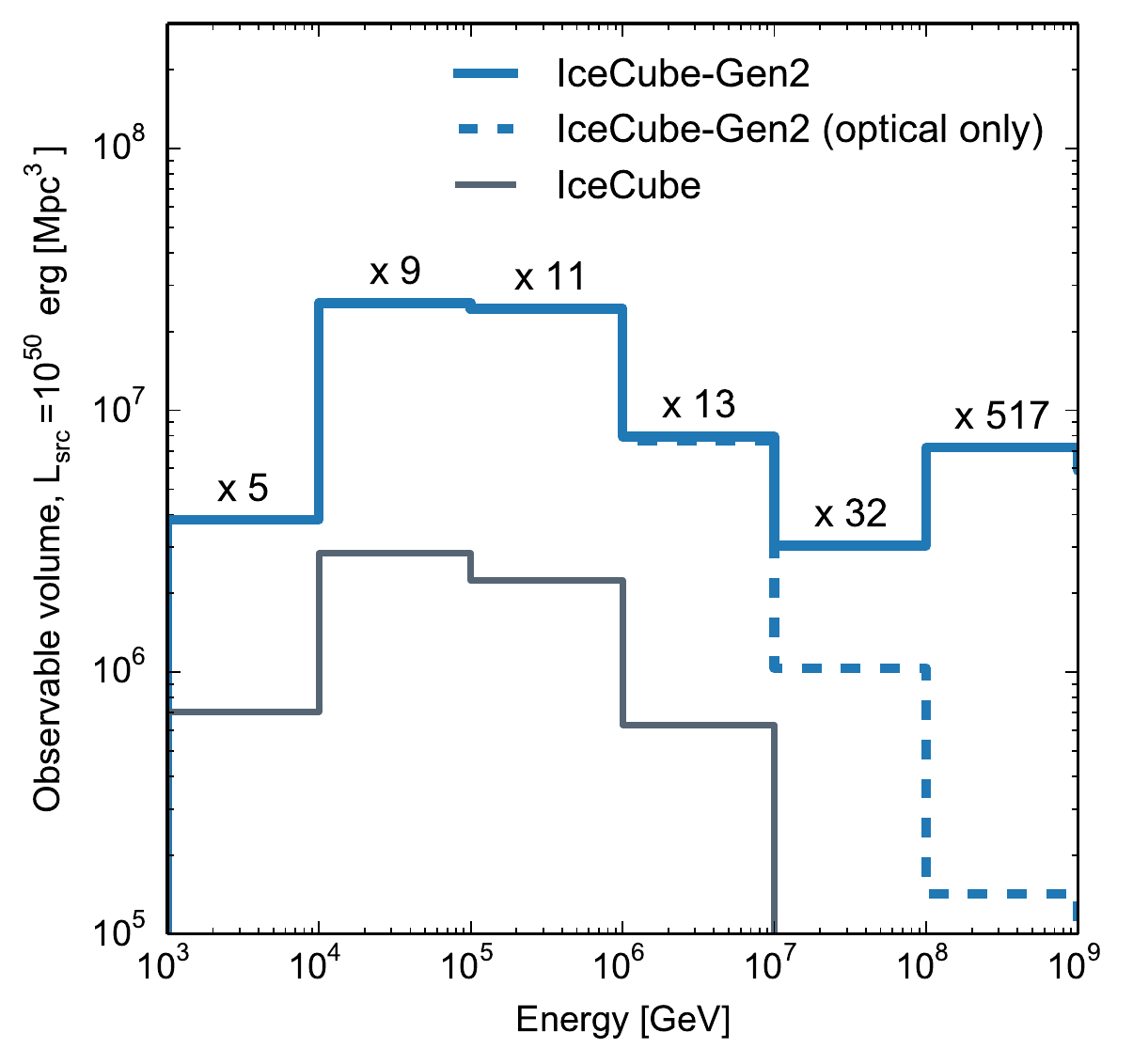}
\caption{Observable volume of IceCube and IceCube-Gen2 for a generic 100~s burst  with equivalent isotropic emission of $10^{50}$~erg in neutrinos. The observable volumes are calculated separately for each decade in energy, assuming a neutrino spectrum of $dN/dE \propto E^{-2}$ in that decade only and an equal flux of neutrinos in all flavors.}
\label{fig:gen2_transsens}
\end{figure}

An important performance measure for transient events is the volume within the universe in which they can be observed. Figure \ref{fig:gen2_transsens} shows the observable volume of the universe for IceCube-Gen2 in comparison to IceCube for a generic 100~s burst with equivalent isotropic emission of $10^{50}$~erg in neutrinos as a function of energy. An order-of-magnitude increase in observable volume is expected for energies up to 10~PeV compared to IceCube, while at energies above 100~PeV the radio array will allow for the first time the observation of a relevant portion of the universe. The observable volume of up to few times $10^7$~Mpc$^3$ for such a burst is similar to the one that gravitational wave detectors will reach in the next decade for the detection of binary neutron star mergers~\cite{LIGOScientific:2019vkc}.

\subsubsection{Detectability of source populations}
\label{sec:SourcePopulations}

\begin{figure}[tb]
\centering\includegraphics[width=0.49\linewidth]{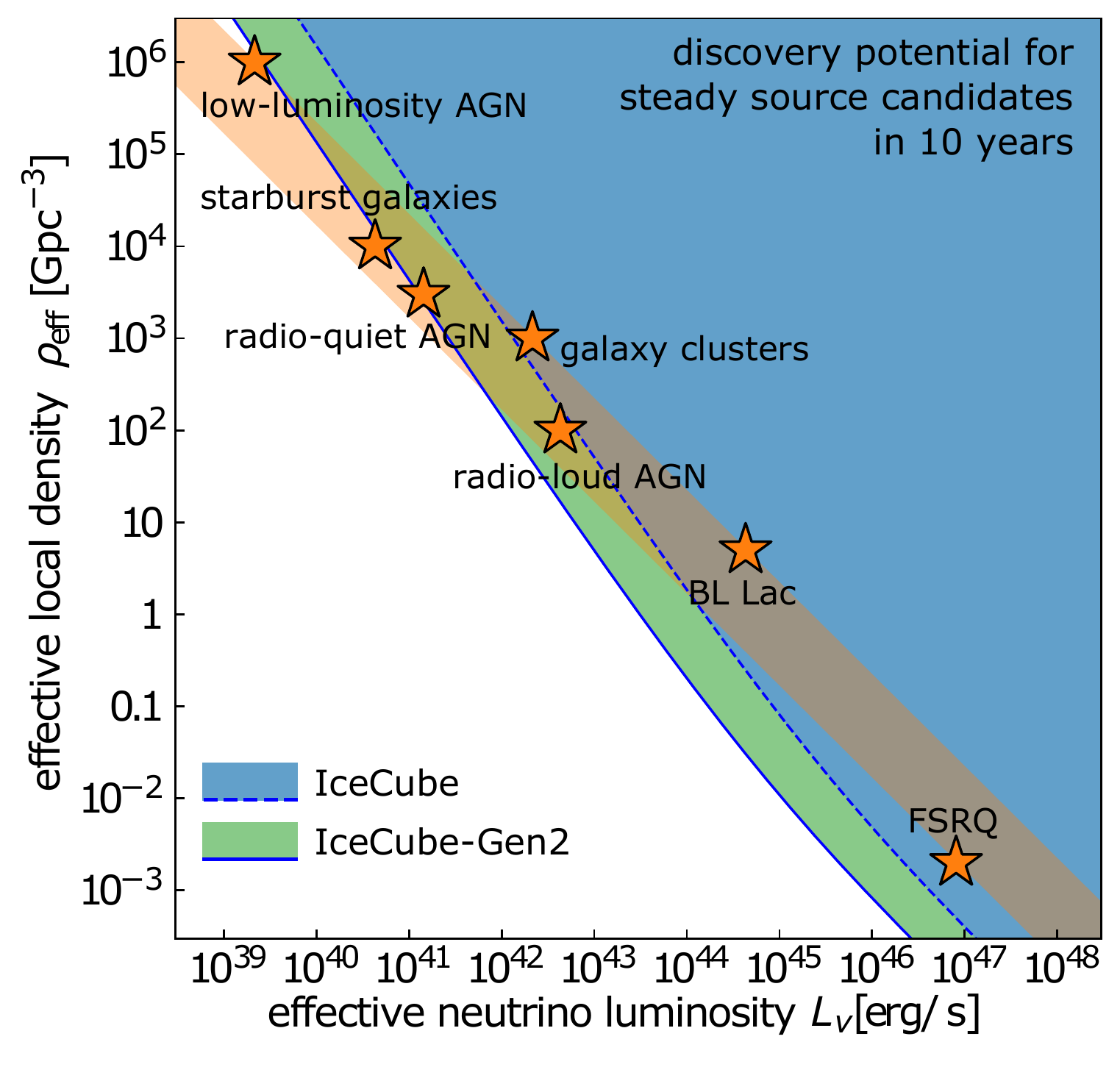}
\includegraphics[width=0.49\linewidth]{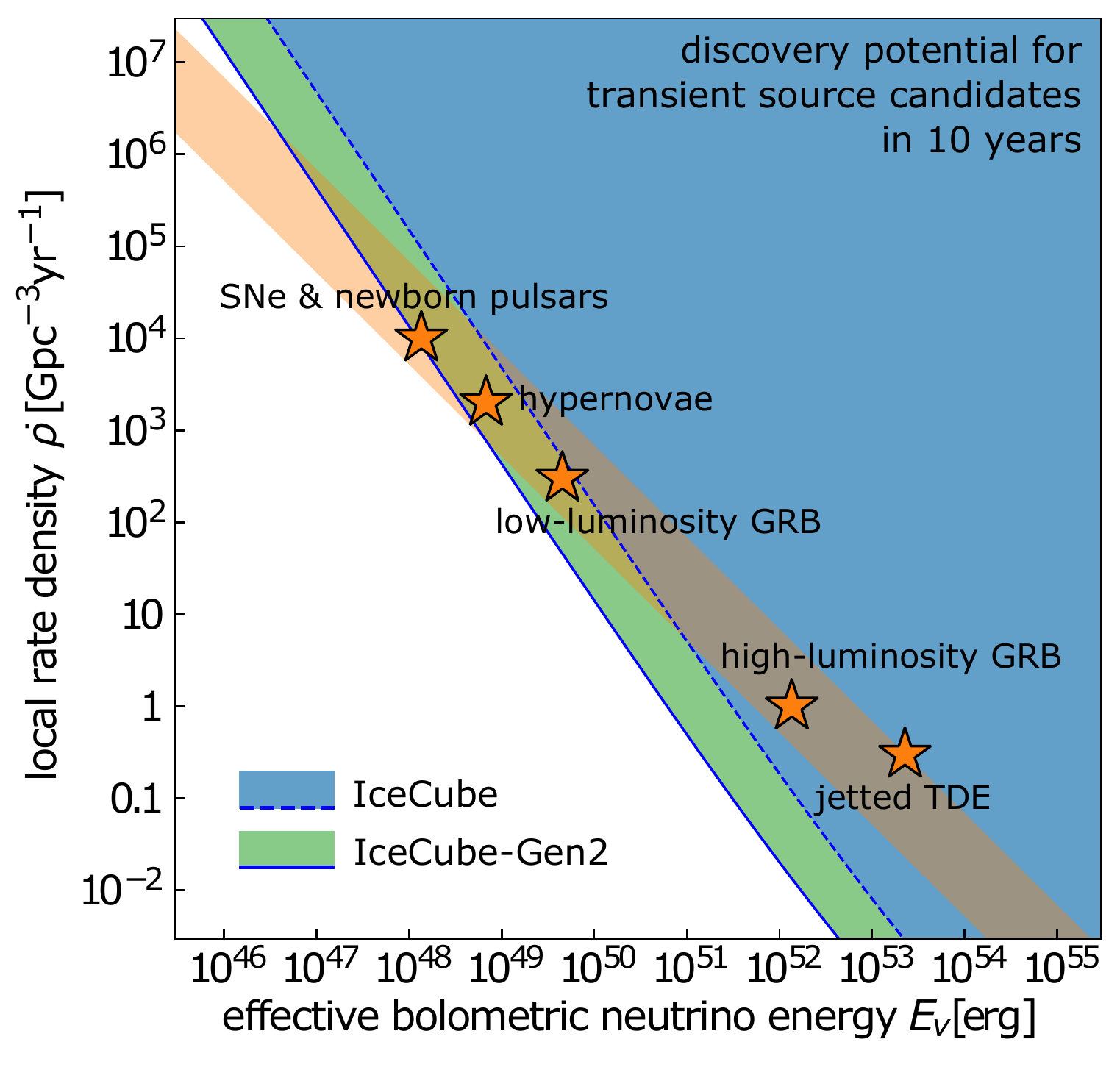}
\caption{{\bf Left:} Comparison of the effective local density and luminosity of extragalactic neutrino source populations to the discovery potential of IceCube and IceCube-Gen2. We indicate several candidate populations ($\medwhitestar$) by the required neutrino luminosity to account for the full diffuse flux~\cite{Murase:2016gly} (see also \cite{Silvestri:2009xb}). The orange band indicates the luminosity / density range that is compatible with the total observed diffuse neutrino flux. The lower (upper) edge of the band assumes rapid (no) redshift evolution. The shaded regions indicate IceCube's (blue, dashed line) and IceCube-Gen2's (green, solid line) ability to discover one or more sources of the population ($E^2\phi_{\nu_\mu+\bar\nu_\mu}\simeq 10^{-12}~{\rm TeV}/{\rm cm}^2/{\rm s}$ in the Northern Hemisphere~\cite{Aartsen:2018ywr}). {\bf Right:} The same comparison for transient neutrino sources parametrized by their local rate density and bolometric energy~\cite{Murase:2018utn}. The discovery potential for the closest source is based on 10 years of livetime ($E^2F_{\nu_\mu+\bar\nu_\mu}\simeq 0.1~{\rm GeV}/{\rm cm}^2$ in the Northern Hemisphere~\cite{Meagher:2017htr}). Only the IceCube-Gen2 optical array has been considered for this figure.}
\label{fig:population_sensitivity_ma}
\end{figure}

IceCube performed the first step towards identifying the sources of astrophysical neutrinos, by associating high-energy neutrinos with the highly luminous blazar TXS~0506+056. IceCube's capability of identifying sources is limited to high-luminosity neutrino sources that have a low density in the local universe, such as blazars, and neutrino transients with a low rate, such as GRBs. Accordingly, IceCube has set stringent constraints on the contribution of these two source populations to the observed cosmic neutrino flux (cf. Section \ref{sec:identify_sources} and references therein), thus establishing that rather lower-luminosity / higher-density populations must be responsible for the bulk of cosmic neutrinos. The brightest sources of such populations would still be below the detection threshold of IceCube and can only be identified with a more sensitive instrument. 

Figure \ref{fig:population_sensitivity_ma} compares the identification capabilities of IceCube and IceCube-Gen2 for the most common neutrino source and transient candidates. If sources like radio-quiet and/or low-luminosity AGNs, galaxy clusters, starburst galaxies, or transients like CCSNe produce the majority of cosmic neutrinos, they can be identified only with a detector with a five times better sensitivity such as IceCube-Gen2. In combination with correlation or stacking searches, IceCube-Gen2 can identify a cumulative signal from populations where the closest sources have up to 20 times fainter neutrino fluxes than point sources detectable by IceCube. So their signal remains in reach, even if several of the candidate populations contribute similar fractions to the total observed neutrino flux.

\subsection{Understanding cosmic particle acceleration through multi-messenger observations}
\label{sec:UnderstandingSources}

Multi-messenger astronomy, the combination of astrophysical observations in CR, neutrinos, photons, and gravitational waves, is a powerful new program to identify the physical processes driving the high-energy universe.  Astrophysical neutrinos can provide an unobstructed view deep into the processes powering cosmic accelerators.  Unlike their counterparts in photons and charged CR, their small cross section and absence of electric charge allow neutrinos to travel the cosmological distances necessary to reach Earth from their sources without absorption or deflection. High-energy astrophysical neutrinos are a smoking-gun signal of hadronic interactions, and will point the way to the sources of the high-energy CR.

\subsubsection{Probing particle acceleration in active galaxies}

\begin{figure}[tb]
\centering\includegraphics[width=0.99\linewidth]{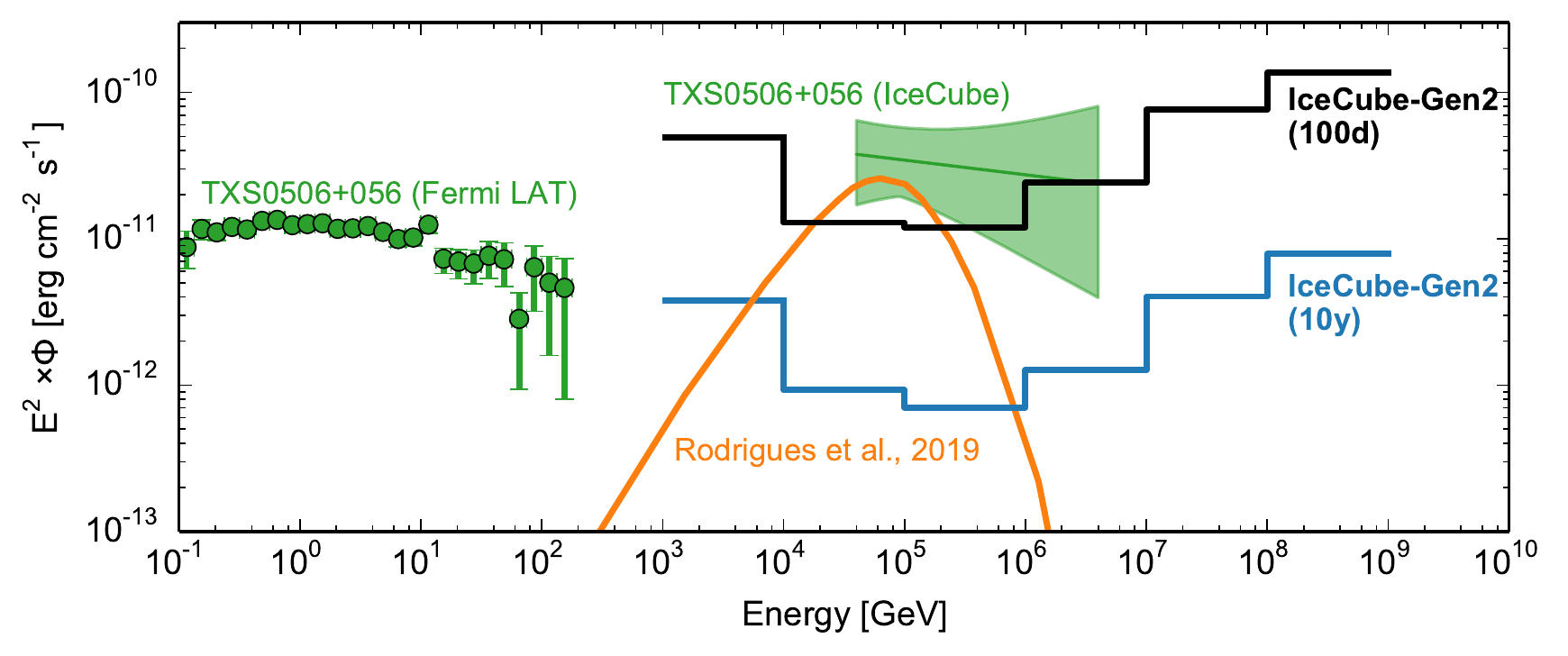}
\caption{$5\sigma$ discovery potential of IceCube-Gen2 for a flux of muon neutrinos in relation to observations of the Blazar TXS0506+056. The black and blue curves correspond to 100 days and 10 years of observations and
indicate the sensitivity for neutrino flares and the time-averaged neutrino emission, respectively. The best-fit muon neutrino flux during the 2014-2015 activity period for TXS 0506+056~\citep{IceCube:2018cha} is shown as a green band, while the green markers show the average $\gamma$-ray flux of TXS 0506+056 between 2008 to 2018 observed by \emph{Fermi}~LAT~\cite{Aartsen:2019gxs}. The orange curve corresponds to the predicted neutrino flux from modeling the multi-messenger emission during the flare period in~\cite{Rodrigues:2018tku}. Only the sensitivity of the IceCube-Gen2 optical array has been considered for this figure.}
\label{fig:agn_nu}
\end{figure}

The electromagnetic emission from the high-energy extragalactic sky is dominated by blazars, a subclass of radio-loud active galactic nuclei (AGN) powered by supermassive black holes that display relativistic jets, with one jet pointed near the line of sight of the Earth. 
The high-energy photon emission from blazars could be explained by the decay of neutral pions from energetic hadronic interactions. 
Given this extreme luminosity and the potential hadronic origin of their high-energy emission, blazars (and more generally, AGN) have long been believed to be sources of neutrinos and CR~\cite{1993A&A...269...67M, Mannheim:1995mm, 1997ApJ...488..669H}. Indeed, the observations of high-energy neutrinos from the direction of the blazar TXS~0506+056~\cite{IceCube:2018dnn} (cf. Section \ref{sec:identify_sources}) with IceCube provided strong evidence for a scenario in which CR are accelerated in AGN jets. However, much remains unknown about blazar physics that further observations with neutrinos could help answer, such as the location of the CR acceleration region in the jet, and the underlying mechanisms driving this acceleration (see, e.g.,~\cite{2019Galax...7...20B} for a recent review on modeling particle acceleration and multi-messenger emission in blazars).

\begin{figure}[tb]
\centering\includegraphics[width=0.7\linewidth]{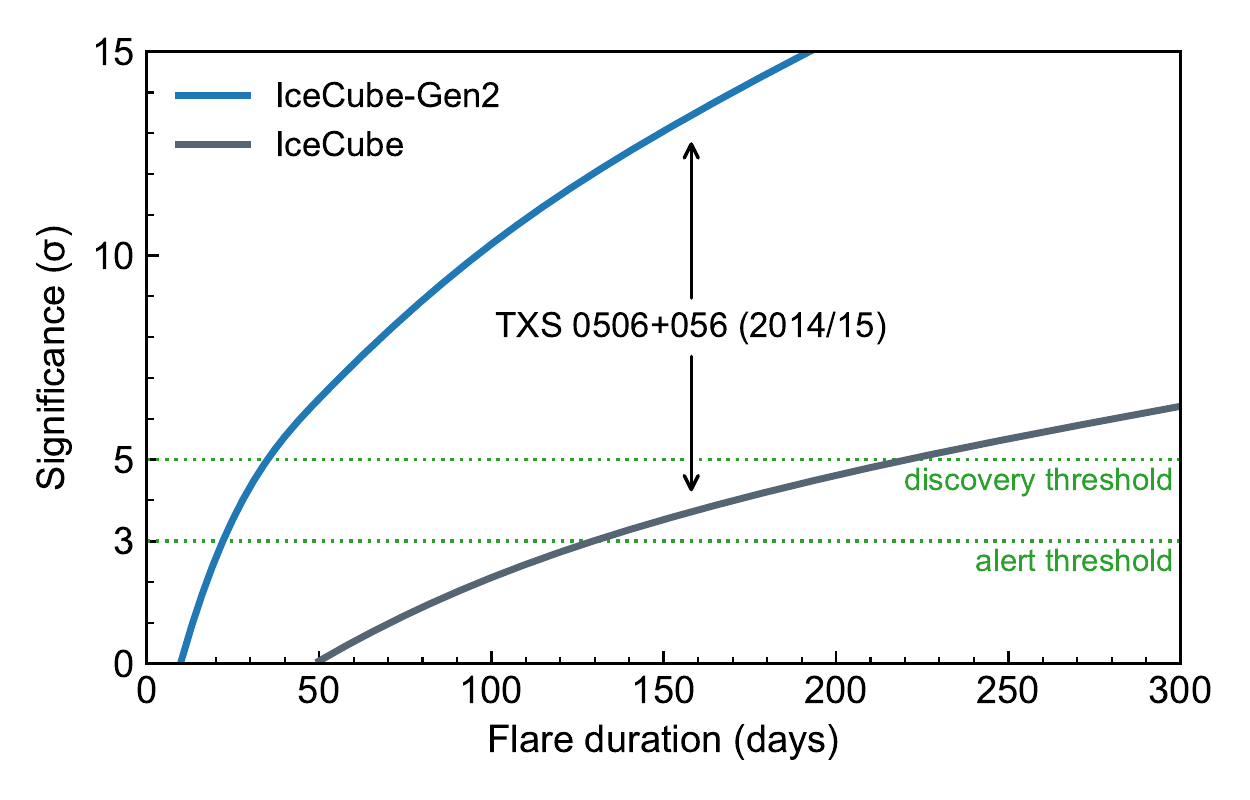}
\caption{Discovery potential of IceCube and IceCube-Gen2 for neutrino flares similar to the one observed for TXS0506+056 in 2014/15 which lasted 158 days. Shown is the projected significance of the observation as a function of the flare duration. The flux and spectral index of the assumed flare are the ones observed for TXS0506+056 (see Figure~\ref{fig:agn_nu}) and assumed constant within the flare duration, i.e., the neutrino fluence increases with flare duration.
Green dotted lines mark the 5$\sigma$ discovery threshold, as well as the lower threshold for sending alerts to partner telescopes for follow-up observations.}
\label{fig:flare_sensitivity}
\end{figure}

The larger astrophysical neutrino samples provided by IceCube-Gen2 will enable definitive detection of multiple neutrino flares from a population of blazars. For a TXS 0506+056-type flare with the same best-fit spectral and temporal characteristics as the one identified in 2014--2015, 38 muon-neutrino events would be expected by IceCube-Gen2, compared to the 13$\pm$5 events identified in IceCube. As shown in Figures~\ref{fig:agn_nu} and~\ref{fig:flare_sensitivity}, the improvement in sensitivity would result in a $>$~5$\sigma$ significance detection of such a flare. A time-averaged or quiescent phase neutrino emission could be detected in 10 years of IceCube-Gen2 if its power is $\gtrsim$~10\% of the average $\gamma$-ray emission observed by \emph{Fermi} LAT. Other blazars usually suggested as neutrino sources, such as Markarian 421 or 1ES 1959+650~\cite{2005ApJ...630..186R} will also be within the sensitivity of the IceCube-Gen2 detector during extended flares. Figure~\ref{fig:flare_sensitivity} demonstrates the improvements for flare detections expected with IceCube-Gen2. Using again the flux and spectral index of the TXS~0506+056 flare as a template the significance of the detection as a function of the duration of such a flare is shown (the flux is assumed constant during the duration of the flare, i.e., the neutrino fluence increases with flare duration). While the modeled flare would need to last for 220 days to be detected with IceCube at 5$\sigma$ significance, IceCube-Gen2 can already detect the modeled flare if it lasts only 36 days. Hence, IceCube-Gen2 is sensitive to neutrino flares with 6 times lower fluence than IceCube, greatly increasing the range of potentially observable medium term transients.

The blazar detections provided by IceCube-Gen2 will allow the characterization of the neutrino spectrum, and therefore the nature and maximum energy of the particles being accelerated at the source. The detections will also help in determining CR acceleration efficiency for the various sub-classes of blazars. Multi-messenger observations can be used to constrain the spectrum and density of the target photon population involved in the p$\gamma$ interactions that produce the neutrino emission, the contribution from EM cascading to the observed blazar emission, the structure and Doppler factor of the jet, and other physical parameters~\cite{Rodrigues:2018tku,2019ApJ...881...46R,2018ApJ...864...84K, 2018ApJ...863L..10A, 2019NatAs...3...88G}. As illustrated in these models, electromagnetic broadband observations that are simultaneous with the neutrino detections are crucial in understanding the hadronic emission process. In particular, X-ray and $\gamma$-ray observations will be the most sensitive probes for these types of correlated studies. 

While the highest-energy neutrinos might originate in the jets of AGNs and be observable predominantly from blazars, a substantial fraction of the observed sub-PeV and PeV neutrinos could be emitted by AGN cores~\cite{Tjus:2014dna, Kalashev:2015,Kimura:2015}. Strong thermal radiation fields can turn the cores opaque to GeV $\gamma$~rays, thus solving the puzzle that the extragalactic TeV neutrino flux is inconsistent with the extragalactic GeV $\gamma$-ray flux if the sources are transparent \cite{Murase:2015xka}. IceCube-Gen2 will be able to identify if the observed neutrino flux originates from AGN cores and/or jets via source and cross-correlation searches. 
The precise spectrum and flavor ratio measurements (see Section \ref{Sec:Spectrum}) will also enable the study of the acceleration processes and environmental conditions in AGN cores or jets, even in regions that are opaque to high-energy EM radiation.

\subsubsection{Neutrinos from gamma-ray bursts}

\begin{figure}[tb]
\centering \includegraphics[width=.75\textwidth]{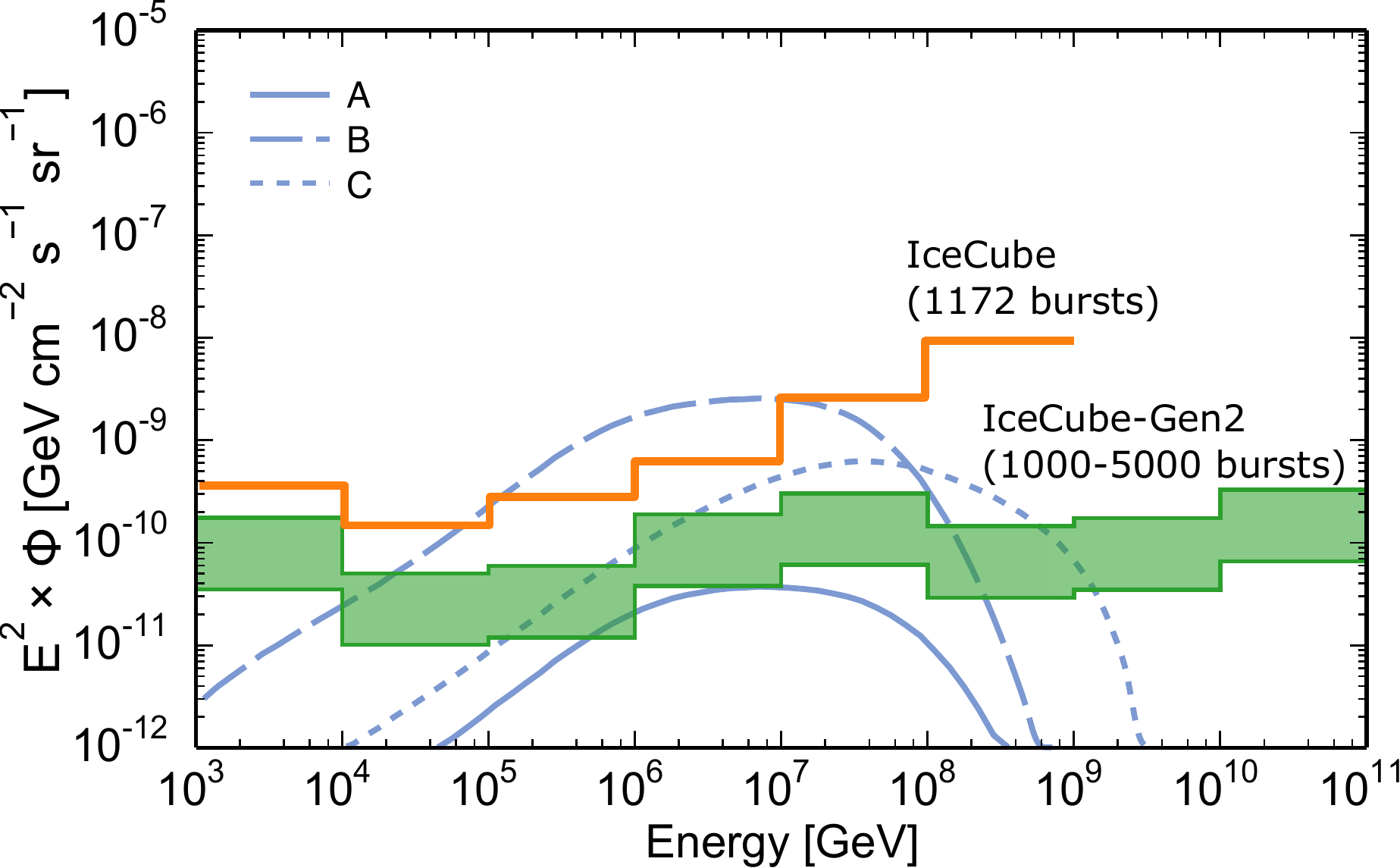}
\caption{Upper limits from IceCube (rescaled to one energy bin per decade from~\cite{2017ApJ...843..112A}) and sensitivity of IceCube-Gen2 to the diffuse neutrino flux from GRB. Also shown are three scenarios from~\cite{2018A&A...611A.101B} in which GRBs produce the UHE cosmic rays (see also ~\cite{Baerwald:2013pu,Baerwald:2014zga,Globus:2014fka,Bustamante:2014oka}).}
\label{fig:grb_limits}
\end{figure}

GRBs, either short (lasting $<$2~s) or long, have been suggested as sources of the UHE cosmic rays and high-energy neutrinos~\cite{Waxman:1995vg,Waxman:1997ti}; a prediction later revised by, e.g., \cite{2012PhRvL.108w1101H}. An alternative sub-photospheric dissipation mechanism for GRBs that also results in neutrino emission has also been proposed~\cite{2010MNRAS.407.1033B,2013PhRvL.110x1101B,1999ApJ...521..640D}. 
Long GRBs are associated with CCSNe that develop relativistic jets and short GRBs are associated with the merger of compact objects --- two neutron stars (NS-NS) and/or a neutron star and a black hole (NS-BH) --- that also develop these jets. IceCube has studied 1,172 GRBs and has not found coincident neutrino emission \cite{2017ApJ...843..112A}. This implies that GRBs contribute no more than $\sim$1\% of the diffuse neutrino flux~\cite{aartsen16}. Furthermore, in a wide range of scenarios, GRBs are constrained as a source of UHE cosmic rays~\cite{2018A&A...611A.101B}. However, these 1,172 GRBs were initially detected by satellites and are subject to selection effects: e.g., only the most luminous are found. Low-luminosity GRBs are potential neutrino and UHE cosmic-ray sources \cite{Zhang:2017moz}. IceCube-Gen2 will be able to probe the remaining viable scenarios for neutrino production by GRBs of all types. Figure~\ref{fig:grb_limits} shows the current best upper limits of IceCube and the expected sensitivity for IceCube-Gen2 on the diffuse flux from GRBs after following 1000-5000 GRBs (assuming 667 bursts/year). This can be compared to three models that assume UHE cosmic rays are produced by GRBs~\cite{2018A&A...611A.101B}).

\subsubsection{Multi-messenger sources of high-energy neutrinos and gravitational waves}

LIGO~\cite{2015CQGra..32g4001L} and VIRGO~\cite{2015CQGra..32b4001A} have revolutionized multi-messenger astrophysics with their detection of gravitational waves. The most spectacular observation to date were the joint detections of GW170817 and GRB170817A by LIGO/VIRGO and Fermi-GBM respectively, which confirmed the association of the merger of binary neutron stars with short GRBs. Interestingly, GRB170817A was probably seen off-axis with respect to the relativistic jet.
As already discussed, NS-NS and NS-BH mergers are expected to be neutrino sources.
The most promising emission scenario from short GRBs is related to their extended emission observed in $\gamma$-rays and X-rays that can last up to several hundred seconds~\cite{2017ApJ...848L...4K}.
The expected emission of high-energy neutrinos from neutron star mergers 
may be higher than inferred from $\gamma$-ray observations if the production sites are partially or fully opaque to $\gamma$-rays. This can be the case for neutron-star mergers where the dynamical/wind ejecta that produce kilonova emission absorb some of the $\gamma$-rays~\cite{2018PhRvD..98d3020K}, or for core-collapse events where the stellar envelope allows neutrinos to escape but blocks $\gamma$-rays~\cite{2001PhRvL..87q1102M,2016PhRvD..93h3003S,2012PhRvD..86h3007B}.
Within a few years, GW from 
NS-NS mergers will be detectable out to $\sim$325\,Mpc~\cite{LIGOScientific:2019vkc}, with a detection rate of 15--500~yr$^{-1}$~\cite{2018arXiv181112907T}. A study of neutrinos in coincidence with GW events complements studies of neutrinos in coincidence with GRBs observed by satellites because the observational biases are different. Compact binary mergers observed with GW may well be relatively dim in $\gamma$-rays, but they would be closer to Earth, potentially favoring neutrino detection. A Galactic core-collapse supernova is certain to be a source of MeV neutrinos and is also speculated to be a source of GW and TeV neutrinos. 

IceCube and other neutrino detectors such as ANTARES and the Pierre Auger Observatory have searched for neutrino emission in a time window of $\pm500\,$s~\cite{2016PhRvD..93l2010A,2017PhRvD..96b2005A,2017ApJ...850L..35A, 2014PhRvD..90j2002A,Albert:2018jnn,2012PhRvD..85j3004B,Bartos:2018jco}. Starting on November 2016, IceCube conducts GW-neutrino coincident searches in near-real time~\cite{2017PhRvD..96b2005A,2019arXiv190105486C,Aartsen:2020mla}.
Low-latency search for a high-energy neutrino counterpart enables well-localized GW-coincident neutrinos to reduce the sky area to be surveyed by electromagnetic telescopes in the follow-up of a GW observation.

\begin{figure}[tbp]
   \centering  
   \includegraphics[width=0.7\textwidth]{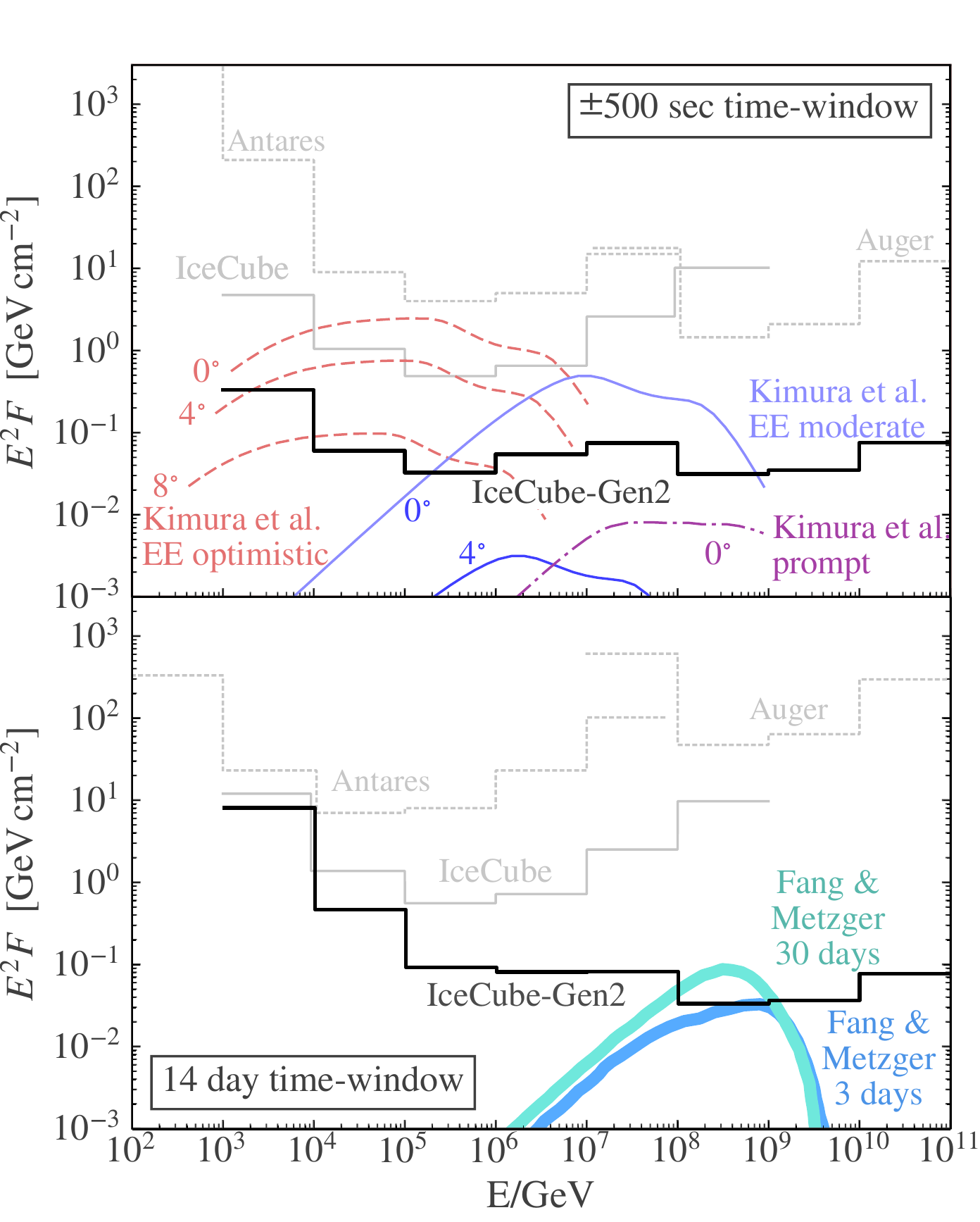}
\caption{Upper limits (at 90\% CL) from various instruments on the neutrino spectral fluence from GW170817 during a $\pm$500 s window centered on the gravitational wave (GW) trigger time (top panel), and a 14-day window following the trigger (bottom panel). For comparison, the sensitivity of IceCube-Gen2 (at 90\% CL) to an event at a similar position on the sky (solid black line) 
is presented. Also shown are several predictions by neutrino emission models \cite{2017ApJ...849..153F,2018PhRvD..98d3020K} scaled to a distance of 40 Mpc. Separate curves are displayed for different components of the emission (prompt and extended (EE)), observation angles relative to the jet axis, and time scales of the emission. See \cite{2017ApJ...850L..35A} for details. Limits and sensitivities are calculated separately for each energy decade, assuming a spectral fluence $F(E) \propto E^{-2}$ in that decade only. All fluences are shown as the per-flavor sum of neutrino and anti-neutrino fluence, assuming equal fluence in all flavors, as expected for standard neutrino oscillation parameters. Figure adopted from \cite{2017ApJ...850L..35A}.}
\label{fig:GW170817}
\end{figure}

IceCube-Gen2 will be able to probe a range of neutrino production scenarios in gravitational wave sources that IceCube is insensitive to, and provide regular multi-messenger detections for some of the (more optimistic) emission channels~\cite{2017ApJ...849..153F,2018PhRvD..98d3020K}. It will be particularly interesting to observe `gamma-dark' high-energy transients where gravitational waves and neutrinos are the only messengers to escape (e.g., \cite{2012PhRvD..86h3007B}). In Figure~\ref{fig:GW170817} we show high-energy neutrino observational constraints for the NS-NS merger GW170817, obtained by IceCube, ANTARES and the Pierre Auger Observatory. We also show the results scaled to IceCube-Gen2's sensitivity --- an improvement of over an order of magnitude with respect to IceCube. The observation of sources similar to GW170817 during IceCube-Gen2 operation will enable us to probe a broader range of models (e.g. the ``moderate" model in~\cite{2017ApJ...848L...4K}) and determine the model parameters.

The near future will see KAGRA and LIGO-India come on-line. The construction of IceCube-Gen2 will coincide with substantial development for gravitational wave detectors such as LIGOA+~\cite{LIGOAplus} and Voyager, with gravitational wave sensitivity range extending up to 700\,Mpc \cite{LIGOwhitepaper}, further increasing the detection rate by an order of magnitude.

\subsubsection{Cosmic rays from core-collapse supernovae and tidal disruption events}

Even though neutrinos from high-luminosity GRBs have not been observed so far, a large population of low-luminosity GRBs could contribute significantly to the cosmic neutrino flux~\citep{murase06,Murase:2013ffa}. It is speculated that "choked'' jets, where the relativistic jet fails to penetrate the progenitor star, may explain relativistic SNe and low-luminosity GRBs, providing a unified picture of GRBs and SNe~\citep{2001PhRvL..87q1102M,2016PhRvD..93h3003S}.  This scenario could be physically probed by the detection of high-energy neutrinos in coincidence with SNe containing relativistic jets~\citep{Kowalski:2007xb,esmaili18}. Such neutrino emission is expected in a relatively short time window ($\sim$100 s) after core-collapse. Thus, this scenario predicts a high-energy neutrino signal followed by the appearance of a CCSN.

Another proposed transient source of high-energy CR and neutrinos is the tidal disruption of stars by supermassive black holes~\cite{biehl18,daifang16,murase16,winter17}.  Such TDEs occur when a star is disintegrated by  strong gravitational forces as it spirals towards the black hole.  TDEs have been detected across a range of wavelengths, and, in some cases, have been observed to launch relativistic particle jets. While constraints from radio observations suggest that most TDEs do not form such relativistic jets, some models predict neutrino emission without observed jets, invoking relativistic outflows or choked jets~\citep{2016PhRvD..93h3003S}.

Two complementary search strategies can be applied to identify neutrino emission from transient populations like TDE and CCSN. First, the high-energy neutrino alerts released by IceCube's realtime program~\citep{aartsen2016b} can be followed up with optical instruments to search for potential optical counterparts of the signatures described above. Second, a catalog of optically detected SNe and TDEs, from instruments such as the All-Sky Automated Survey for Supernovae (ASAS-SN) and the Zwicky Transient Facility (ZTF) \citep{Shappee:2013mna,bellm19}, can be used to search for the combined neutrino signal from the entire source populations. 
Both strategies are currently being applied in parallel by the IceCube collaboration~\citep{Kankare:2019bzi}.

\begin{figure}[tb]
\centering \includegraphics[width=.75\textwidth]{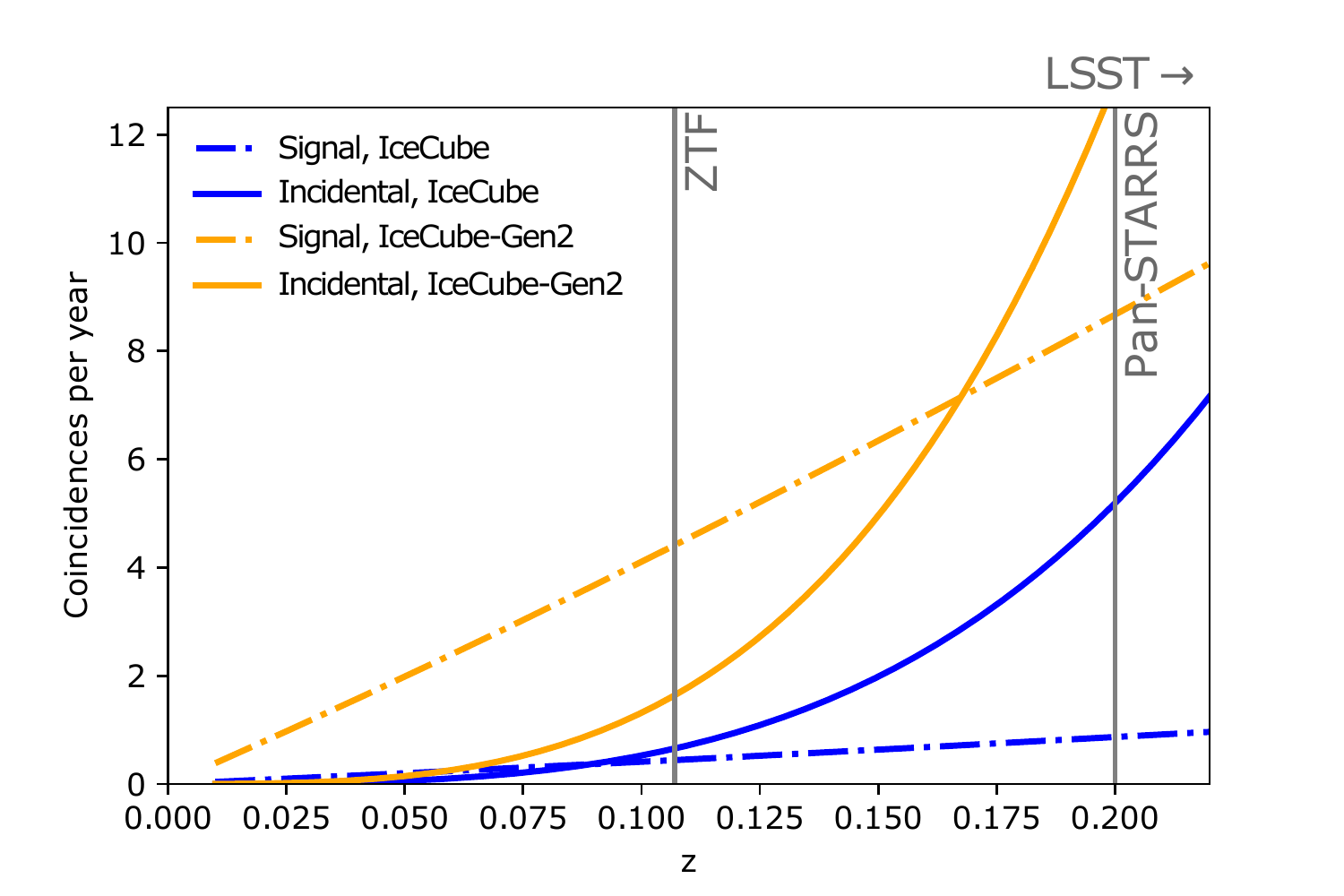}
\caption{Number of coincidences expected in a target-of-opportunity (ToO) program for IceCube (blue) and IceCube-Gen2 (orange). Assuming that the redshift evolution of the sources follows the star formation rate, the true coincidences are shown as dashed lines, while the random coincidences are shown as solid lines.}
\label{fig:ToO}
\end{figure}

IceCube-Gen2 will yield about 5 times more alerts from high-energy track-like neutrino events with improved angular resolution compared to IceCube. This will greatly reduce the fraction of alerts due to chance coincidences between neutrinos and causally unconnected optical transients. Figure~\ref{fig:ToO} shows the expected number of real and background neutrino-optical counterpart coincidences as a function of source redshift. Here, the redshift range and number of detectable coincidences is greatly increased by IceCube-Gen2. Up to 6 coincident detections of high-energy neutrinos and CCSNe can be expected per year from sources with a redshift below $z=$~0.15.
The vast improvements offered by IceCube-Gen2 should not be considered in isolation. The next generation of optical all-sky survey facilities, such as ZTF~\citep{bellm19} and the Vera C. Rubin Observatory~\citep{Abell:2009aa}, will significantly increase the sensitivity of these combined searches. High-cadence all-sky observations will reduce the uncertainty for resolving the SN explosion time from the current $\sim$30~day window to $\sim$3~days, further reducing the chance coincidence background rate.

\subsubsection{Low-energy neutrinos from core-collapse supernovae}

In a CCSN, 99\% of the gravitational binding energy of the stellar remnant is converted into $\mathcal{O}(10-20)$ MeV neutrinos \cite{Scholberg:2012id}, and the neutrinos are thought to be important drivers of the explosion.
Neutrinos from each stage of the explosion provide insight into the physics of the SN and equation of state of the resulting proto-neutron star. In the most massive stars, a core-collapse is expected to create a black hole, causing a sudden cutoff in the neutrino emission within $\sim$1~s of the core-collapse. Observation of the cutoff would give the first evidence for the formation of a black hole in real-time and enable crude localization of the black hole. Finally, SN neutrinos can constrain not only fundamental neutrino properties such as the neutrino mass hierarchy \cite{Scholberg:2017czd}, but also physics beyond the Standard Model~\cite{Kolb:1987qy,Farzan:2014gza,Fischer:2016cyd,Shalgar:2019rqe}.

While IceCube-Gen2 is designed to reconstruct neutrinos with energies at TeV energies and above, its optical array will be sensitive to the $\sim$20~MeV $\bar{\nu}_e$ produced in the accretion and cooling phases of a CCSN. A $\bar{\nu}_e$ interacting in the ice will produce a 20~MeV positron via inverse beta decay, and the Cherenkov emission from each positron will create much less than one photoelectron per optical sensor on average. Although this light level is insufficient to reconstruct individual neutrinos, in aggregate the burst of $\bar{\nu}_e$ from a CCSN causes a significant correlated rise in the noise rates of the optical sensors in the detector. The features of this correlated noise signal have been extensively studied for IceCube~\cite{Abbasi:2011ss}. IceCube-Gen2 will thus be able to record a detailed ``light curve'' of the neutrinos from such a SN in the vicinity of the Milky Way, giving excellent sensitivity to the neutrino mass hierarchy \cite{Abbasi:2011ss}, black hole formation \cite{Baum:2015drl}, and new physics such as axion-nucleon interactions~\cite{Benzvi:2017ncw}. 

Monitoring for a burst of MeV neutrinos in IceCube-Gen2 will be a powerful tool in the search for CCSNe in our galaxy, including CCSNe which are not visible in optical telescopes due to obscuration by dust or black hole formation. IceCube has an up-time in excess of 99.7\% for SNe monitoring, and similar values are expected for the optical array of IceCube-Gen2. Since CCSNe neutrinos can arrive hours to days before the first detectable photons, IceCube provides crucial early warning for optical follow-up. Therefore it is a participant in the SuperNova Early Warning System (SNEWS) \cite{SNEWS04}, a network of neutrino detectors monitoring the neutrino sky for CCSNe. A combined measurement of SN neutrinos with IceCube or in the future IceCube-Gen2 and several other detectors could be used to localize the position of the SN to an area between 100 and 1000~deg$^2$ \cite{Brdar:2018zds, Linzer:2019swe}, an area easily covered by the next generation of wide-field optical transient surveys.

With the additional instrumentation of IceCube-Gen2, the detection horizon for CCSNe will grow from $\sim$80~kpc in IceCube to $\sim$300~kpc. IceCube-Gen2 thus achieves a substantial gain in sensitivity to CCSNe neutrinos from the Milky Way and the Magellanic Clouds, independent of the mass of the progenitor and models of neutrino production in the explosion. The multi-PMT design of the IceCube-Gen2 instrumentation will also greatly improve the sensitivity to the average energy and spectral shape of the CCSNe energy spectrum, improving the resolution of the average energy from $>$25\% \cite{Kopke:2017req} to $\sim$5\% \cite{Lozano:2018}.

\subsubsection{Spectrum and flavor composition of the astrophysical neutrino flux}
\label{Sec:Spectrum}
\label{Sec:Flavor}

Given our current limited understanding of the sources of the cosmic neutrino and the UHE cosmic-ray flux, the processes and environments that lead to such dramatic acceleration of particles remain speculative.  However, some key properties of these sources are imprinted upon the observed diffuse flux of astrophysical neutrinos detected by IceCube.  The examination of the neutrino spectrum and flavor composition of this signal, in addition to comparisons to signals observed in high-energy $\gamma$-rays and CR, can thus be used to elucidate the acceleration mechanisms at work and the environments where the neutrinos are created. 

\begin{figure}[th]
\centering\includegraphics[width=0.99\linewidth]{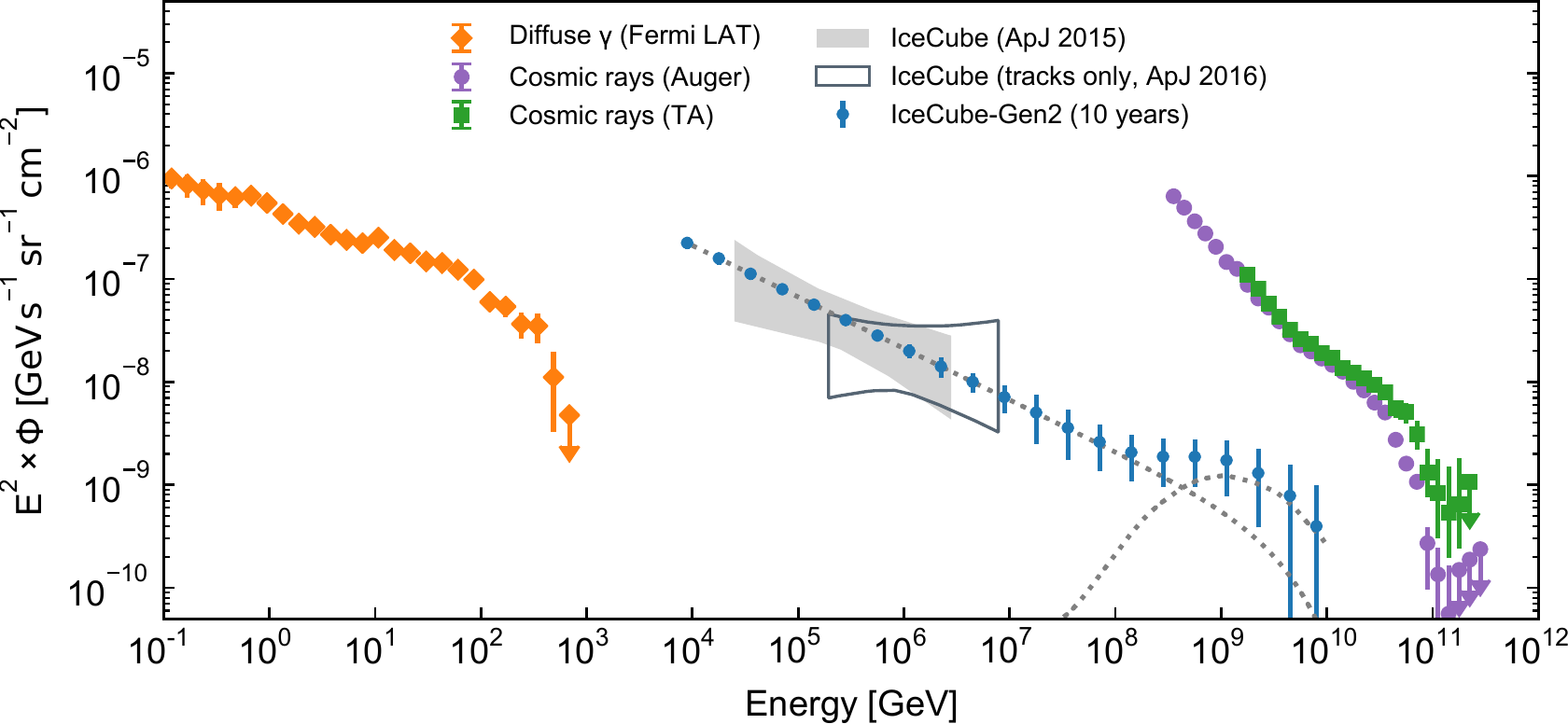}
 \caption{The high-energy astrophysical neutrino spectrum, compared to the extragalactic $\gamma$-ray spectrum measured by Fermi-LAT \cite{Ackermann:2014usa} and the highest energy CR spectrum measured by the Telescope Array~\cite{AbuZayyad:2012ru} and the Pierre Auger Observatory \cite{ThePierreAuger:2013eja}. The grey band represents the range of neutrino fluxes obtained in~\cite{Aartsen:2015knd,Aartsen:2016xlq}. The blue points are the median flux levels and 68\% confidence intervals that would be obtained from 10 years of IceCube-Gen2 data, assuming that the flux from cosmic neutrino sources continues as $\Phi \propto E^{-2.5}$, and, in addition, a cosmogenic neutrino flux (10\% proton fraction in the UHE cosmic rays) as described in \cite{vanVliet:2019nse} (both indicated by gray dotted lines). Neutrino fluxes are shown as the all-flavor sum of neutrino and anti-neutrino flux, assuming an equal flux in all flavors.}
 \label{fig:spectrum}
\end{figure}

A measurement of the astrophysical neutrino energy spectrum from IceCube along with a projection for IceCube-Gen2 are shown in Figure~\ref{fig:spectrum}, together with the extragalactic CR and $\gamma$-ray spectra. Currently, the extrapolation of the neutrino spectrum observed by IceCube up to about 10~PeV, to the energy range of the extragalactic CR with energies of tens of EeV, is not sufficiently precise to firmly establish a link between the two. The combination of optical and radio-detection methods in IceCube-Gen2 will close this gap in energy,  allowing us to measure the energy spectrum with significantly better precision up to energies three orders of magnitude higher than IceCube.
 
Such a precise measurement of the spectrum and composition of the diffuse flux of astrophysical neutrinos reveals details about the environment in which CR are accelerated and neutrinos produced.  The environment can have significant impact on the spectrum and flavor composition of extraterrestrial neutrinos~\cite{Kashti:2005qa,Lipari:2007su,Bustamante:2015waa}, as the presence of sufficiently strong magnetic fields leads to a damping of muons and thus suppression of the flux of electron neutrinos above a critical energy. Interestingly, very high accelerating gradients would have the opposite effect \cite{Koers:2007je,Klein:2012ug}.  Currently, the constraints derived from IceCube data~\cite{Aartsen:2015ivb,Aartsen:2015knd,Aartsen:2018vez} indicate consistency with the benchmark prediction from complete pion decay of $\nu_e:\nu_\mu:\nu_\tau = 1:2:0$ which is transformed to approximately 1:1:1 by neutrino oscillations over astronomical distances~\cite{Beacom:2003nh}. While the IceCube constraints are sufficiently strong to rule out that the neutrinos are produced via neutron decay, they are insufficient to probe muon-damping scenarios~\cite{Aartsen:2015knd,Bustamante:2019sdb}.

\begin{figure}[tbp]
\centering
\includegraphics[width=0.97\textwidth]{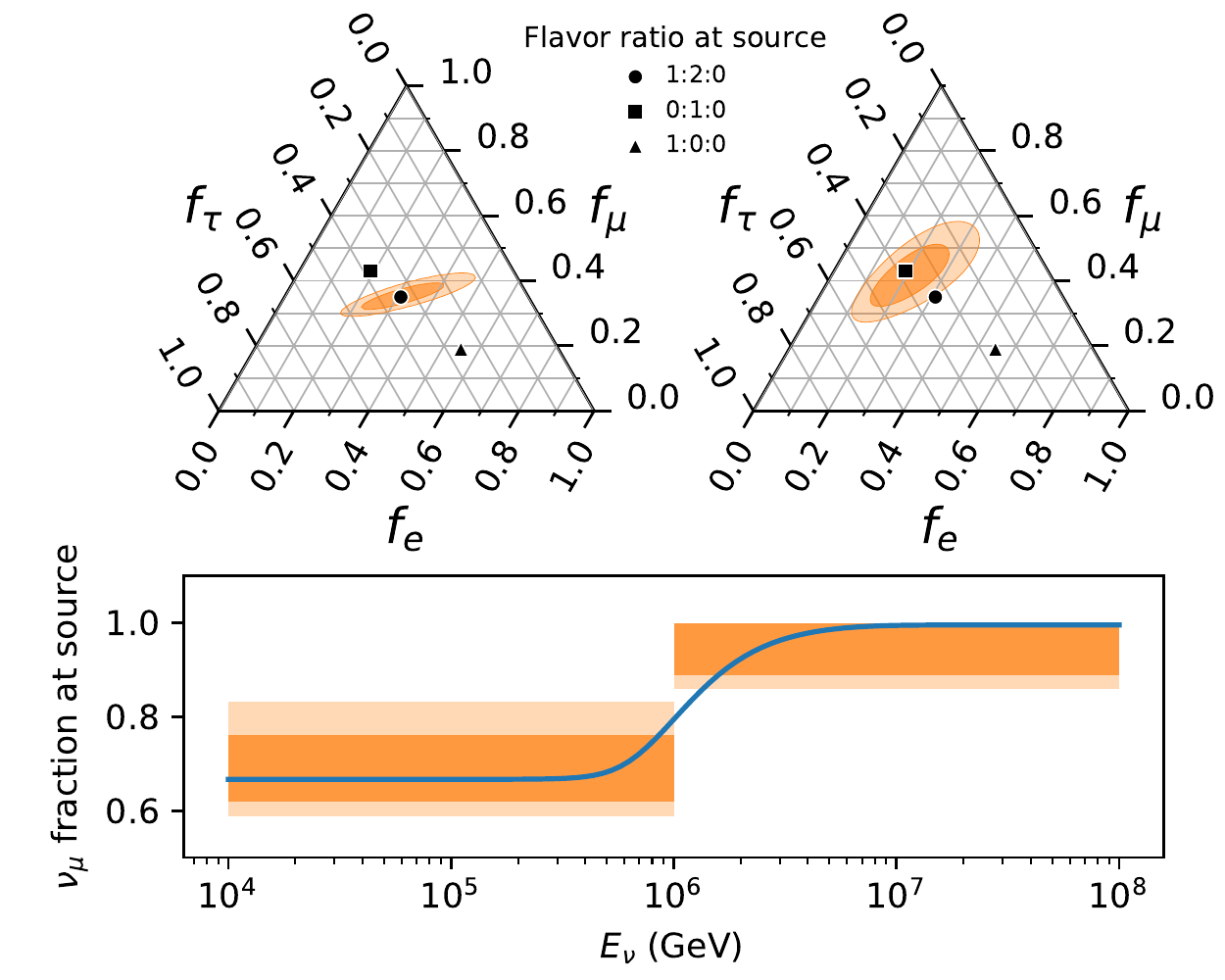}
\caption{Sensitivity to muon cooling in the sources with IceCube-Gen2. Above a critical energy, the decay time for secondary muons from pion decay exceeds the cooling time, and the flavor ratio at the source changes from 1:2:0 to 0:1:0 \cite{Kashti:2005qa}. The lower panel shows the $\nu_{\mu}$ fraction at the source as a function of energy, assuming that the muon critical energy is 2 PeV. The error bars show the 68\% CL constraints on the $\nu_{\mu}$ fraction below and above 1 PeV from the flavor composition at Earth observed with IceCube-Gen2, assuming standard oscillations \cite{Gonzalez-Garcia:2014bfa}. 
The dashed (solid) contours in the upper panels show the corresponding 68\% (95\%) constraints without any assumptions about the mixing matrix.}
\label{fig:flavor}
\end{figure}

The measurement of the diffuse spectrum will allow us to firmly establish the connection between high-energy neutrinos and extragalactic CR by matching their spectra, which would imply that the neutrino sources identified by IceCube-Gen2 also represent the dominant sources of the extragalactic CR.  Moreover, the large samples of neutrinos of all flavors in IceCube-Gen2 will allow us to observe the energy dependence of the flavor ratio over a large energy range, as shown in Figure~\ref{fig:flavor}. The sensitivity to detect a changing flavor composition as a function of neutrino energy will allow to distinguish different acceleration scenarios and source environments expected within GRBs, AGN cores, or AGN jets~\cite{Kashti:2005qa}.
 
The unique interaction and decay signatures generated by high-energy tau neutrinos interacting within IceCube's instrumented volume allow for additional handles on the flavor composition of the astrophysical neutrino flux.  The primary method of identifying $\nu_\tau$ neutrinos is to search for high-energy charged-current events with a "double-bang" structure of two nearby cascades, the first one due to the hadronic shower at the interaction vertex and the second one due to the decay of the tau lepton. The density of instrumentation in the IceCube detector limits the rate of observable $\nu_\tau$ events since the distance between the two cascades is most often much smaller than the distance between two DOMs. So far, only two candidate events could be identified~\cite{Stachurska:2019srh}. The measurement of the $\nu_\tau$ fraction plays an important role in constraining source physics and in ruling out/discovering beyond the Standard Model physics, should the observed fraction be outside of the range expected by standard oscillations.  With its much increased size, Icecube-Gen2 will be able to observe a significantly higher rate of these events, leading to stronger constraints on the flavor ratio.  Studies extrapolating the recent identification of high-energy $\nu_\tau$ candidates to IceCube-Gen2 show that above an energy of 300 TeV, the yearly rate of identified tau neutrinos would be equal or greater than 1 event/year in the optical array, a significant increase compared to the current rate of 0.2 events/year in IceCube. Additional tau neutrino identifications at EeV energies might be provided by the radio array at EeV energies. The flavor composition constraints in Figure~\ref{fig:flavor} include the effects of the predicted $\nu_\tau$ identification performance.

The fraction of $\bar{\nu}_e$ events observed is another important measurement for understanding conditions within astrophysical sources. When neutrinos are produced in $pp$ collisions, the fraction of $\bar{\nu}_e$ events is expected to be 1/6, while in case of $p\gamma$ collisions the fraction is smaller. If only the pion production via the $\Delta$-resonance is considered and the target is optically thin, it is $\sim$1/14~\cite{Bhattacharya:2011qu}. In a realistic case, however, the fraction of $\bar{\nu}_e$ from $p\gamma$ collisions depends on several parameters, the contributions from multi-pion production, the optical depth of the photon target and the chemical composition of the accelerated beam~\cite{2017JCAP...01..033B}. 

Unique sensitivity to the $\bar{\nu}_e$ flux in IceCube and IceCube-Gen2 is available through the `Glashow resonance' centered at a neutrino energy of 6.3 PeV. The necessary exposure to distinguish production of neutrinos in pp, $p\gamma$, Fe-$\gamma$ collisions, low and high opacity targets and other scenarios is discussed in detail in~\cite{2017JCAP...01..033B}: Even though a first candidate Glashow resonance event has recently been observed in IceCube~\cite{UHECRGlashowTalk,EPSGlashowTalk}, many of these production scenarios would require more than 100 years of IceCube exposure to distinguish between them, while a large number of them is accessible to IceCube-Gen2 in $\leq$ 15 years of observations. 

\subsection{Revealing the sources and propagation of the highest energy particles in the universe}
\label{sec:UHECRSources}

There is general consensus that the flux of CR below a few PeV originates from Galactic sources. Around 3~PeV the CR spectrum shows a significant break, the so-called CR {\it knee}. This spectral feature could be due to Galactic CR sources cutting off at their maximal acceleration energy or to CR escape from the Milky Way becoming more efficient. The energy range from PeV to EeV energies marks the transition region from CR of Galactic origin to CR of extragalactic origin. The nature of the CR in this transition region is not well understood, with potential contributions from heavy nuclei accelerated in the Milky Way, from Galactic super-accelerators, as well as a rising fraction of CR from extragalactic sources that dominate the CR flux above a few EeV.

CR sources, both Galactic and extragalactic, can produce neutrinos at various stages: during their acceleration in the source, while they escape from the source environment and long afterwards during CR propagation in the ubiquitous magnetic fields. Point-source neutrino emission is expected from direct interaction of CR in the sources or from interaction with close-by gas targets (e.g., molecular clouds or galaxy cluster gas). Diffuse emission is expected from the interactions of CR during  propagation, which should be correlated with the gas distribution in the Milky Way for Galactic CR, but isotropic for CR of extragalactic origin. 

\subsubsection{Galactic CR sources and neutrino emission}
\label{sec:galacticSources}
Many sources in our Galaxy show high-energy (GeV--TeV) $\gamma$-ray emission. Such emission has been observed in association with supernova remnants and their interaction with near-by molecular clouds, pulsars and their nebulae, binary systems, and massive star clusters. 
The observations of high-energy photons indicate the presence of a population of particles in these sources that emit $\gamma$-rays via leptonic and/or hadronic processes. The latter implies that the sources accelerate protons and/or nuclei, thus contributing to the Galactic CR. Observation of neutrino emission from these sources would be a diagnostic for such hadronic processes that are often difficult to identify based on $\gamma$-rays alone. Indeed, all the above mentioned sources of high-energy $\gamma$-rays have also been speculated to be continuous or transient neutrino emitters; see, e.g.,~\cite{Bednarek:2004ky,Kistler:2006hp}.

It has long been speculated that Galactic CCSNe could be responsible for the majority of the observed CR~\cite{Baade1934}. These events produce ejecta with kinetic energy of the order of $10^{51}$~erg per supernova (SN) explosion, at a rate of about 3 per century. Diffuse shock acceleration taking place in remnant shocks could convert a significant fraction of $\mathcal{O}(0.1)$ of this kinetic energy into a non-thermal population of CR. The {\it Fermi}~LAT has observed features in the gamma-ray spectra of the shell-type supernova remnants IC 443 and W44~\cite{Ackermann:2013wqa} that can be related to hadronic processes. Neutrino emission from CR acceleration and interaction in supernova remnants has been studied in~\cite{Drury:1993pd,Gaisser:1996qe,AlvarezMuniz:2002tn,Costantini:2004ap}. 

Pulsars and their nebulae have also been considered as potential sites for CR acceleration~\cite{Protheroe:1997er,Blasi:2000xm,Beall:2001pa,Fang:2012rx,Fang:2013cba}. However, many details of the proposed mechanisms leading to extraction of rotational energy and acceleration of charged particles at these sites are vague. As with SNRs, the relation between TeV $\gamma$-ray and neutrino emission can be exploited to estimate the neutrino flux from pulsar wind nebulae~\cite{AlvarezMuniz:2002tn,Guetta:2002hv,Amato:2003kw}.

Pulsars (and other CR sources) born in massive star clusters like Cygnus OB2 or in the Galactic Center region would inject their CR into the near environment, which has enhanced gas densities and an enhanced magnetic field strength compared to the Galactic average. This could enhance locally the neutrino signal from CR interactions with the interstellar gas ~\cite{Bednarek:2001ir,Bednarek:2003cx}.

\begin{figure}[tb]
\begin{center}
\includegraphics[width=0.99\textwidth]{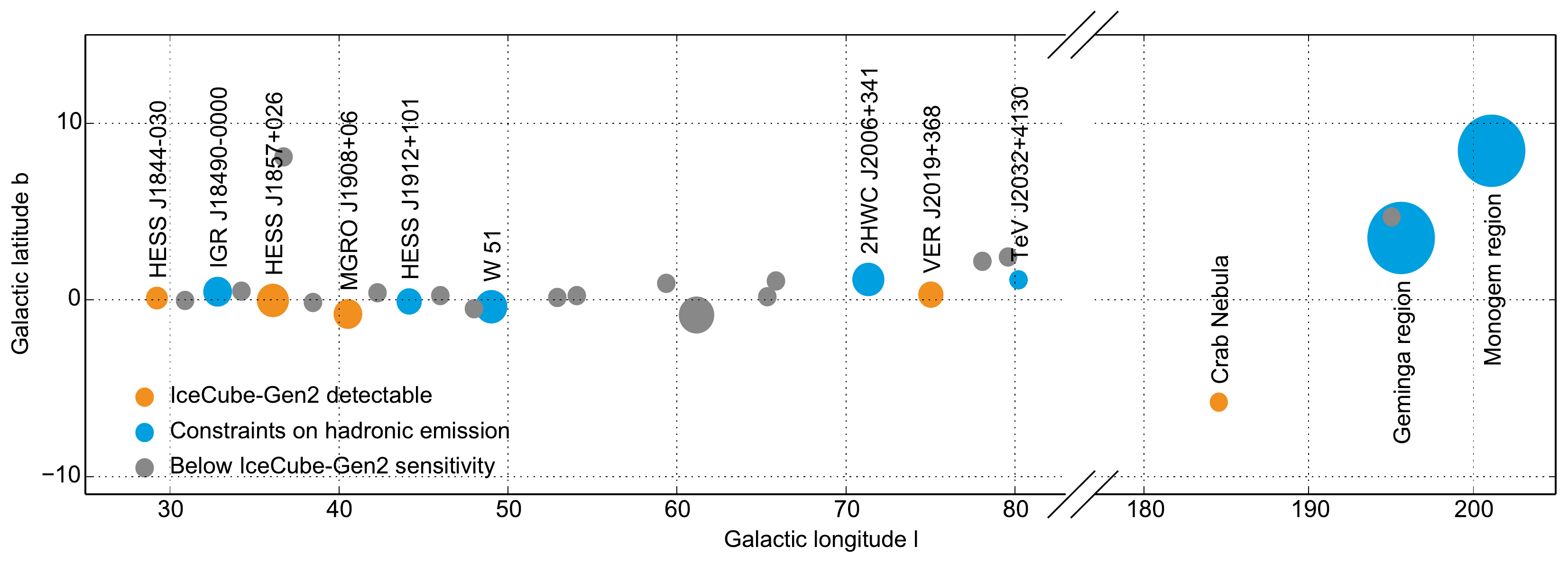}
\caption[]{Sensitivity of IceCube-Gen2 to hadronic emission of neutrinos by Galactic sources. Shown are TeV $\gamma$-ray sources detected by HAWC near the Galactic plane and in the Northern hemisphere (declination $\delta \ge -5^{\circ}$). The size of the source marker corresponds to the source extension reported by HAWC for sources that are larger than 0.5$^{\circ}$. Sources marked in orange will be detected by IceCube-Gen2 if their $\gamma$-ray emission originates solely from hadronic processes. For sources marked in blue, the contribution of hadronic emission can be constrained at the 90\% confidence level, while sources marked in gray are below the sensitivity of IceCube-Gen2.}\label{fig:HAWCsources}
\end{center}
\end{figure}

Figure \ref{fig:HAWCsources} shows a map of Galactic sources detected by the High Altitude Water Cherenkov (HAWC) detector above few TeV in energy. The size of the disk indicates the spatial extension of the source measured by HAWC, for sources with an extension larger than 0.5$^{\circ}$. The brightest of these sources will be detectable by IceCube-Gen2 in case all of the observed $\gamma$-ray emission originates from hadronic processes. For several more sources, the contribution of hadronic processes to the $\gamma$-ray emission can be constrained.  In many of these sources the origin of the TeV $\gamma$-ray emission is not understood well, sometimes spatially extended, and might arise from different contributions from PWNs, local CR over-densities and molecular cloud interactions. The detection of neutrinos will give essential clues for understanding CR acceleration processes in these sources and regions.

\subsubsection{Diffuse Galactic emission and the propagation of cosmic rays}

The expected diffuse Galactic neutrino flux from interactions of CR with interstellar gas and unresolved CR sources is, averaged over the whole sky, about an order of magnitude lower than the isotropic flux observed in IceCube for $E\simeq$~100~TeV~\cite{Ahlers:2013xia,Joshi:2013aua,Kachelriess:2014oma,Ahlers:2015moa}. 
The diffuse flux estimate relies however on the assumption that the local CR flux is a good approximation for the average Galactic CR density. This is not necessarily the case in more complex diffusion scenarios~\cite{Casse:2001be,Effenberger:2012jc,Gaggero:2015xza}, and/or for strongly inhomogeneous source and target distributions in the Galaxy~\cite{Gaggero:2013rya,Werner:2014sya,Crocker:2004nk,Candia:2005nw}, as well as time-dependent local CR injection episodes~\cite{Neronov:2013lza}. Such scenarios often predict an enhancement of the hadronic $\gamma$-ray and neutrino emissions in the multi-TeV region that can be tested with neutrino telescopes. 

The distribution of gas and CR in our Galaxy is strongly correlated with the Galactic plane, so Galactic diffuse neutrino emission can be identified in IceCube and IceCube-Gen2 data via correlation of the neutrino events with the Galactic plane. 
IceCube is just sensitive enough to marginally constrain the most optimistic of the aforementioned scenarios \cite{Aartsen:2017ujz}, but IceCube-Gen2 will enable the Galactic plane emission to be probed (at 90\%~CL) down to the level expected from the local CR flux. It will be able to detect any 
strong enhancements of the multi-TeV diffuse Galactic neutrino emission such as is predicted in \cite{Gaggero:2015xza}, at $>$5$\sigma$ significance, and distinguish between various propagation and source distribution scenarios.

\subsubsection{Cosmogenic neutrinos}
\label{sec:cosmogenic}
At EeV energies all CR are believed to be of extragalactic origin \cite{AugerAnisotropy}. From CR measurements themselves, we neither know what the accelerators of these UHE cosmic rays are, nor what maximum energies they can reach. Also, particle composition measurements at the highest energies still leave room for various scenarios ranging from only protons to a heavy composition with no protons at all \cite{Auger2014,TA2018}. Limited knowledge about the Galactic and extragalactic magnetic fields that influence the propagation, further complicates the identification of sources based on CR observations alone. 

\begin{figure}[t]
\centering\includegraphics[width=0.7\linewidth]{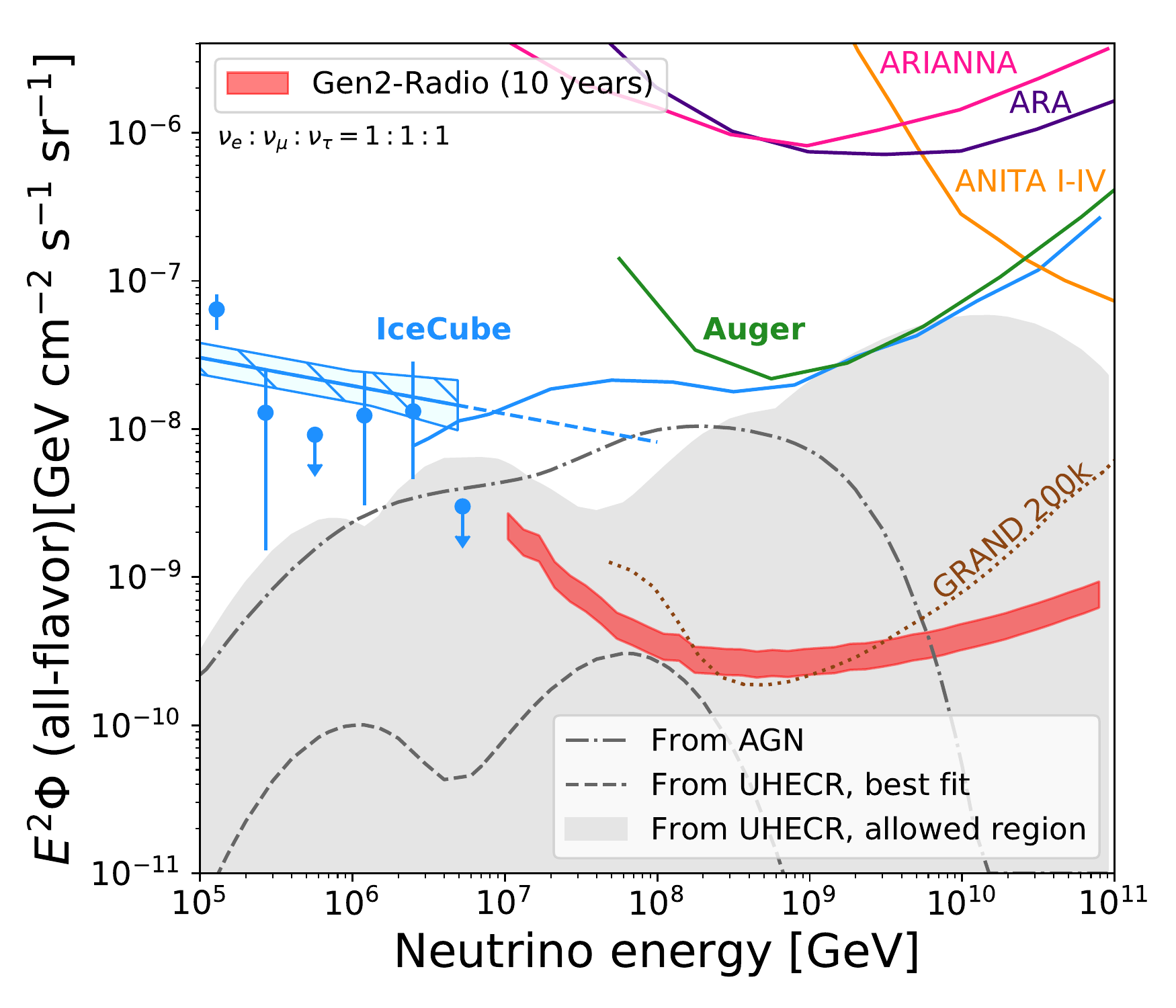}
\caption{Sensitivity of the IceCube-Gen2 radio array at the highest energies in comparison to models \cite{Murase:2014foa,vanVliet:2019nse,Heinze:2019jou}, existing upper limits \cite{Aartsen:2018vtx,Aab:2019auo,Gorham:2019guw,Allison:2015eky,Anker:2019rzo}, and the 10 year sensitivity of the proposed GRAND array of 200,000 antennas \cite{Alvarez-Muniz:2018bhp}. The uncertainties on the IceCube-Gen2 radio array sensitivity are $\pm 20\%$, which are uncertainties in the estimated sensitivity of the array, e.g. due to remaining design decisions.}
\label{fig:Gen2_cosmogenics}
\end{figure}

Detecting neutrinos above PeV energies can resolve these open questions. Above EeV energies the CR interactions with the cosmic microwave background and the EBL can produce neutrinos~\cite{Berezinsky:1969qj}. In addition, unstable atomic nuclei, produced during photo-disintegration or photo-pion production of CR, can produce neutrinos when decaying. All of these neutrinos are referred to as \emph{cosmogenic neutrinos}. So while this secondary flux of neutrinos is extremely well motivated, its level depends strongly on the composition of the CR \cite{Hooper:2004jc,Ave:2004uj}, the cosmic evolution of the sources, the spectral index of the sources, and their maximum acceleration energy \cite{Kotera:2010yn,Ahlers:2010fw}. IceCube can already exclude scenarios with a very strong evolution of the CR sources with cosmic redshift \cite{Aartsen:2016ngq,Aartsen:2018vtx}, but to draw firm conclusions much larger exposures are needed. 

The sweet spot for cosmogenic neutrino detection is at 10$^{18}$ eV, since the flux at these energies depends less strongly on the maximum acceleration energy and spectral index than at the highest energies \cite{Yoshida:2012gf}. The flux at this energy is primarily a function of the proton fraction and even the most conservative flux estimates peak here \cite{vanVliet:2019nse}. Conveniently, this is the energy region where in-ice neutrino detectors, and the radio array of IceCube-Gen2 will be most sensitive to the energy flux of neutrinos (see Fig.~\ref{fig:Gen2_cosmogenics}). 

With the predicted sensitivity, IceCube-Gen2 will also be able to provide independent evidence for whether the observed cut-off in the flux of UHE cosmic rays is due to the GZK suppression or just due to reaching the limit of acceleration in the sources \cite{Aab:2016zth}.  The neutrino flux at energies above 10 EeV depends primarily on the maximum energy of CR protons and the spectral index of their power-law spectrum rather than the source evolution parameters \cite{Kotera:2010yn}. Detecting the corresponding neutrino flux will be a measurement independent of the uncertainties in the modelling of the hadronic interactions in extensive air showers and, thus, complementary to results from air shower arrays.

The IceCube-Gen2 sensitivity would reach the current best-fit models to CR data, assuming sources identical in CR luminosity, spectrum and composition, as well as a rigidity-dependent cut-off and thereby essentially no protons at the highest-energies \cite{Heinze:2019jou,AlvesBatista:2018zui}. In an only slightly more favorable scenario of 10\% protons, IceCube-Gen2 will detect at least 3 events per year above $\sim$100~PeV. Only the radio array has been considered for this sensitivity calculation. The sensitivity of the IceCube-Gen2 optical array will be about 5 times better than the IceCube upper limit shown in Figure~\ref{fig:Gen2_cosmogenics} and therefore not contribute substantially to the performance of the observatory in this energy range. In the case that no cosmogenic neutrinos are discovered by IceCube-Gen2, the observation would exclude all redshift evolution scenarios\footnote{using the customary $(1+z)^{m}$ parametrization of redshift evolution} with $m>0$ for a proton-fraction of more than 20\%, thereby excluding many source populations that evolve with the star formation rate, various AGN models, and GRBs as sources of UHE cosmic rays.

\subsection{Probing fundamental physics with high-energy neutrinos}
\label{sec:BSMPhysics}

IceCube has been extremely successful in searches for dark matter and other new physics beyond the Standard Model~\cite{Ahlers:2018mkf}~(BSM physics). IceCube-Gen2 will provide  new  opportunities  to  study  particle physics at energies and baselines well above those accessible at terrestrial accelerators and local natural sources due to increased statistics, extended energy reach and improved flavor identification. This section highlights some of the opportunities that this observatory will provide in probing fundamental physics and searches for new particles.

\subsubsection{Neutrino cross sections at high energies}
The neutrino-nucleon cross section in the TeV--PeV range was measured for the first time using astrophysical and atmospheric neutrinos in~\cite{Aartsen:2017kpd,Bustamante:2017xuy,Aartsen:2018vez}, extending previous results~\cite{Berezinsky:1974kz, Hooper:2002yq, Hussain:2006wg, Borriello:2007cs, Hussain:2007ba} and measurements that used GeV neutrinos from accelerators\ \cite{Conrad:1997ne, Formaggio:2013kya, Tanabashi:2018oca}.  The measurements agree with high-precision Standard Model predictions~\cite{CooperSarkar:2011pa}. Future measurements in the EeV range could probe BSM modifications of the cross section at center-of-momentum energies of up to 100~TeV~\cite{Kusenko:2001gj,AlvarezMuniz:2001mk,Anchordoqui:2001cg,Cornet:2001gy,Kowalski:2002gb,AlvarezMuniz:2002ga,Anchordoqui:2005pn,Marfatia:2015hva,Ellis:2016dgb,Anchordoqui:2018qom} and test the structure of nucleons with a sensitivity comparable with colliders~\cite{Henley:2005ms, Armesto:2007tg,  Illarionov:2011wc}. 

IceCube has measured the neutrino-nucleon cross-section with 30-40\% uncertainty up to 1~PeV, using only one year of data~\cite{Aartsen:2017kpd}. Each year of data from IceCube-Gen2 will yield roughly one order of magnitude more statistics. Therefore with this new instrument it will be possible to study the cross-section with significantly higher precision and, in multiple energy bins, to energies beyond 10~PeV. This sensitivity will allow unique tests of BSM physics involving extra dimensions, leptoquarks or sphalerons~\cite{Romero:2009vu,Hooper:2002yq,Klein:2013xoa,Dutta:2015dka,Ellis:2016dgb,Klein:2019nbu}. It would also probe the QCD parton distribution functions (PDFs) at large momentum transfers $Q^2$ ($Q^2 \approx M_W^2$) and Bjorken-x values down to $\sim 10^{-4}$ \cite{CooperSarkar:2011pa,Bertone:2018dse,Connolly:2011vc}, where non-perturbative QCD effects are expected to start becoming important.  This data will complement results from the FASERnu experiment at CERN, which will use forward neutrinos from LHC interactions to measure neutrino cross-sections at energies centered around 1 TeV \cite{Abreu:2020ddv}.

At the EeV scale, measuring the cross section to within an order of magnitude could distinguish between Standard Model predictions and BSM modifications. This target is achievable with tens of events in the hundreds of PeV--EeV energy range. In the scenario presented in Section~\ref{sec:cosmogenic}, IceCube-Gen2 would record (mostly via its radio detection component) about 3~events/year above 100~PeV just from cosmogenic neutrinos, bringing into reach a test of the Standard Model at these extreme energies~\cite{Connolly:2011vc}. 

IceCube-Gen2 will enable a measurement of the inelasticity of neutrino interactions (the fraction of energy transferred to the target) over a wider energy range and with higher precision than IceCube~\cite{Aartsen:2018vez}, using a larger sample of events. This capability will allow us to study a number of new physics topics since the inelasticity is sensitive to new phenomena, including non-standard types of interactions, such as `diffractive' neutrino interactions~\cite{Seckel:1997kk} or BSM physics, e.g., the production of heavy vector bosons. Furthermore, the inelasticity distribution is sensitive to more conventional physics, e.g., the evolution of the PDFs, including nuclear effects like shadowing, which affect both the inelasticity and the cross-section \cite{Klein:2020nuk}.

\begin{figure}[bt]
   \centering
    \includegraphics[width=0.60\textwidth]{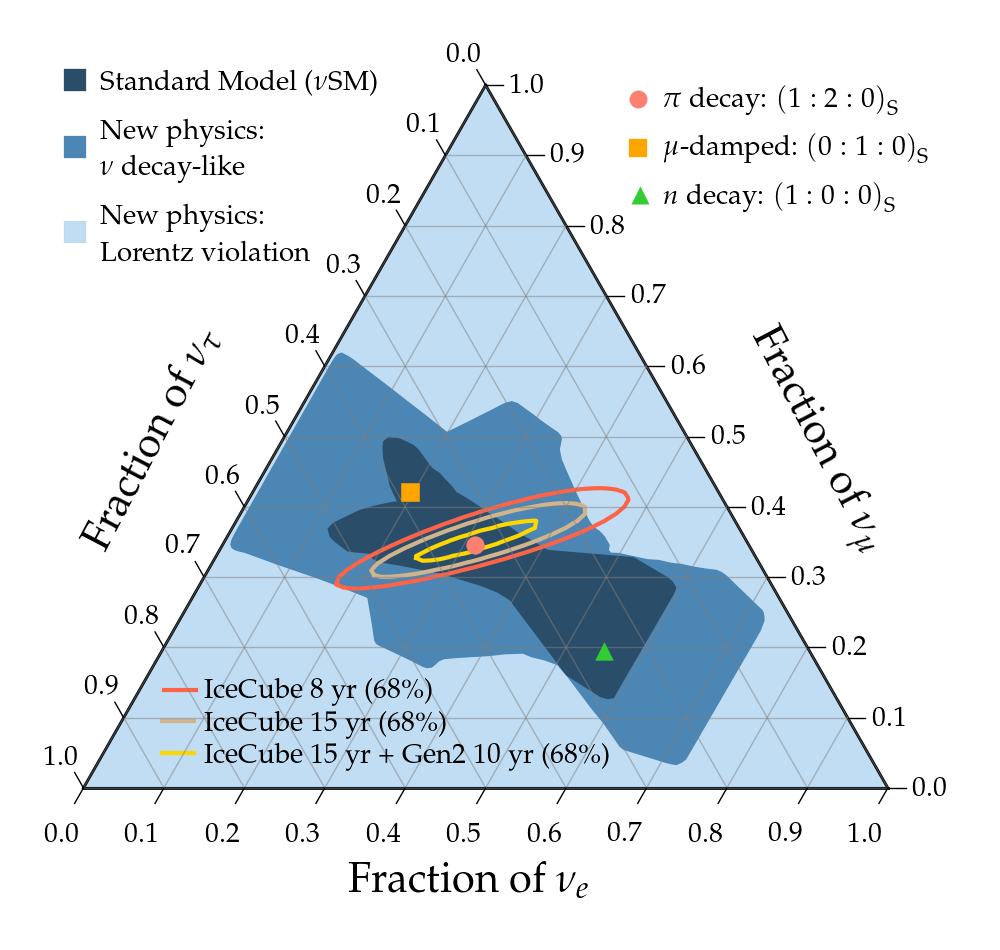}

  \caption{Flavor composition at Earth of high-energy cosmic neutrinos, indicating the ``theoretically palatable"\ \cite{Bustamante:2015waa} regions accessible with standard oscillations, with new physics similar to neutrino decay, and with new physics similar to Lorentz-invariance violation~\cite{Arguelles:2015dca,Rasmussen:2017ert}. To generate the colored regions, the neutrino mixing parameters are generously varied within their uncertainties at $3\sigma$, and the flavor ratios at the sources are varied over all possible values. The colored contours indicated the expected constraints from IceCube and IceCube-Gen2, if neutrinos are produced with ratios of ($\nu_{e}:\nu_{\mu}:\nu_{\tau}$) = ($1:2:0$) at the source and assuming Standard Model neutrino oscillations.}
   \label{fig:flavor_ratio_constraints}
\end{figure}

\subsubsection{New physics constraints from flavor mixing}
\label{sec:flavor_new_physics}

As high-energy cosmic neutrinos oscillate on their way to Earth, the allowed range of each flavor's fractional contribution to the total measured flux is small (see Figure~\ref{fig:flavor_ratio_constraints}), even after accounting for uncertainties in the parameters that drive the oscillations and in the neutrino production process~\cite{Bustamante:2015waa}. That is, given a flavor composition at the source and the standard oscillation scenario, the expected flavor ratio at the Earth is quite restricted. 
However, mixing remains untested at high energies and over cosmological propagation baselines~\cite{Learned:1994wg}. Even small BSM effects could affect flavor mixing, vastly expanding the allowed region of flavor ratios at the Earth and making the flavor ratio measurement a very sensitive probe of BSM physics~\cite{Beacom:2003nh, Pakvasa:2007dc, Bustamante:2010bf, Bustamante:2010nq, Mehta:2011qb, Bustamante:2015waa, Arguelles:2015dca, Shoemaker:2015qul,Gonzalez-Garcia:2016gpq, Rasmussen:2017ert, Ahlers:2018yom, Bustamante:2018mzu}. Figure~\ref{fig:flavor_ratio_constraints} shows two examples ($\nu$-decay~\cite{Bustamante:2015waa} and Lorentz invariance violation~\cite{Arguelles:2015dca,Rasmussen:2017ert}) of how BSM physics effects extend the region of possible flavor composition after cosmological distances. The figure also shows the sensitivities of IceCube and IceCube-Gen2 to put constraints on the flavor composition of cosmic neutrinos, assuming standard neutrino oscillations and a 1:2:0 production ratio at the sources. IceCube-Gen2 will allow substantially more sensitive searches for, and constraints on, BSM effects using the flavor ratio than the current generation of neutrino telescopes. The improved statistics will also permit searches for a potential energy dependence of mixing (cf. Section~\ref{Sec:Flavor}), which could also point to the presence of BSM effects~\cite{Mehta:2011qb, Bustamante:2015waa}.

\subsubsection{Sterile neutrinos}

A detector of the size of IceCube-Gen2 opens the possibility to search for heavy sterile neutrino decay. For example, an active, light neutrino might scatter off a target nucleon producing a heavy, sterile neutrino ~\cite{Coloma:2017ppo}, which has a variety of interesting decay channels~\cite{Bertuzzo:2018itn,Bertuzzo:2018ftf,Ballett:2018ynz}. Several of these decay signatures will only be observable in the larger instrumented volume of IceCube-Gen2. One such example is the sterile neutrino decaying into pairs of muons; characteristically, this muon pair would be emitted from a secondary vertex that is displaced from the initial neutrino-nucleon interaction.
 
Such double muon tracks can also point to new physics via neutrino trident  interactions~\cite{Ge:2017poy,Zhou:2019vxt,Beacom:2019pzs}. This very rare process is  mediated by the W, Z, or a virtual photon in the Standard Model. But if it is additionally mediated by a new vector or scalar boson, the final state of a neutrino interaction in the detector---a double muon track and a particle cascade simultaneously produced from the same vertex---would be a very distinctive signal (although several other interesting signals are also expected~\cite{Beacom:2019pzs}). As in the previous case, the size of IceCube-Gen2 will benefit the reconstruction of the double track, or to measure the unusual light yield of a single track over a sufficiently long lever arm, if the two tracks are not individually reconstructed.

\subsubsection {Searches for unexpected neutrino properties and additional species}

The extremely long distances that astrophysical neutrinos travel make them a good probe of fundamental neutrino properties. The observation of non-zero neutrino masses opens the possibility of neutrino decay and astrophysical neutrinos can place strong constraints on decay  scenarios~\cite{Beacom:2002vi,Baerwald:2012kc,Bustamante:2016ciw,Denton:2018aml,Bustamante:2020niz,Abdullahi:2020rge}. Additionally, unlike other new physics scenarios, neutrino decay imprints a specific correlation between the energy distribution of the events and the flavor composition~\cite{Shoemaker:2015qul}. This makes it one of the most predictive signatures of new physics. 

The existence of additional neutrino species is of great interest to the neutrino community, specially due to the recent claims by the MiniBooNE collaboration~\cite{Aguilar-Arevalo:2018gpe}. Global data allow non-unitarity in the neutrino mixing (PMNS) matrix~\cite{Parke:2015goa}, which may indicate potential mixings of standard neutrinos and sterile neutrinos. 
The effect of additional sterile neutrino species on the flavor ratio was discussed in~\cite{Brdar:2016thq}. If the initial flavor composition does not contain a significant sterile component its effect is small. 
This is not the case when the neutrino flux has a large initial sterile component~\cite{Arguelles:2019tum} (which can happen for example in the case of decaying dark matter~\cite{Brdar:2016thq}). In this situation the astrophysical neutrino flavor can appear in the tau corner in Fig~\ref{fig:flavor_ratio_constraints}. This is a very striking signature of non-standard neutrino physics as this corner is forbidden by unitary constraints under the assumption of standard production mechanisms~\cite{Arguelles:2015dca,Ahlers:2018yom}. The IceCube-Gen2 flavor composition measurement will be competitive with and complementary to terrestrial high-precision oscillation experiments to search for sterile neutrinos~\cite{Arguelles:2019tum}.

\subsubsection{Tests of fundamental symmetries}

Theories allowing for a new spacetime structure at the Planck scale, like quantum gravity, can accommodate violations of symmetries that are taken as fundamental in the Standard Model, like CPT and Lorentz invariance\ \cite{Colladay:1998fq}.  These effects can be detected in a neutrino telescope as an anomalous oscillation effect proportional to the neutrino energy, instead of inversely proportional to energy 
as in standard oscillations~\cite{Kostelecky:2003cr, Hooper:2005jp,Kostelecky:2008ts, Kostelecky:2011gq, Gorham:2012qs, Borriello:2013ala, Stecker:2014xja, Anchordoqui:2014hua, Tomar:2015fha, Amelino-Camelia:2015nqa, Liao:2017yuy, Anchordoqui:2006wc}. Currently, the strongest constraints on Lorentz invariance violation (LIV) with neutrinos come from IceCube, using the high-energy atmospheric neutrino flux\ \cite{Aartsen:2017ibm}. IceCube-Gen2 can add both an increase in statistics of high-energy atmospheric neutrinos, as well as a high-statistics observation of cosmic neutrinos that could provide unprecedented sensitivity to the parameters quantifying the deviation from Standard Model physics \cite{AmelinoCamelia:1997gz, Hooper:2005jp, GonzalezGarcia:2005xw, Anchordoqui:2005gj, Bazo:2009en, Bustamante:2010nq, Kostelecky:2011gq, Diaz:2013wia, Stecker:2014oxa, Stecker:2014xja, Tomar:2015fha, Ellis:2018ogq, Laha:2018hsh}. 
 The detailed characterization of the extragalactic neutrino flux with the high statistics available with IceCube-Gen2 can be used to set limits to Lorentz invariance violation which are complementary to those obtained from the distortion of the oscillation pattern at lower energies. Indeed, the first PeV events detected by IceCube have already been used to set such a limit, under the assumption that they are extragalactic. The bound obtained from these events is $\delta < 10^{-18}$, where $\delta$ is the deviation from the speed of light, orders of magnitude smaller than the previous best limit of $10^{-13}$~\cite{Borriello:2013ala, Diaz:2013wia}. Additionally, the detection of neutrino flares in cosmic sources, with additional assumptions of the mechanism of neutrino emission, can be used as precise time-of-flight measurements. This exercise was performed on the candidate  source TXS 0506+056~\cite{Ellis:2018ogq,Laha:2018hsh} resulting in very strong constraints on LIV under the assumption of simultaneous $\gamma$-ray and neutrino production; a similar study was performed on a previous candidate source yielding similar strong constraints~\cite{Wang:2016lne}. As IceCube-Gen2 discovers more high-energy neutrino sources, and our understanding of the emission of $\gamma$-rays and neutrinos improves, such studies will yield significantly more robust bounds.

Lorentz violation can also be broken in such as way as to produce subluminal particles. In this case, neutrino bremsstrahlung would not happen, but the expected flavor composition of ultra-high-energy neutrinos would be modified. This can be modelled as a change in the neutrino potential due to its interaction with an ambient LIV field~\cite{Kostelecky:2003xn}. In general, any new interaction that is not diagonal in flavor will modify the expected flavor composition~\cite{Arguelles:2015dca}.  

At the moment, the astrophysical neutrino constraints have a strong dependence on the details on the production mechanism and the initial flavor composition. It is remarkable, though, that already in some scenarios the astrophysical neutrino constraints are several orders of magnitude stronger than the terrestrial ones. 
These effective operator bounds apply not only to interactions between neutrinos and the ambient Lorentz violating field, but can be adapted to other kinds of interaction~\cite{Rasmussen:2017ert}. 
Other interesting BSM scenarios are coherent interactions between neutrinos and dark matter~\cite{Arguelles:2017atb,Capozzi:2018bps,Farzan:2018pnk,Pandey:2018wvh,Alvey:2019jzx,Koren:2019wwi,Choi:2019ixb}; neutrinos and dark energy~\cite{Klop:2017dim}; very-long-range $L_e-L_\mu$ and $L_e-L_\tau$ gauged interactions sourced by the universe's electron content~\cite{He:1990pn,He:1991qd,Foot:1994vd,Bustamante:2018mzu}; and new interactions between high-energy cosmic neutrinos and low-energy relic neutrinos~\cite{Ng:2014pca,Ioka:2014kca,Ibe:2014pja,Kamada:2015era,DiFranzo:2015qea,Murase:2019xqi,Kelly:2018tyg,Bustamante:2020mep}. 

\subsubsection{Dark matter searches} 
Dark matter searches with IceCube
have focused on generic thermal relic {\it Weakly Interacting Massive Particles} (WIMPs), by looking for neutrinos from annihilations (or decays) of
dark matter captured in the Sun, Earth, or in the Galactic halo or galaxy clusters~\cite{losHeros:2017otx}. 
Dark matter candidates with a mass beyond the typical WIMP scale of a few GeV to TeV may also have been non-thermally produced in
the early universe~\cite{Chung:2001cb,Chung:1998zb,Chung:1998ua}. 
Neutrinos offer many advantages to search for annihilation or decays of heavy dark matter 
with masses beyond about 100~TeV~\cite{Gondolo:1991rn,Murase:2012xs,Rott:2014kfa}. The neutrino interaction cross section increases with energy~\cite{CooperSarkar:2011pa} such that the event rates predicted for heavy decaying dark matter per volume of ice remains constant as function of the particle mass up to several tens of TeV. By contrast, $\gamma$-ray signals fall by a factor of $1/m_{\chi}$ and are further attenuated by the interstellar radiation field~\cite{Moskalenko:2005ng}. 

Currently, dark matter lifetimes at the level of 10$^{28}$~s~\cite{Aartsen:2018mxl} are being constrained by IceCube's highest energy neutrinos, resulting in the strongest constraints on the dark matter lifetime above 100~TeV.
These bounds are statistically limited and lifetimes exceeding 10$^{29}$~s could be tested for decay modes which involve neutrinos in the final state.
These searches are expected to be significantly improved by the order-of-magnitude increase in statistics that IceCube-Gen2 will provide.

Beyond the traditional, model-agnostic, WIMP scenario IceCube-Gen2 will also have power to study dark-portal scenarios, where the dark matter is part of a secluded sector~\cite{Pospelov:2007mp}. In these scenarios, the dark sector often couples to the Standard Model either via kinetic mixing or neutrino mass mixing.  These secluded models can increase the expected neutrino flux from dark matter annihilations in the Sun, reducing the attenuation of the neutrino flux since, for favourable model parameters, the mediator can leave the dense solar interior and decay outside the Sun~\cite{Leane:2017vag,Ardid:2017lry}. 

It has recently been shown that such dark portal scenarios imprint signatures on the diffuse astrophysical neutrino spectrum~\cite{Arguelles:2017atb}, as well as on the expected attenuation from a single neutrino source~\cite{Kelly:2018tyg}. With the improved characterization of the astrophysical diffuse flux and observation of new neutrino sources expected from IceCube-Gen2, such dark portal scenarios can be further investigated.

\subsubsection{Other particle physics searches and exotica}

With its unparalleled size, IceCube-Gen2 will have unparalleled sensitivity to searches for new BSM particles and exotica.  We saw that IceCube has produced important physics results in many areas that were either not considered during the proposal, or were too speculative to formally propose.  We expect the same for IceCube-Gen2. Here, we list a few planned new-particle and exotica searches, but these are just a sample of what is possible. New particles can be detected through either their passage through the detector, or by producing unique signatures from interactions either within the detector or outside of it.  Magnetic monopoles are one example of the former; relativistic monopoles are notable for their large, but roughly constant $dE/dx$, without large stochastic fluctuations~\cite{Aartsen:2015exf}.  Some models of supersymmetry or Kaluza-Klein particles lead to the latter, through the production of pairs of laterally separated particles which may travel upward through the detector~\cite{Kopper:2016hhb}. 

An intriguing possibility is the search for low-scale gravity effects through micro black hole production. If the center-of-mass energy of the interaction of a neutrino with a nucleon exceeds the Planck scale, a microscopic black hole can be produced~\cite{Kowalski:2002gb,AlvarezMuniz:2002ga,Anchordoqui:2006fn}.
However, in our 4-dimensional world, the Planck scale lies at energies $M_P \sim 10^{19}$~GeV, while the largest accelerators only reach TeV center-of-mass energies. But in 4+D space-time dimensions the Planck scale may be much lower, and the interaction of a ultra-high-energy neutrino with a nucleus inside the detector could produce a micro black hole. The evaporation of the black hole through Hawking radiation (in $\sim$~10$^{-27}$~s) will produce a burst of Standard Model particles that can be detected in a neutrino telescope. Although the free parameters of extra-dimension models are many and the uncertainties in the predictions large, a detectable signal can be expected in a large volume detector in the most favourable scenarios, taking into account the already existing limits on the ultra-high-energy neutrino flux. Even if the original energy of the incoming neutrinos is not high enough to form a black hole, elastic neutrino-parton scattering through exchange of D-dimensional gravitons could be possible, another feature of low-energy gravity models. In such case the neutrino is not destroyed in the interaction, as in black hole production, but it continues on its way ready for another elastic interaction after a mean free path which, for a given energy, depends on the number of extra dimensions. The energy lost in each interaction goes into a hadronic shower, producing a very unusual signature in the detector: multiple particle showers without a lepton track between them. Current calculations predict that a large detector could detect a handful of events per year, being able to probe extra-dimension models with $D$ up to six~\cite{Illana:2020jpi}. The larger size of IceCube-Gen2 will allow to follow such events over a larger lever arm, increasing the identification efficiency. IceCube is just too small to detect several correlated showers.

The atmosphere, acting as a target for UHE cosmic rays, can also be a useful source for searches for physics beyond the Standard Model~\cite{Illana:2007ra, Illana:2005pu, Albuquerque:2006am, Reynoso:2007zb}. The interaction of a high-energy CR with a nucleon in the atmosphere can take place at a much higher center-of-mass energy than that achievable in man-made accelerators. Supersymmetric particles can be produced in pairs and, except for the lightest one, they can be charged. Even if unstable, due to the boost in the interaction, they can reach the depths of the detector and emit Cherenkov light as they traverse the array. The signature is two minimum ionizing, parallel, coincident tracks separated by a distance of over 100 meters \cite{Albuquerque:2009vk}.  
The  interactions of CR with the atmosphere can also be used to probe non-standard neutrino interactions due to TeV gravity effects. At high energies neutrino interactions with matter may become stronger, and the atmosphere can become opaque to neutrinos of energies above a few PeV. A signature in IceCube-Gen2 would be an absence of neutrinos above such energy,  accompanied by an 
excess of muon bundles from the neutrino interactions, mainly at horizontal zenith angles, where the atmospheric depth is larger. A large array is of paramount importance for these signatures in order to be able to efficiently reconstruct parallel tracks and/or muon bundles.

\newpage

%

\newpage

\section{IceCube-Gen2 design}
\label{sec:gen2_design}

The IceCube-Gen2 facility has to meet several performance requirements, outlined in Section \ref{sec:objectives}, to accomplish its science goals discussed in Section \ref{sec:Gen2Science}. We describe a preliminary baseline design that  meets these requirements in Section \ref{sec:baseline}. We also discuss the physical and practical considerations that have led to the design. All detector sensitivities used to illustrate the science capabilities of IceCube-Gen2 in Section \ref{sec:Gen2Science} are  based on this design. In Section \ref{sec:performance} we discuss basic performance indicators such as effective area and angular resolution, while in Section \ref{sec:RnD} we present the ongoing R\&D efforts that aim at improving the performance or reducing the costs and logistics footprint beyond that of the baseline design.

\subsection{Design considerations}
\label{sec:baseline}

\begin{figure}[tb]
\begin{center}
\includegraphics[width=1.0\textwidth]{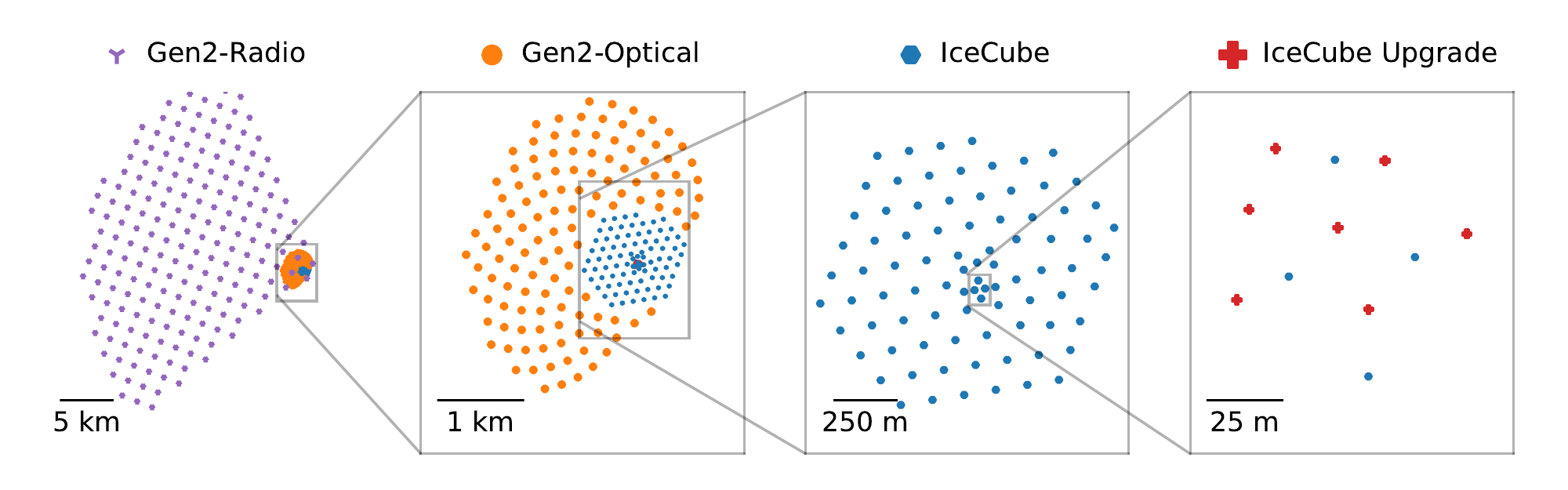}
\end{center}

\caption{Top view of the envisioned IceCube-Gen2 Neutrino Observatory facility at the South Pole station, Antarctica. From left to right: The radio array consisting of 200 stations.  IceCube-Gen2 strings in the optical high-energy array. 120 new strings (shown as orange points) are spaced 240 m apart and instrumented with 80 optical modules (mDOMs) each, over a vertical length of 1.25 km. The total instrumented volume in this design is 7.9 times larger than the current IceCube detector array (blue points). On the far right, the layout for the seven IceCube Upgrade strings relative to existing IceCube strings is shown.}
\label{fig:topview}
\end{figure}

To accomplish its science goals, the IceCube-Gen2 facility will encompass the operating IceCube detector and four new components: an in-ice optical array, complemented by a low-energy core, a surface air shower array, and an extended radio detector array. As a first step, IceCube's DeepCore is to be extended by 7 new strings which will be deployed in the near future in the IceCube Upgrade. Figure \ref{fig:topview} shows a top view of the IceCube-Gen2 facility, with its various components, each utilizing optimized technologies for the targeted energy ranges. The surface array will be installed on the footprint of the optical array.

In the following, we review the most important factors that influence the sensitivity of a large-volume optical-Cherenkov and a radio-based neutrino detector. First, we consider how the constraints that determined the final design of IceCube can be overcome to create a more sensitive facility at comparable cost. Second, we describe the design considerations for the large-volume radio array and developments that inform the design. 

\subsubsection{Optical array}

High-energy neutrinos are detected via the Cherenkov radiation induced by the relativistic, charged particles produced when the neutrinos interact with nucleons and electrons. 

Charged-current interactions of electron and tau neutrinos, as well as neutral-current interactions of all flavors of neutrinos produce secondary particles that either deposit all their energy over short distances (electromagnetic or hadronic showers) or decay quickly (tau leptons of low energies).  Compared to the average distance between sensors, the Cherenkov light emission region will appear point-like. The rate of events that can be detected is proportional to the instrumented volume of the detection.Tau neutrinos at higher energies produce a double-bang signature, where the interaction and decay vertex --- separated on average by ($E_\tau$~/~1~PeV)~$\times$~50~m  --- can potentially be resolved. With a larger instrumented volume, the number of contained double-bang events will increase compared to IceCube at energies $\gg$1~PeV, where the separation distance starts to approach the geometrical dimensions of the array.

When a muon neutrino undergoes a charged-current interaction, it produces a muon that can be detected far from the interaction vertex. For example, a 10 PeV (100 TeV) muon will travel on average 21\,km (13\,km)  before its energy drops to 10\% of its initial energy. This makes the active volume for muon neutrino detection significantly larger than the instrumented volume, and the observable muon-neutrino rate proportional to the projected geometric area of the detector (e.g. the cross-section of the detector for a beam of muons). Muon neutrino events can be separated from the background of penetrating atmospheric muons by direction and energy, and - in the case of astrophysical neutrinos from transients - also by timing. In case of down-going neutrino events with their vertex inside the detector, one can further use the outer layers of the optical array, or a CR surface array with sufficiently low threshold, as a veto against atmospheric neutrinos \cite{Schonert:2008is, Gaisser:2014bja}. 

{\bf Geometry considerations:}
We can increase the neutrino event rate by expanding the detector. One way to do this is to deploy longer strings. While the strong scattering of light in the South Pole ice at shallow depths is prohibitive, in-situ measurements \cite{Aartsen:2013rt} suggest that the 125\,m of ice above and below the IceCube instrumented volume have adequate optical properties to allow neutrino detection, leading to an 25\% increase in the geometric area for horizontal track events. The same gain is obtained for the contained volume for cascade events. 

Significantly larger gains in effective area and volume are potentially obtained by increasing the string spacing. However, it needs to be ensured that this does not hamper the detector's ability to achieve its scientific goals. For IceCube
one of the requirements was that geometry and timing calibration could be done with flashers, a set of LEDs in the upper hemisphere of each DOM that can produce short light pulses. This requirement, along with the requirement to have adequate sensitivity for muons 
of TeV energy, favored a string spacing of 125\,m.  For IceCube-Gen2 the limit on the string spacing does not apply any longer for the following reasons. First, additional calibration strategies have become available, including using CR muons~\cite{Aartsen:2013vja}, acoustic calibration modules~\cite{Heinen:2019lwd} and camera systems~\cite{Kang:2019giw} to calibrate the detector geometry and orientation of the sensors in the boreholes; Second, the baseline geometry for the optical array, shown in Figure~\ref{fig:topview}, results in an energy threshold for through-going muons of 10--30 TeV.  This suppresses the atmospheric contribution that vastly dominates the neutrino flux below 10 TeV at a negligible cost on the sensitivity to the astrophysical flux since the brighter events at higher energies can be reconstructed well even with the larger string spacing.

Typical absorption lengths at a wavelength of $\sim$400~nm are between 50~m and 200~m in the upper half of the current detector, and often exceed 200~m in the lower half.
Although the optical properties vary with the layered structure of the ice, the average absorption and scattering lengths dictate the distance the strings of sensors can be spaced apart without impacting the  response of the detector. 
Studies show that spacings of $\sim$200--300~m allow us to maintain high efficiency for detecting astrophysical neutrinos and a roughly constant sensitivity to neutrino point sources, where the advantages of a larger volume for the 300 m vs 200~m spacing are offset by a higher energy threshold and reduced directional resolution. 

A spacing of 240 m allows us to verify experimentally the performance of IceCube-Gen2 using a pruned version of IceCube, where light recorded on only 1 out of 4 strings is used for performing an analysis.  It allows us to comprehensively test the performance of such a sparse array using tools optimized for IceCube. Accordingly, we use the 240~m string spacing for the baseline design discussed in this manuscript. 

{\bf Preliminary baseline design:} The baseline design for the optical component encompasses 120 new strings that are added to the existing IceCube strings with an average horizontal spacing of 240 m. Each string hosts 80 modules, totaling 9600 new modules, between 1325 m and 2575 m below the surface. Vertical spacing between modules amounts to 16~m, resulting in an instrumented geometric volume of 7.9~km$^{3}$. Each module on the string is assumed to collect nearly three times as many photons as an IceCube DOM. 

Higher photo-collection area can be achieved via different routes, as discussed further in Section~\ref{sec:RnD}.  Already designed for the IceCube Upgrade, the mDOM would achieve the required photon collection performance if all of its 24 3" PMTs would be of high quantum-efficiency ($\gtrsim$35\%) type. Equivalent alternatives with approximately the same performance would be to deploy up to 110 mDOMs with standard quantum efficiency ($\sim$25\%) per string (instead of the 80 of the baseline design) or using slightly larger PMTs with 3.5" photocathode diameter. Being equivalent in performance, the future choice will mostly be based on overall costs. Other variants discussed in Section \ref{sec:RnD} can provide further advantage in terms of costs or sensitivity but still need to be developed. The baseline sensor assumes pixelization, i.e., using many small PMTs per sensor that provide for directional resolution of the detected photons, and hence improved reconstruction performance~\cite{KittlerTalk}.

\begin{figure}[ht]
\begin{center}
\includegraphics[width=0.5\textwidth]{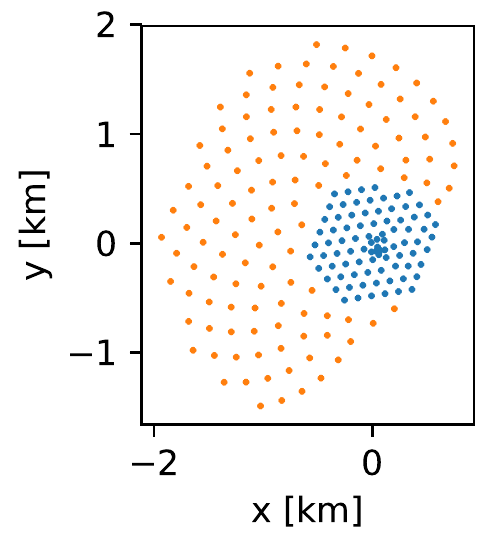}
\end{center}
\caption{The reference detector string layout (sunflower geometry) has a uniform string spacing of $\sim$~240~m, with an instrumented volume of   
7.9~km${^3}$. }
\label{fig:geometries}
\end{figure}

Figure~\ref{fig:geometries} shows the baseline geometry for the optical in-ice component. Based on the experience with IceCube, the sunflower geometry was chosen to avoid 
straight line rows and columns of strings in order to avoid "corridors" through which muons can enter and mimic starting events from neutrinos (see, e.g., \cite{Aartsen:2017bap}).  
In total, one obtains a contained volume that is 7.9 times larger compared to IceCube, while the geometrical cross-section from the side increases by a factor 3-5, depending on the incidence angle (cf. Figure \ref{fig:design:muonperformance}).

{\bf Surface veto:} \label{sec:gen2_surface}
Down-going $\gtrsim$PeV neutrinos can be identified as starting events. But it is also possible to use a surface veto to separate astrophysical neutrinos from background events. A surface veto with air shower detectors eliminates both, penetrating muons and down-going atmospheric neutrinos, as the latter are part of an air shower~\cite{Schonert:2008is,Gaisser:2014bja} as well. Downg-going events identified with a surface veto have a very high astrophysical purity. Studies with IceCube and its surface component IceTop~\cite{IceCube:2012nn} suggest that above an energy of 200 to 300\,TeV, background can be suppressed to a level lower than the astrophysical flux observed with IceCube, albeit only within the small aperture of 0.26~km$^2$sr, covering about 5\% of the sky, where the footprints of the detectors overlap. With the larger area of the IceCube-Gen2 high-energy array, the acceptance for coincident events that can be vetoed increases to $\sim$10~km$^2$sr, 40 times higher than with the current surface array. It would also cover at least 20\% of the sky. More details about the surface array and a typical surface veto station that would be deployed at the top of each IceCube-Gen2 string are presented in Section \ref{sec:surface}.

\subsubsection{Radio array}
\label{sec:radio_array}
Radio emission is generated in ice by particle showers through the \textit{Askaryan effect} \cite{Askaryan1962}. The electromagnetic component of the shower evolves over time as additional electrons are up-scattered from the ice mostly through the Compton effect and positrons are depleted by in-flight annihilation. This leads to a relativistically moving negative charge excess in the shower front and a charge separation along the shower axis. Macroscopically speaking, a dipole is formed that changes as the shower develops and thereby emits radiation. Due to coherence effects, the emission is only strong at angles close to the Cherenkov angle, where all emitted frequencies arrive at the same time at an observer. The emitted frequency range is governed by the shower geometry and the spectrum typically shows the strongest contribution between 100 MHz and 1 GHz \cite{AlvarezMuniz:2011ya}. In the time-domain, the emission corresponds to a broad-band nanosecond-scale radio pulse, which has been observed both at accelerator experiments \cite{Saltzberg:2000bk,Gorham:2004ny,Gorham:2006fy} and in air showers \cite{Aab:2014esa,Schellart:2014oaa,Scholten:2016gmj}. The fact that radio emission is generated by both purely electromagnetic showers and the electromagnetic component of hadronic showers means that a radio detector is sensitive to all flavors, albeit with different sensitivities. While hadronic showers initiated by all neutrino flavors and secondaries are detected with the same efficiency, due to the LPM-effect \cite{Landau:1953um,Migdal:1956tc} there is a preference for $\nu_e$-induced, purely electromagnetic, showers only below EeV energies.  The LPM-effect significantly stretches electromagnetic showers at EeV energies, which suppresses the high frequency emission and introduces measurable shower to shower fluctuations \cite{AlvarezMuniz:2010ty}, although, at higher energies, the presence of photonuclear and electronuclear interactions (not subject to LPM suppression) moderate the increase \cite{Gerhardt:2010bj}.
Given a suitable geometry a radio detector will detect both showers following the interaction of a $\nu_{\tau}$ and the $\tau$ decay, as well as stochastic energy losses of high-energy $\tau$'s or $\mu$'s \cite{Dutta:2000hh}. Taking into account secondary interactions, the radio detection method exhibits roughly equal effective volumes for all three neutrino flavors at the highest energies \cite{Garcia-Fernandez:2020dhb}.

The signal amplitude scales linearly with the shower energy, giving radio detection an energy threshold above the thermal noise floor of a couple of PeV per shower, depending in detail on trigger and antenna gain. The attenuation length of radio waves in ice depends on depth and temperature. At the South Pole and other locations with cold ice it is roughly of the order of one kilometer \cite{Barwick:2005zz}, meaning that showers can be viewed from afar. It should be noted that the index-of-refraction profile of naturally occurring ice-sheets follows a typical behavior \cite{kravchenko_2004}. Snow accumulating at the surface is compacted and recrystallizes into the so-called \textit{firn} where the density increases with depth. This results in an exponentially increasing index of refraction from roughly 1.30 at the surface to 1.78 at the transition to the ice, reaching 1.77 at roughly a depth of 150 m at South Pole \cite{Barwick:2018rsp,Abdul:2017luv}. Due to the exponential profile, the radio signals do not travel in straight lines in the firn, but bend downwards. This simple model of the firn structure results in zones close to the surface that a signal can never reach, as no ray-tracing solution exists, so-called \textit{shadow zones} \cite{kravchenko_2004,Barwick:2005zz,Kravchenko_2006}. Also, the downward refraction of upward moving signals and reflection at the firn-air boundary increase the effective volume available to a detector at a certain depth \cite{Glaser:2019cws,Kelley:2017xjj}.

In order to reconstruct the energy of a shower, resolving the vertex location is key. The amplitude of the radio signal drops as $1/r$ with distance to the vertex. The signal at the detector is furthermore reduced by the attenuation over the path traversed from vertex to detector. To reconstruct the vertex location, several routes can be followed: timing in the wave-front can be used to determine the vertex as the origin of a spherical wave. This will be easier the closer the vertex is to the detector. Alternatively, both the direct emission to the receiver and the one reflected at the surface could be measured. The amplitude ratio and the time difference of the two pulses uniquely tag the distance of the vertex \cite{Kelley:2017xjj,Anker:2019zcx}. 

In order to reconstruct the arrival direction of the neutrino, one needs both the arrival direction of the signal at the antenna and information regarding where on the Cherenkov cone the signal was detected. The signal arrival direction can be obtained from timing, given enough antennas, sub-nanosecond time resolution, and a suitable distance between antennas. Using this information alone, the neutrino arrival direction can only be limited to a ring-like region on sky, a projection of the Cherenkov cone, limited only by the field of view of the detector. But the electric field provides, apart from its amplitude, also information about the polarization and frequency spectrum. Using the fact that the polarization always points towards the shower axis and that the frequency content is determined by the viewing angle to the shower axis (and Cherenkov angle), one can determine the arrival direction of the neutrino to within degree-scale precision, given a suitable resolution on both parameters \cite{Glaser:2019rxw,Glaser:2019kjh}.

{\bf Station geometry considerations:}
As the radio signal travels long distances in the ice, most prototype radio detectors have chosen a compact station design. A station typically consists of a cluster of antennas and acts independently in both triggering and reconstructing the shower. This is different from the approach in the optical detector.  For an array, radio stations are placed far apart from each other so that they have almost independent effective volumes thereby maximizing the total effective volume. 

When optimizing a single station, various considerations play a role: one would like to use antennas with the highest gain to optimize sensitivity for low-amplitude signals, which both lowers the energy threshold and increases the distance at which interactions can be detected. Also, one desires the best polarization sensitivity and the broadest frequency response to optimize the input for the arrival direction reconstruction, as discussed above. Such antennas are typically too large to be deployed in deep holes obtainable by drilling. One can therefore either stay at the surface and remain flexible with respect to the antenna type, or go deep and accept the limitations given by the borehole geometry. In the latter case, the lack in antenna gain can be overcome by phased-array techniques that add signals from several antennas, thereby digitally mimicking an antenna with higher gain \cite{Vieregg:2015baa,Avva:2016ggs,Allison:2018ynt}. A higher antenna gain lowers the energy threshold of a given detector and increases the effective volume, as signals of the same strength can be detected at larger distances. 

It has been argued that air shower signals may create a background for radio detectors of neutrinos, either by catastrophic energy losses from muons, refraction of the signal into the ice or from not fully developed air showers \cite{Garcia-Fernandez:2020dhb,deVries:2015oda}. As such signals are due to the same emission mechanisms, they are in principle indistinguishable from neutrino-induced signals when detected with a single antenna. Adding antennas sensitive to signals arriving from above at the surface provides every station with a self-veto. Furthermore, the footprints of radio signals from air showers are much wider than those from neutrinos (different index-of-refraction profile and (almost) no attenuation in air) and are therefore often detected in multiple stations, which improves the veto efficiency. The signals from air showers also provide detector calibration opportunity and possibly CR science \cite{Barwick:2016mxm}. 

{\bf Station depth considerations:}
At South Pole, the deeper the antennas of a station, the larger its effective neutrino volume and sky coverage, as more and cleaner signal trajectories will be able to reach the detector. This effect can be understood in the context of ray bending in the changing index of refraction in the firn. The advantages of going deep have to be balanced against the opportunity for energy reconstruction in detecting a direct and a reflected signal at intermediate depths \cite{Anker:2019zcx}, the costs of drilling and the restrictions in antenna geometry, with narrow antennas typically providing a poorer polarization sensitivity and lower gain. Still, in comparison to the optical array, there is little to be gained from great depths and the radio array can always remain comparatively close to the surface.  

{\bf Station spacing considerations:}
To maximize the total effective volume, stations should be placed as far apart from each other as needed to no longer have overlapping effective volumes. As shown in \cite{Glaser:2019cws} the minimal distance is a function of energy and may be as large as 2 km, depending on the depth of the station. The total effective volume then scales linearly with the number of stations. However, large distances place a heavier burden on deployment and operations logistics and one may consider coincidences between stations helpful in improving confidence in reconstruction methods to obtain events of extremely high quality.

\begin{figure}[tb]
    \centering
    \includegraphics[width=0.55\textwidth]{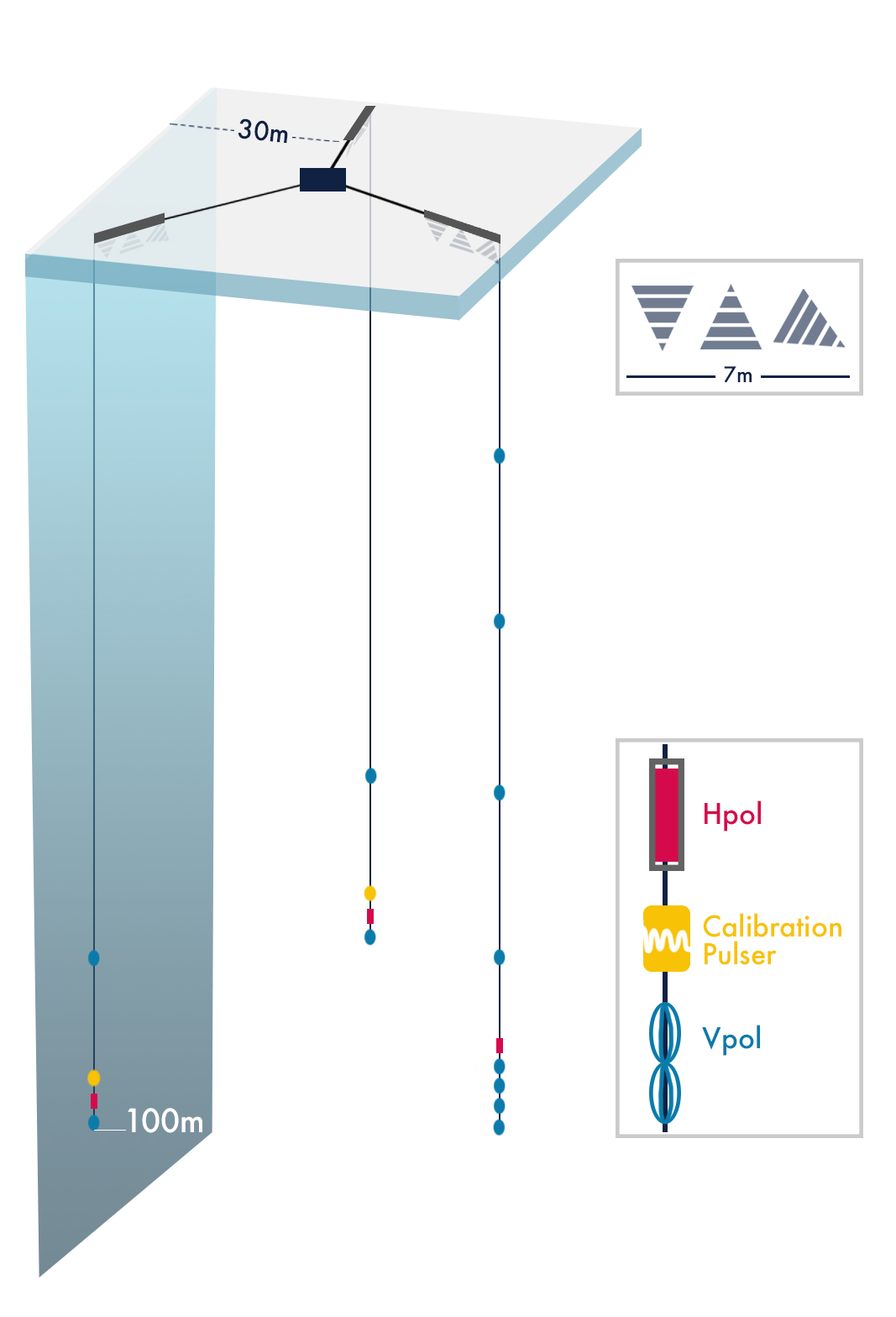}
    \caption{Station layout of the preliminary baseline design. The station consists of three strings to the depth of 100 meters. One string hosts the phased array trigger and additional antennas to reconstruct the vertex position. Antennas on the two other strings allow the reconstruction of the arrival direction and the electric field. Two strings host calibration pulsers to triangulate the antenna positions. At the surface, 9 log-periodic dipole antennas are deployed in 3 slots each to provide additional reconstruction information, as well as an air-shower veto.}
    \label{fig:Radio_component}
\end{figure}

{\bf Preliminary baseline design:}
The baseline design for the radio component of IceCube-Gen2 is informed by both simulations and the experience gained from operating the pilot arrays ARA and ARIANNA \cite{Allison:2011wk,Allison:2015eky,Barwick:2014rca,Barwick:2014pca} and assembled to meet the science goals outlined in Section~\ref{sec:objectives}.

The autonomous stations combine shallow, sub-surface log-periodic dipole antennas with cylindrical dipole antennas and slot antennas at depths of down to 100 m. The main trigger is provided by a phased-array at 100 m. The shallow antennas offer improved signal sensitivity, both with respect to the frequency range and polarization, as well as veto-capabilities against air showers. The deeper antennas provide a larger effective volume and sky coverage. A fraction of neutrino events will be detected in all detector components, marking a sample of events with the best reconstructed properties. 
In order to reach the overall sensitivity goal, 200 stations would be deployed at the South Pole over an area of 500 km$^2$. 

Simulations for this design, which inform the performance studies and science capabilities described in Section~\ref{sec:Gen2Science}, use the trigger level sensitivities of the phased array\footnote{As shown in \cite{Allison:2018ynt}, the phased array can be approximated by a $1.5\sigma$ amplitude over RMS noise threshold in a single dipole, which significantly speeds up the simulations and presents an average performance over all viewing angles of the hardware implementation of the phased array.} at a depth of 100 m.

\subsection{Instrument performance}
\label{sec:performance}
Section~\ref{sec:Gen2Science} discussed in detail the capabilities of the Icecube-Gen2 facility to address critical science questions in neutrino and multi-messenger astronomy. Here, the underlying instrument performance that enable the science is discussed. We describe the performance of the optical 
and radio component separately first and conclude with a discussion about the advantages of a combination of both techniques in a single location that goes beyond the sum of the individual parts. 

\subsubsection{Optical array}

\begin{figure}[tb]
\begin{center}
\includegraphics[width=\textwidth]{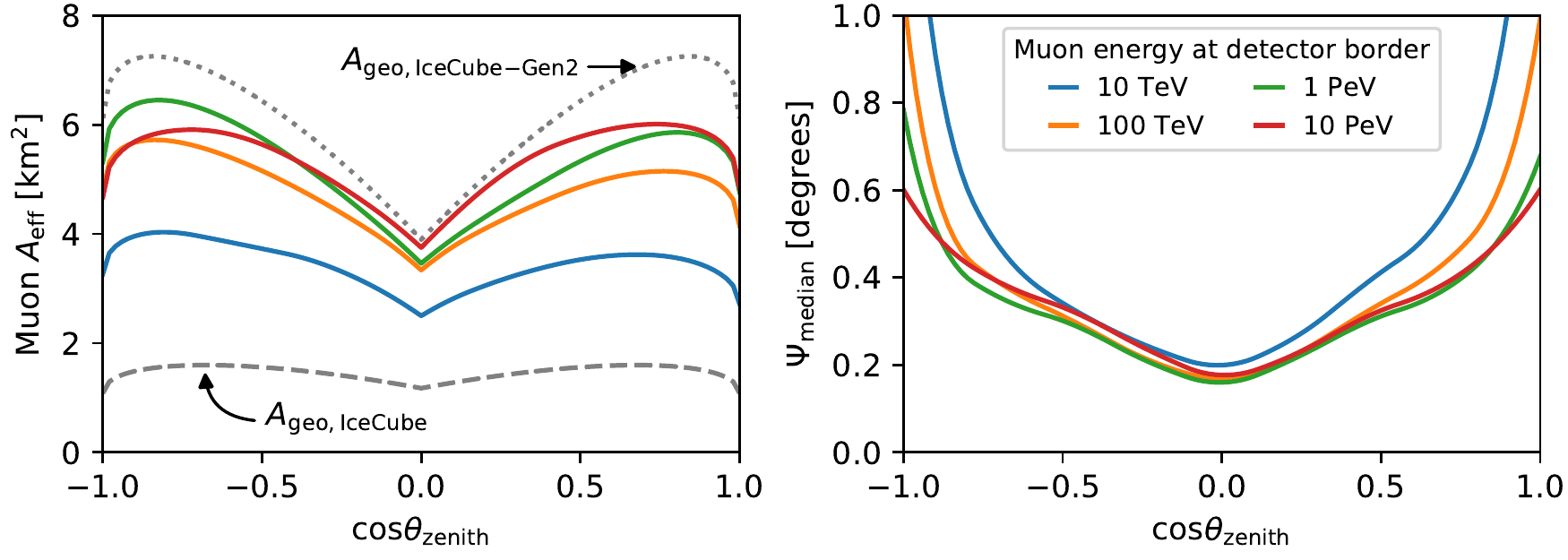}
\end{center}
\caption{Muon reconstruction performance of the IceCube-Gen2 optical array with 240 m string spacing. {\it Left:} Effective area for single muons after quality cuts as a function of zenith angle for different muon energies. {\it Right:} Median angular error of the directional reconstruction for muons evaluated at the same energies.  }
\label{fig:design:muonperformance}
\end{figure}

We characterize the performance of the baseline design as shown in Figure~\ref{fig:geometries} by studying simulation and IceCube experimental data. These studies focus on two neutrino detection channels: muons entering the instrumented volume from the outside and isolated cascades within the volume.  To estimate the detector performance for through-going muons, an isotropic flux of muons entering the detector is simulated, using the standard IceCube simulation chain. IceCube-like sensors are assumed, however each simulated sensor features a 3 times larger PMT photocathode area than the actual IceCube DOM (approximating the photon collection efficiency of the IceCube-Gen2 baseline sensor). Each muon's direction and energy are reconstructed using IceCube's reconstruction algorithms. We then apply a series of quality cuts to these simulated events to remove any whose direction cannot be reconstructed reliably. From these simulated data we derive the muon effective area as a function of energy and zenith angle, and the point spread function (PSF) as shown for the 240\,m sunflower layout in Figure~\ref{fig:design:muonperformance}. These quality cuts are responsible for the difference between the effective and geometrical areas, as, e.g., corner-clipping tracks that cannot be well reconstructed are removed. Note that the PSF is scaled by a factor of 0.8 to account for reconstruction improvements related to pixelated sensors that are not included in the standard IceCube reconstruction chain, and have been studied separately~\cite{KittlerTalk}.

\begin{figure}[tb]
\begin{center}
\includegraphics[width=0.49\textwidth]{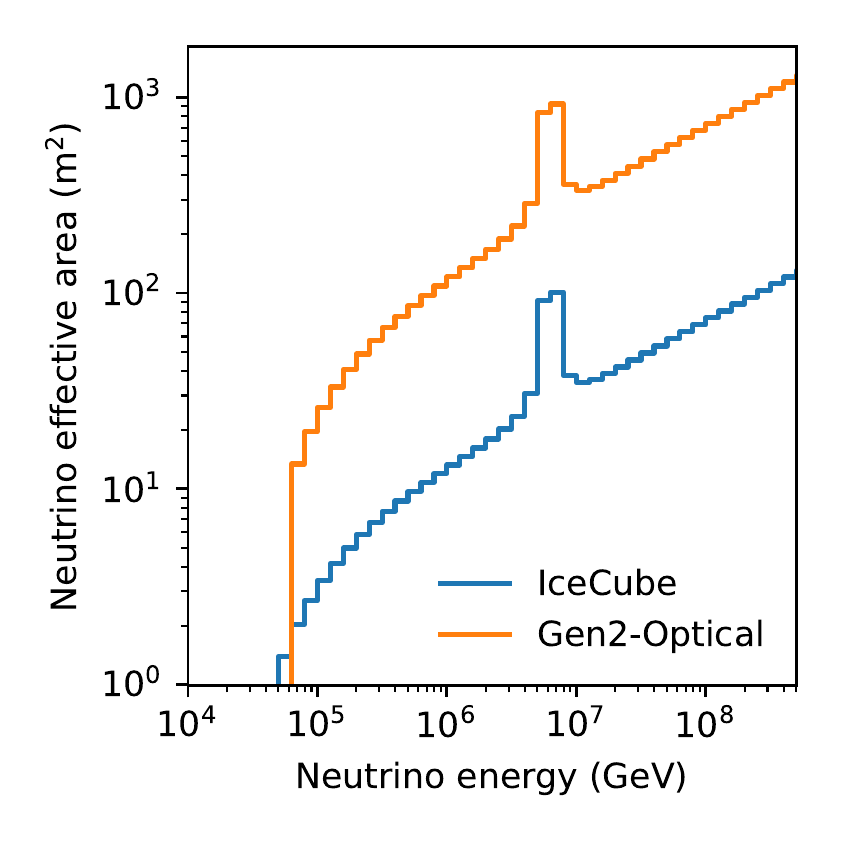}
 \includegraphics[width=0.49\textwidth]{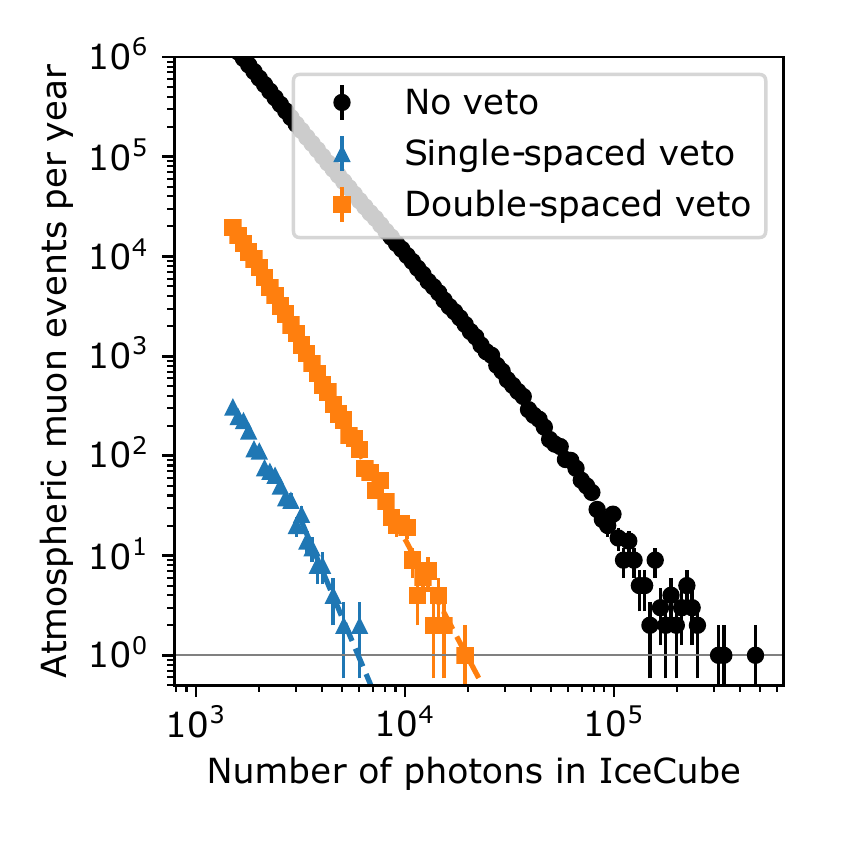}
\end{center}
\caption{Effective area for shower-type events and background rejection performance of IceCube and the IceCube-Gen2 in-ice detector with 240 m string spacing. {\it Left:} Effective area for cascades as a function of neutrino energy, after cuts that ensure a reliable vertex reconstruction. {\it Right:} Penetrating muon rejection capabilities of IceCube derived from experimental data and using the information from all strings (blue points) and only 25\% of the strings (orange points) equivalent to a detector with $\sim$~240~m string spacing. The number of collected photons is a proxy for deposited energy. The wider string spacing implies a factor 3 higher energy threshold to ensure an equivalent rejection performance for atmospheric muons as for IceCube. }
\label{fig:design:contained}
\end{figure}

We characterize the performance for cascades in a similar way. Instead of simulating a flux of incoming muons, though, we simulate single electromagnetic cascades distributed evenly throughout the instrumented volume, and reconstruct each cascade's position and energy. The quality cuts for cascades ensure that the position reconstruction is reliable, and thus that the associated energy estimate is as well. Figure~\ref{fig:design:contained} shows the effective areas for cascade detection compared to IceCube.

To fully characterize the sensitivity to neutrinos, we must also consider the background from penetrating atmospheric muons. IceCube has already demonstrated several successful strategies for removing this background~\cite{Aartsen:2013jdh,Aartsen:2014muf,Aartsen:2016xlq,Aartsen:2020aqd}. Here we focus on how to extrapolate these techniques to IceCube-Gen2, with emphasis on the resulting energy threshold. Entering muons from below the horizon are relatively simple to separate from atmospheric muons. Correctly reconstructed muons can only be neutrino-induced, and cuts on the angular reconstruction quality are sufficient to remove all atmospheric muons.

Neutrino interactions inside the detector volume are slightly more difficult to isolate, requiring that the outer layer of the detector be used as a veto. We estimate the effective energy threshold of such a veto strategy using a variant of the technique used in \cite{Aartsen:2013jdh},
where the observation of Cherenkov light in the outer string layer of IceCube above a certain threshold leads to the rejection of the event.  The results are shown in Figure~\ref{fig:design:contained}. First, we identify penetrating atmospheric muon events that first trigger the outer layer of IceCube. We then apply the veto using only the inner detector, with three out of four strings removed to model a detector with twice the string spacing. Compared with a veto using all strings, the energy threshold increases by a factor of $\sim$3. From this we conclude that a high-energy starting event selection in a detector with $\sim$240~m - 250~m string spacing will be fully efficient and have negligible penetrating muon background above a deposited energy of ~200 TeV, rather than 60 TeV as in IceCube. 

IceCube-Gen2's efficiency to identify tau neutrinos via their double-bang signature is estimated in a similar way, i.e., with the help of a ``sparse'' IceCube dataset where only the information from every 4th string is used to model a detector with 240~m string spacing. This efficiency is included in the estimate of the flavor composition measurement performance discussed in Sections \ref{Sec:Spectrum} and \ref{sec:flavor_new_physics}. Both, the tau neutrino identification efficiency and the energy threshold for negligible penetrating muon background are conservative estimates of the true quantities, as IceCube-Gen2 will collect 3 times more photons per string in comparison to IceCube. Future detailed simulation studies will provide more accurate values. The instrument response functions for muon and shower-type events introduced in this section are the basis for the evaluation of the science potential of the IceCube-Gen2 observatory presented in Section~\ref{sec:Gen2Science}.

\subsubsection{Radio array}

\begin{figure}[tbp]
    \centering
    \includegraphics[width=0.8\textwidth,trim={0cm 1.5cm 0cm 2.5cm},clip=True]{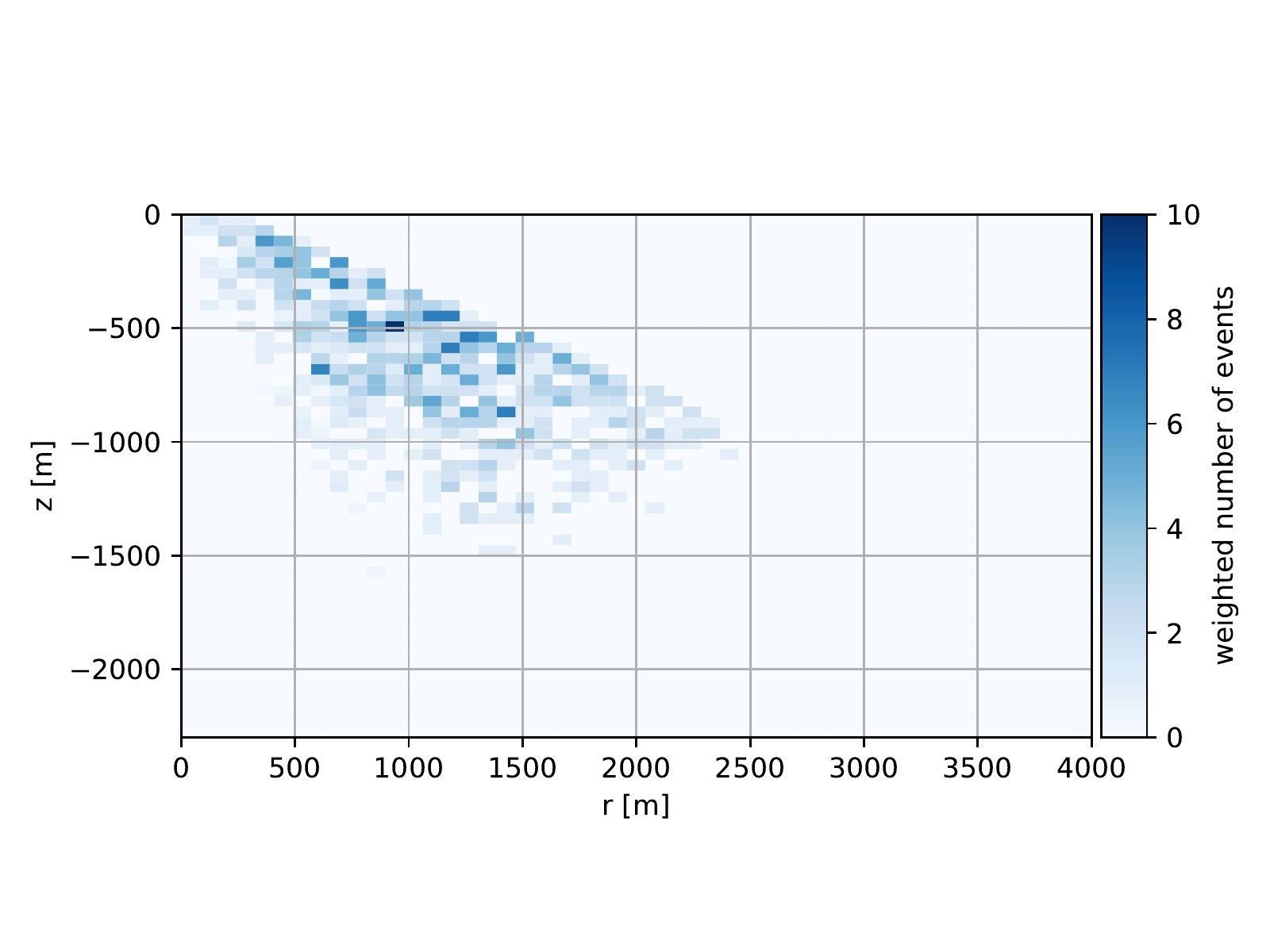}
    \includegraphics[width=0.8\textwidth,trim={0cm 1.5cm 0cm 2.5cm},clip=True]{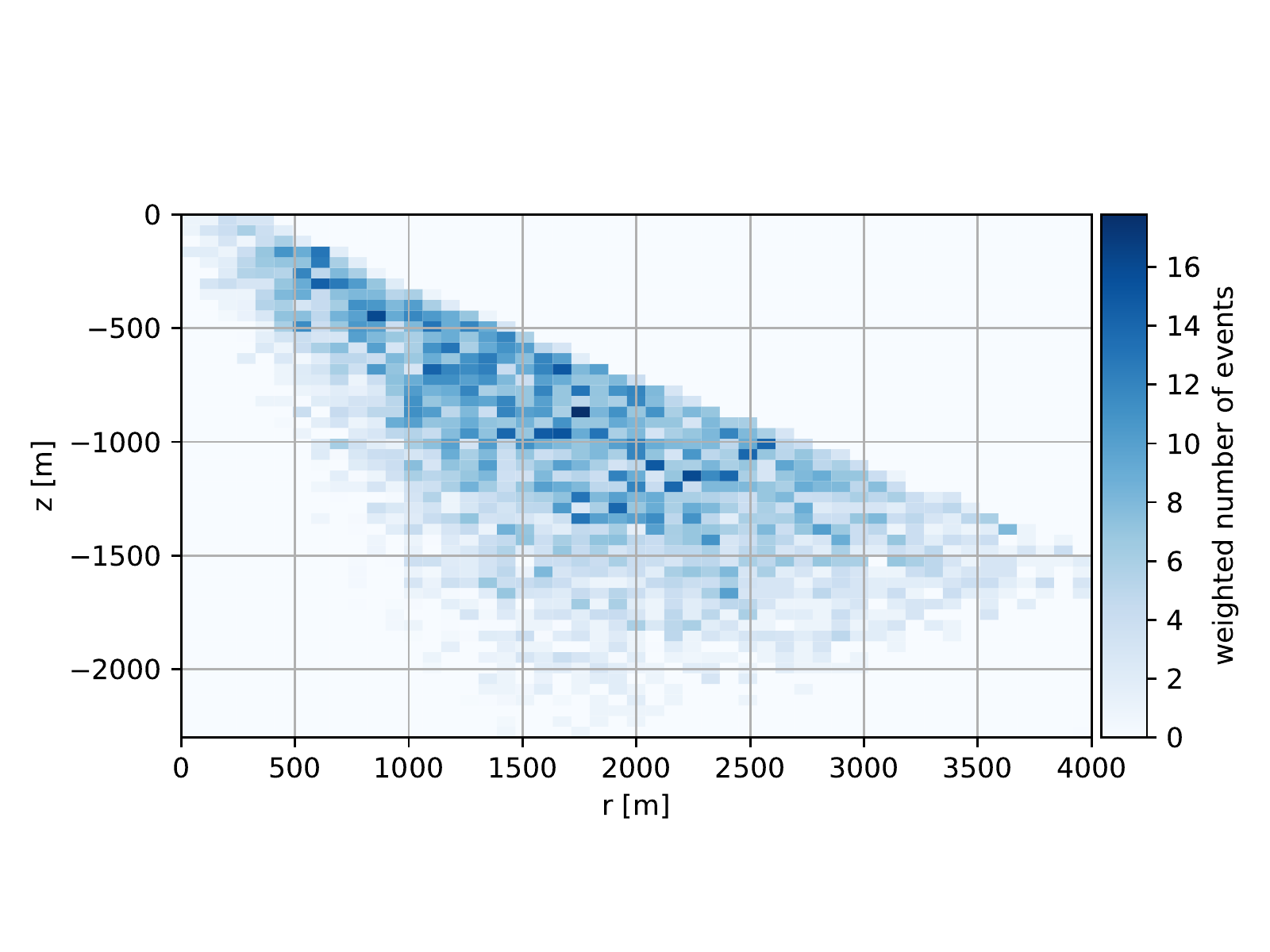}
    \caption{Neutrino vertex locations as function of radial distance $r$ and depth $z$ for a neutrino energy of 10$^{17}$ eV (top) and 10$^{18}$ eV (bottom) that can be detected by a baseline station. The color scale indicates the weighted number of events as obtained from the simulations, where the same number of events are generated per energy bin and weighted according to their arrival direction. The characteristic triangular shape is determined by the ray-tracing solutions arriving at the detector. }
    \label{fig:radio_vertices}
\end{figure}

Based on simulations \cite{Glaser:2019cws} and past experience from prototypes and pilot-stage radio detectors such as RICE \cite{Kravchenko_2006}, ARA \cite{Allison:2011wk} and ARIANNA \cite{Barwick:2014rca}, we extrapolate the performance of the radio array. Simulations show that the baseline stations will be sensitive to emission from neutrino interaction vertices at large distance as depicted in Figure \ref{fig:radio_vertices}, translating to large geometric volumes per station. As the signal amplitude scales with energy, the visible vertex distances are significantly larger at higher energies. Despite being relatively close to the surface, the baseline station will be sensitive to neutrinos interacting close to the bottom of the ice sheet. 

The energy resolution obtainable with the radio detector depends on both the reconstruction of the energy fluence in the electric field and the reconstruction of the vertex distance, as discussed in Section~\ref{sec:radio_array}. The obtainable resolution of the electric field depends in detail on the antenna design chosen. It has been shown with radio detection of air showers that one can reconstruct the detected radio signal to better than 10\% in the absolute energy fluence of the pulse \cite{Aab:2015vta,Aab:2016eeq,Gottowik:2017wio}, with antenna modelling being the driving systematic uncertainty. 

The obtainable resolution of the vertex position will strongly depend on the vertex distance. For close vertices, an uncertainty on the resolved vertex distance of the order of tens of meters is likely obtainable, following the location reconstruction of pulser events \cite{Abdul:2017luv, Allison2019longbaseline,Anker:2019mnx} and early simulation studies. For more distant vertices the reconstruction of the vertex position will be equally good, if both the direct and reflected signal can be measured, as was shown by ARA and ARIANNA using pulser signals \cite{Anker:2019zcx,Allison2019longbaseline}. In general, it is likely that the energy resolution will be dominated by the unknown neutrino energy fraction deposited into the cascade. Systematic uncertainties will stem from ice properties such as the attenuation length and index-of-refraction profile. Both will be addressed in calibration campaigns. 

The resolution of the neutrino arrival direction is determined by the combination of the signal arrival direction, signal polarization, and the viewing angle with respect to the Cherenkov angle.

The accuracy of the reconstruction of the radio signal arrival direction can be extrapolated from the reconstruction of the position of deep pulsers on IceCube strings and those lowered in the SPICE borehole \cite{Lu:2017bcz, Anker2020_pol,Allison2019longbaseline}, as well as from the reconstruction of the radio signal of air showers \cite{Apel:2014usa,Corstanje:2014waa}. It is expected to be better than 1$^{\circ}$. 

The polarization will be reconstructable to better than 10$^{\circ}$, as shown in CR measurements with ARIANNA \cite{Glaser:2019rxw} and likely better than 3$^{\circ}$ as found for deep pulser signals \cite{Anker2020_pol}. The angle to the Cherenkov angle will be reconstructed to better than 2$^{\circ}$, extrapolating from frequency slope measurements of CR~\cite{Barwick:2016mxm,Glaser:2019rxw,Welling:2019scz} and dedicated simulations.
All of the referenced measurements of signal characteristics were obtained with high-gain log-periodic dipole antennas. Preliminary simulations indicate a similar performance for the baseline stations, depending, however, in detail on the number of antennas with signal and the amplitude of the recorded signal.  
 
Combining these uncertainties results in an asymmetric, curved uncertainty region for the neutrino direction of roughly 10$^{\circ} \times$~2$^{\circ}$. It is expected that dedicated reconstruction algorithms will improve on this value, especially for those events detected in more than three antennas or in coincidence of two stations. 

\subsubsection{Combined performance}
There is the option to optimize the positioning of parts of the detector to provide a maximum chance to see coincidences  between the optical and the radio detectors. Several detection channels are possible such as the observation of the first interaction shower with the radio detector and the outgoing muon with the optical component. Such events would provide unprecedented pointing. Achieving a measurable rate of coincidences will require a dedicated optimization of the positioning of the detectors that goes beyond the scope of this white paper. Studies are on-going and will conclude before the final design phase. 

The radio and the optical detector are complementary in their energy range and provide synergistic sensitivity to different detection channels. Combining both techniques will make IceCube-Gen2 a formidable instrument sensitive across a much broader energy range than previously accessible to IceCube. The individual instrument response functions obtained for the radio and optical array, the effective area, the PSF and the energy resolution, are combined to estimate the sensitivity of IceCube-Gen2 to various astrophysical neutrino production scenarios in Section~\ref{sec:Gen2Science}. Potential gains from a coincident detection of a subset of events have not been included in these estimates.

\subsection{Detector development}
\label{sec:RnD}

The design of IceCube-Gen2 builds on the experience of constructing and operating the IceCube detector, the sensor R\&D and construction effort for the IceCube Upgrade, as well as different radio pilot arrays. It profits from a range of recent technical advances in sensor development. These are currently evaluated to determine their feasibility and effectiveness for the IceCube-Gen2 observatory. In the following, we describe current R\&D activities on the sensors, surface detectors, as well as radio components. 

\subsubsection{Optical sensors}

Besides its rich physics program, the IceCube Upgrade project provides also an opportunity to test the optical sensor technology for the IceCube-Gen2 observatory and establish a baseline technology. 

\begin{table}[tb]
\caption{Key parameters of optical sensors designed for the IceCube Upgrade.}
\begin{center}
\begin{tabularx}{0.65\textwidth}{l|ccc}
\hline
Sensor & mDOM & D-Egg  \\
\hline
Number of PMTs & 24 & 2 \\ 
PMT diameter & 3" & 8"  \\ 
Module diameter [cm] & 36 & 30  \\
Effective photocathode area [cm$^2$] & 65 & 43  \\
\hline
\end{tabularx}
\end{center}
\label{tab:mdom_degg_lom}
\end{table}

\begin{figure}[tb]
\begin{center}
\includegraphics[width=0.46\textwidth]{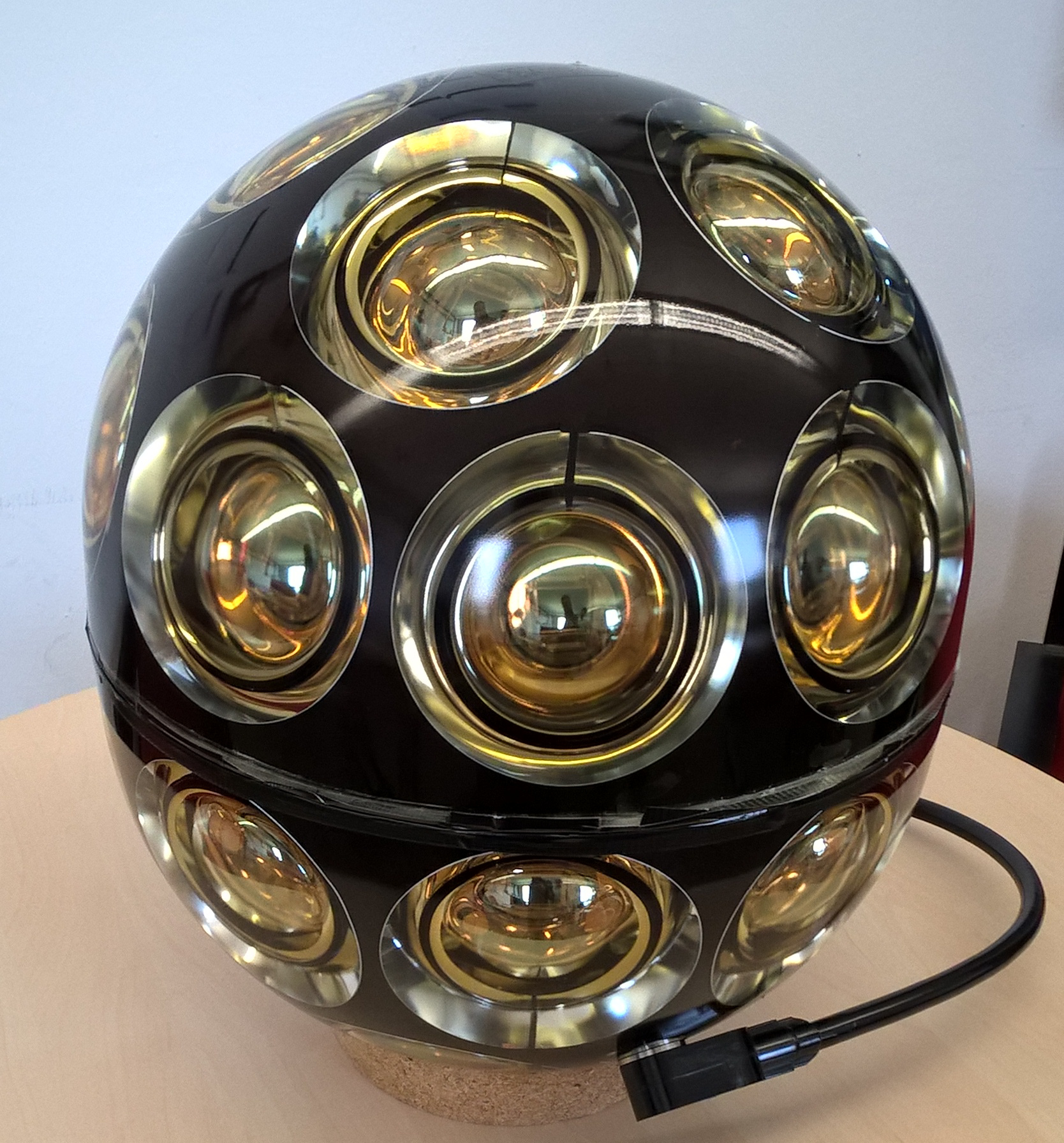}
\hspace{0.02\textwidth}
\includegraphics[width=0.32\textwidth]{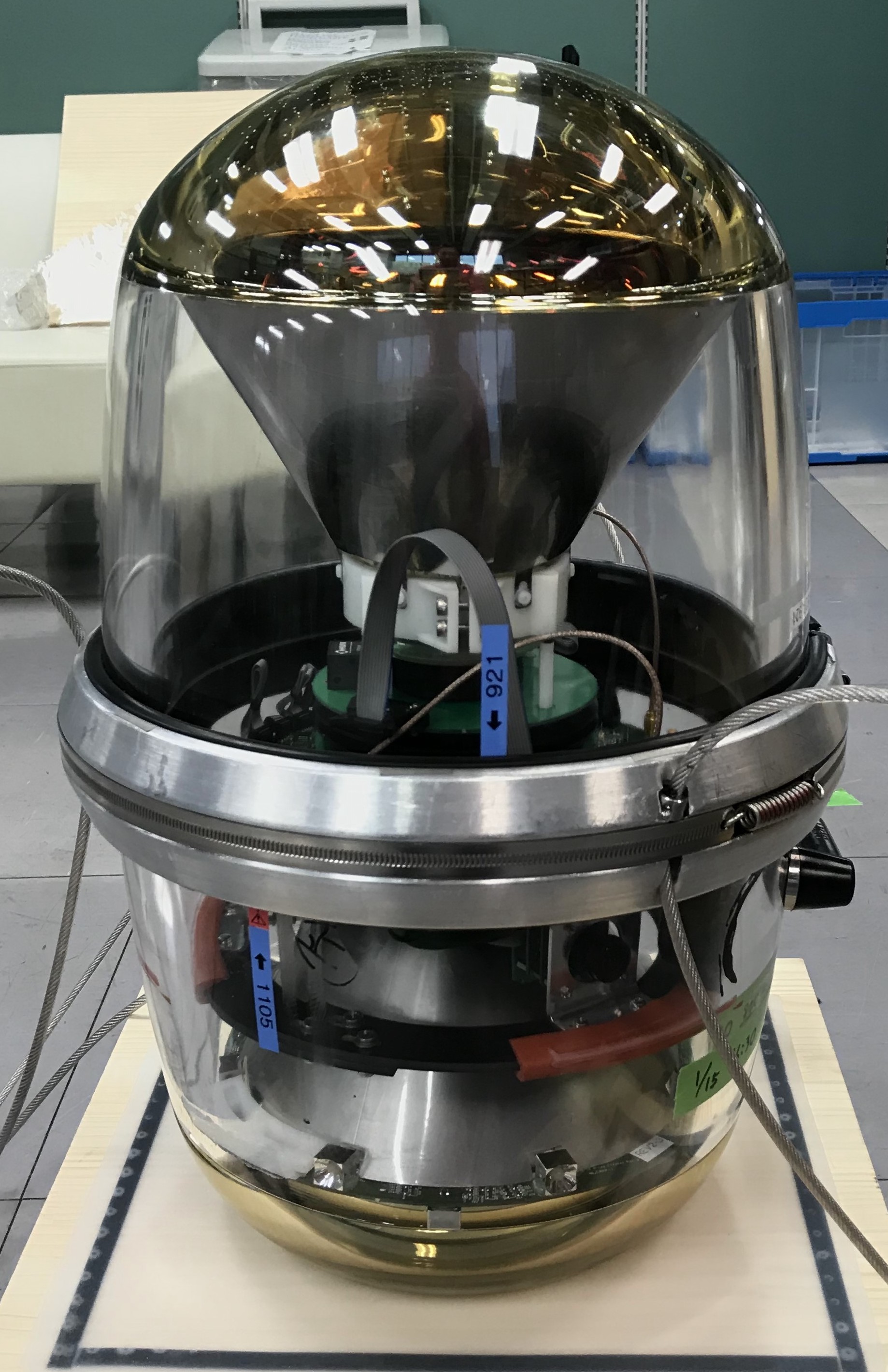}
\end{center}
\caption{Prototypes of an mDOM (left) and an D-Egg (right) for the IceCube Upgrade.}
\label{fig:mdom_degg_lom}
\end{figure}

{\bf mDOM:} The mDOM~\cite{Classen:2017sng,Classen:2019tlb} is one of these sensors. 402 out of 693 sensors to be deployed along the 7 new strings of the IceCube Upgrade will be of this type. The main features are shown in Figure \ref{fig:mdom_degg_lom} and summarized in Table~\ref{tab:mdom_degg_lom}.
In contrast to IceCube's single large 10" PMT, the mDOM used in the IceCube Upgrade consists of 24 smaller 3" PMTs. The key advantages of the mDOM are its factor of 2.2 higher effective photocathode area, the omnidirectional sensitivity and the directional 
information obtained from the individual ``pixels'' (the 24 PMTs). The slightly larger dimension of the pressure housings (14" in comparison to IceCube's 13") implies a small increase ($\sim$~7\%) in the diameter of the drill hole for deployment and hence a minor increase of the per-hole drill costs and time when compared to IceCube.

Due to the maturity of the design and the extensive in-situ testing with a large quantity of sensors to be performed in the IceCube Upgrade, we consider an mDOM-type sensor as the baseline sensor for IceCube-Gen2 for the purpose of evaluating the science potential and budget of the IceCube-Gen2 observatory. However, we continue to study alternative sensor types and evolve designs to optimize the science potential and cost effectiveness of the optical array.
Reducing the diameter of the module's pressure housing is a particularly interesting avenue, as it has impact on the required diameter of the borehole. For IceCube, the bore hole was drilled to a diameter of about 60\,cm.  Drilling a single hole required about 32 hours. Smaller assemblies would allow for a smaller diameter borehole. Based on past drilling simulations and data, a sensor with a diameter reduced by 5\,cm would allow a savings of 15 \% in drill time and fuel consumption. Such a difference could easily result in one less season required for construction (see Section~\ref{sec:schedule} for a construction schedule overview).

{\bf D-Egg:} An alternative design to the mDOM --- following a different approach ---  is the D-Egg~\cite{Ishihara:2017vxn} shown in Figure \ref{fig:mdom_degg_lom} with its properties in Table \ref{tab:mdom_degg_lom}.  277 D-Eggs will be deployed in the IceCube Upgrade. The two 8" PMTs of the D-Egg provide up/down coverage. While this yields only limited directional information, the smaller number of PMTs allows to build modules with a lower power consumption than the mDOM. The transparency of the glass vessel has been improved in the 300--400~nm range, with respect to IceCube's DOM. And, while the D-Egg collects in total less photons per sensor than the mDOM, it's pressure housing with a diameter of slightly below 12" will fit into a smaller borehole, reducing deployment costs. 

The instrumentation costs of a sensor are driven not only by the cost of the PMT, but to a significant degree by the cost for other parts such as electronics and pressure housing (in the case of IceCube, the PMT accounted for only about one-quarter of the cost of a complete DOM). This can offset the savings in drill time and cost from the smaller boreholes. A detailed calculation is necessary to determine which approach --- mDOM or D-Egg --- is the more cost effective. 

A hybrid approach can also be considered, where a D-Egg pressure housing is equipped with multiple small PMTs with a photocathode diameter between 3" and 5". Such or other moderate evolutions of the mDOM and D-Egg design used in the Upgrade are expected for the optical sensors deployed in IceCube-Gen2. This includes the choice and number of PMTs, improvements in the electronics, the mechanical structure and the format of the pressure housing, aiming to obtain the most cost and power-efficient solutions for a large scale array of more than 10000 sensors as planned for IceCube-Gen2. 
\subsubsection{Surface detector array for cosmic rays}
\label{sec:surface}
\begin{figure}[tb]
    \centering
    \includegraphics[height=6cm]{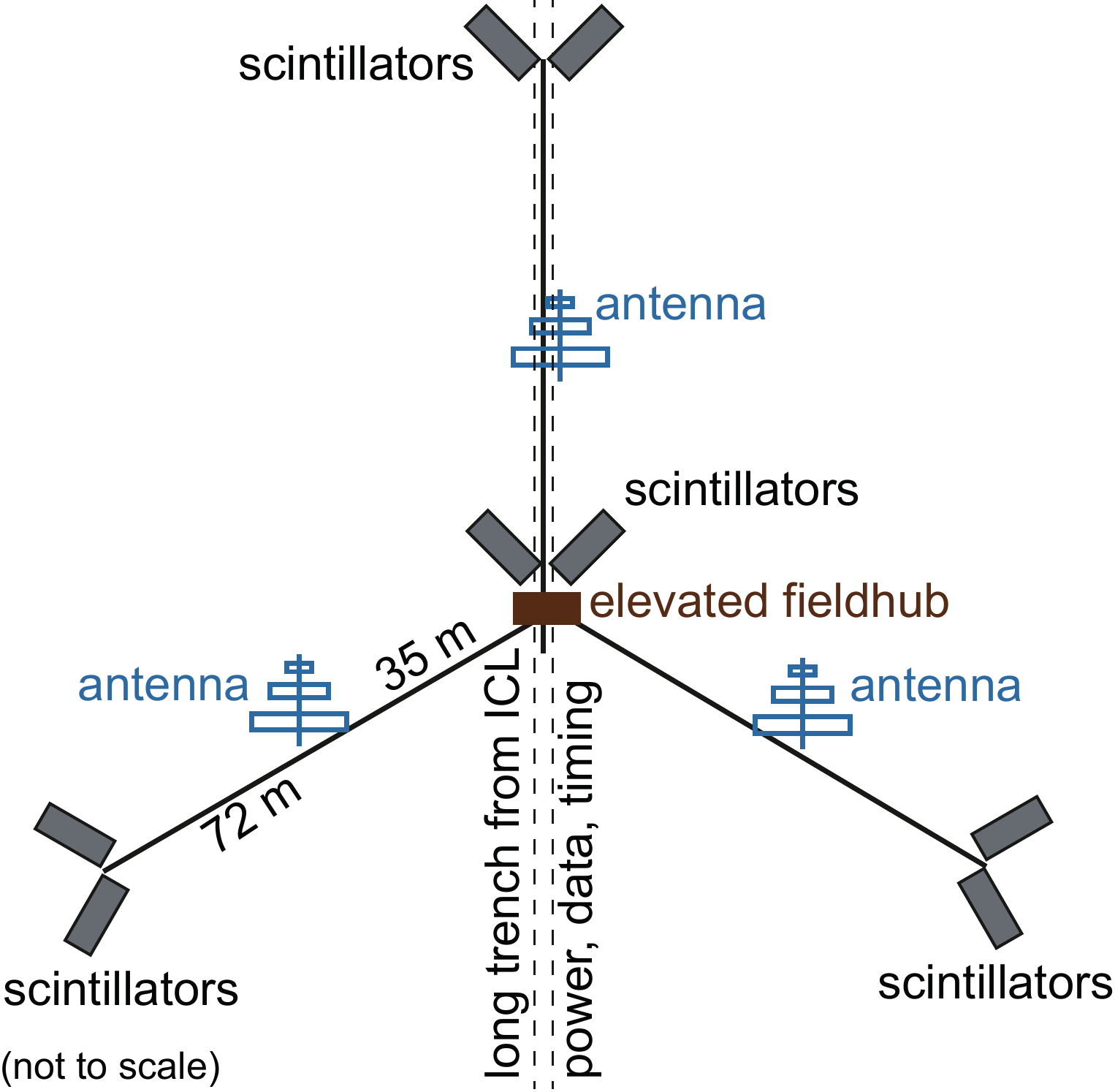}
    \hspace{1cm}
    \includegraphics[height=5cm]{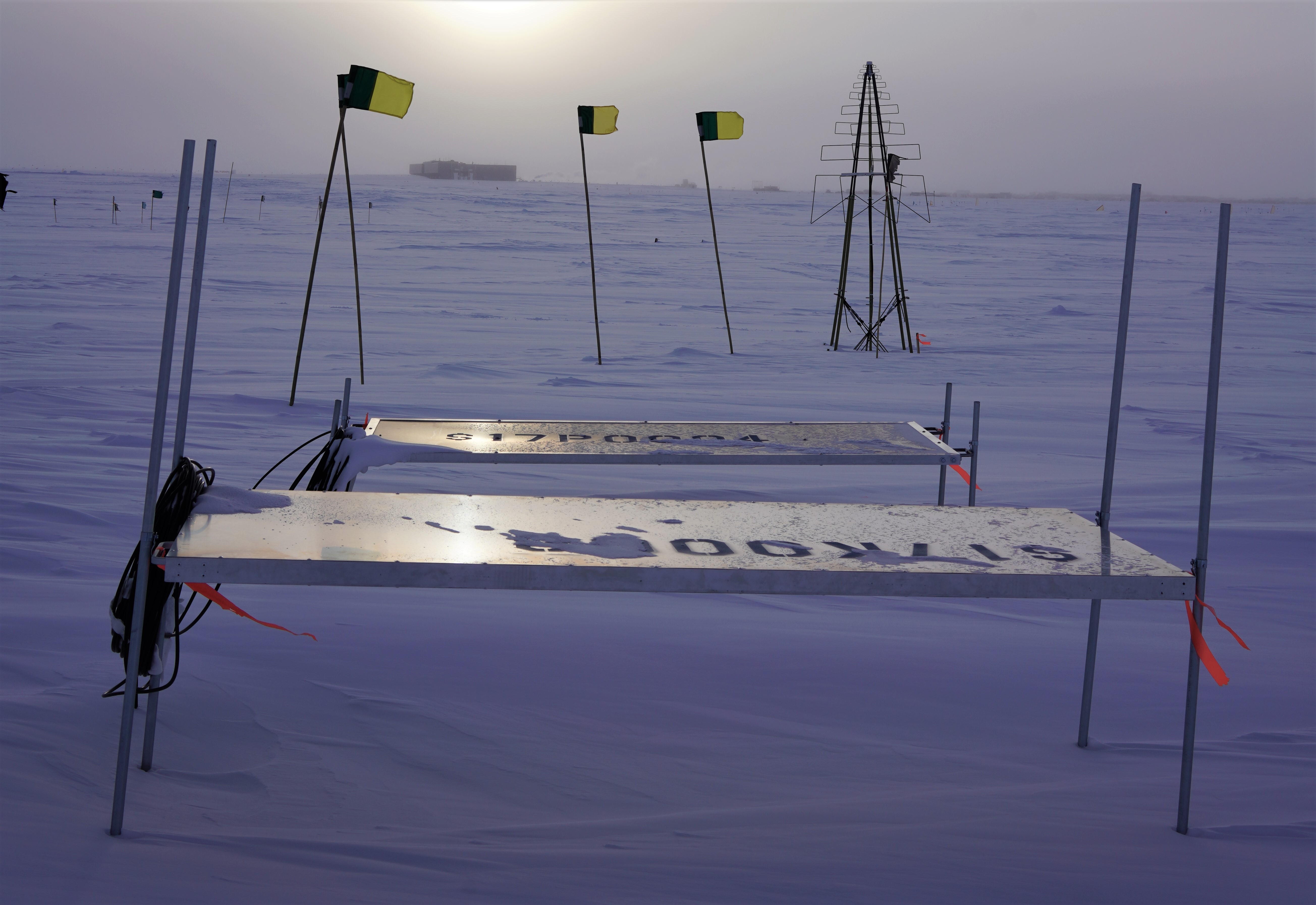}
    \caption{(Left) Layout of a surface station for the enhancement of IceTop, which is the baseline design for the IceCube-Gen2 surface array: a station consists of 4 pairs of scintillation detectors and three radio antennas connected to a common local data-acquisition in the center. (Right) Corresponding prototype detectors at IceTop; both the scintillators and radio antennas are deployed on stands that can be lifted to avoid snow management.}
    \label{fig:surface_layout}
\end{figure} 
IceCube's surface array IceTop~\cite{IceCube:2012nn} has proved to be a very valuable component of IceCube at a modest cost of approximately 5\% of the total investment.  Designed first and foremost for the measurement of the primary CR spectrum and mass composition from 1~PeV up to about an EeV, IceTop is also used to study the anisotropy of the arrival directions of CR, high-pT atmospheric muons, PeV $\gamma$-rays and transient events such as solar flares. Moreover, IceTop provides information that is useful for partial discrimination against CR induced backgrounds to astrophysical neutrinos in the in-ice array~\cite{IceTopVeto_ICRC2019}. Detailed knowledge of the CR spectrum is essential to reduce the systematic uncertainties on the atmospheric backgrounds caused by CR of 10--1000 times higher energy than that of the astrophysical neutrinos~\cite{Aartsen:2013wda,IceCube:2019hmk}.

These are the reasons why the planning for the IceCube-Gen2 facility includes a surface array, with a detector station near the top of each deployed string. The conceptual design will be similar to the maintenance upgrade planned for IceTop~\cite{Haungs:2019ylq, ICRC2019_IceTopUpgrade_Science} --- but with a larger spacing of about 240~m.  The surface array will provide a high-resolution measurement of the primary spectrum and mass composition for energies from 30~PeV to several EeV. A few additional stations between the current IceTop and the new surface array will guarantee a smooth coverage, enabling a consistent analysis of simultaneous measurements by both surface arrays. Moreover, a small overlap with the neutrino radio array (up to the first line of strings next to the optical array) will allow for a cross check of the response to CR induced signals. This overlap will ensure that all detector components of IceCube and IceCube-Gen2 will share the same absolute energy scale by cross-calibration against the same air-shower array on the surface. Figure~\ref{fig:surface_layout} shows the layout of an enhanced IceTop station optimized for easy deployment and low maintenance. This is also the proposed design for a IceCube-Gen2 surface station~\cite{IceScint_ICRC2019, Schroder:2018dvb}. 

With a large-area surface array and the large in-ice volume of the IceCube-Gen2 optical detector, the acceptance for coincident events increases by a factor of about 40, from 0.26~km$^{2}$ sr for the IceTop-IceCube combination to about 10~km$^{2}$~sr for IceCube-Gen2. With hundreds of coincident events per year above one EeV, the planned IceCube-Gen2 detector will allow unprecedented measurement of the evolution of the primary composition in the region where a transition from galactic to extragalactic CR is predicted~\cite{Astro2020_GCR_WhitePaper} --- through comparison of the ratio of the signal from the bundle of TeV muons in the deep detector to the size of the coincident surface shower~\cite{Bai:2007zzm}. The TeV muons will provide information on hadronic interactions in the air showers complementary to the low-energy muons at the surface~\cite{Dembinski:2019uta}. This is important to better understand the flux of atmospheric leptons creating background for the astrophysical neutrino measurements~ \cite{Gaisser:2016obt}. Therefore, the suite of surface detectors at each location should be able to distinguish muons at the surface from the electromagnetic component at the shower front. The design should also allow selection of individual high-energy muons which can be used to constrain the in-ice track reconstruction. 

Understanding the flux of the most-energetic Galactic CR and the transition to extragalactic sources complements IceCube's multi-messenger mission of understanding the origin of CR. Furthermore, the drastically increased aperture for events detected by both the surface and the in-ice array enhances the potential to directly discover nearby sources of PeV $\gamma$-rays, selecting muon-poor events with a large electromagnetic signal at the surface~\cite{Aartsen:2012gka, V.:2017kbm,Aartsen:2019sid}. 

A surface detector with a large aperture increases the possibility of vetoing against CR muons and suppressing the background of atmospheric neutrinos (see Section~\ref{sec:gen2_surface}). For example, a down-going PeV astrophysical neutrino interacting in the ice above the deep array could be distinguished from a CR-induced PeV muon bundle, which would be accompanied by a CR shower of about 10~PeV. Extending the veto capability to the whole hemisphere and/or to lower energies would require a surface footprint that extends significantly beyond the footprint of the optical array~\cite{ICRC:Gen2} and instrumentation deployed more densely between the IceCube-Gen2 strings.

\subsubsection{Radio detectors}
A number of radio instruments have been built at South Pole, most prominently the Askaryan Radio Array (ARA). Similar to the proposed preliminary baseline design for IceCube-Gen2, ARA employs a phased-array and has instrumented strings with two different kinds of antennas as deep as 200 meters. The baseline design also foresees surface antennas to ensure a self-vetoing capability of the array against air showers, a concept piloted in the ARIANNA experiment, also with two dedicated stations at South Pole. In addition to the veto-capabilities, these surface antennas provide better polarization sensitivity than down-hole antennas, which will aid the reconstruction.

As compared to the optical detection technique, the radio detection is not as mature. At the relevant energies, no neutrino candidates have yet been detected, and reconstruction quantities have to be extrapolated from previous but different experiments and dedicated simulations. Additional R\&D and piloting will happen before the construction of IceCube-Gen2. A deployment at Summit station in Greenland (RNO-G) starting in summer 2021 will pilot and test the autonomous baseline design as described above, including the development of targeted electronics and optimized antennas. The site will be used to reduce the burden on South Pole station logistics and allow for a fast turn-around, given the easier accessibility. Is is also currently under discussion to continue an installation on the ARIANNA-site the Ross Ice-Shelf to exploit the reflective properties of the bottom of the shelf ice to increase the field of view \cite{Anker2020}. While characteristics of the site influence science performance of these efforts, technical challenges will be very similar, paving the way for the radio component of IceCube-Gen2.

\subsection{Schedule, costs and logistics}
\label{sec:schedule}
Plans for schedule, costs and logistics for IceCube-Gen2 are primarily based on previous successes in constructing IceCube with dedicated design improvements under consideration.
\subsubsection{Baseline design} 

IceCube demonstrated the ability to deploy 86 strings on time and on budget in a hostile 
environment. Drilling at the South Pole is a formidable challenge for engineering and logistics support.  
The enhanced hot water drill~\cite{Drill:Description} developed for IceCube is capable of drilling 
to 2500\,m depth within about 30 hours. Up to 20 holes have been drilled in a single Antarctic summer season.

For the optical array baseline design with 120 new strings and modules of similar diameter as those in IceCube, we anticipate per hole drill times similar to the ones for IceCube. Due to the larger number of strings we expect a total construction and deployment time of up to 8 austral summer seasons. Similarly to IceCube, data taking will start with a partially completed detector. As the sensor coverage per string is considerably higher than for IceCube, the instrumentation costs will claim a larger fraction of the total budget, while the existing IceCube infrastructure will allow substantial savings on the infrastructure for data acquisition and data systems. The cost per IceCube-Gen2 string is estimated at \$1.2M for the hardware including surface cabling and instrumentation (a single mDOM sensor used in the baseline configuration costs about \$10k, a
single D-Egg sensor about \$8.5k). 

For the radio array, the preliminary baseline design incorporates 200 stations with three strings each. If drilled mechanically the holes can only be 100 meters deep, which is sufficient and efficient for radio stations. The baseline method for drilling is using an ASIG mechanical drill, which is able to drill 5.75'' clear boreholes to 100~m. Switching to RAM drilling technology is considered as an R\&D option to speed-up the drilling procedure for IceCube-Gen2. The hardware costs per station are estimated to \$50k per station not including drilling and deployment. 
Both station design and deployment methods will be tested starting summer 2021 in Greenland. 

The total cost for the facility design outlined in this white paper is anticipated at approximately \$350M, including about \$180M for the instrumentation to be deployed in the optical, surface and radio arrays. This is comparable to the project costs for IceCube of \$279M (with $\sim$50\% used for instrumentation). Note that the IceCube project cost has not been adjusted for inflation. A possible time line for IceCube-Gen2 is shown in Fig.~\ref{fig:gen2_timeline}. Ongoing design efforts are aimed at reducing costs and simplifying logistic impact (see below). After the design phase and the IceCube Upgrade are completed, drill and sensors would go into production and construction at the South Pole would commence.

\begin{figure}[tb]
\centering\includegraphics[width=1.0\linewidth]{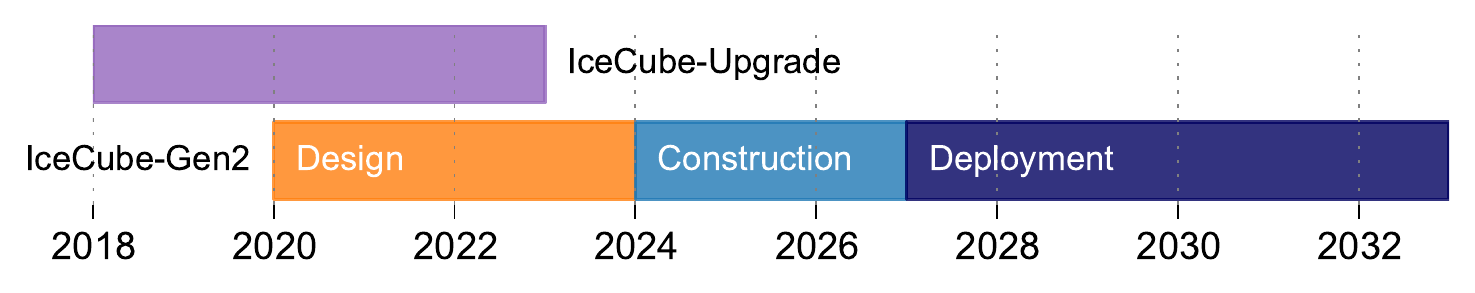}
\caption{Time line for the IceCube Upgrade and projected time line for IceCube-Gen2. \label{fig:gen2_timeline}}
\end{figure}

\subsubsection{Optimization of logistics impact} 
While IceCube has demonstrated how to successfully deploy instrumentation in the Antarctic glacier, it is worthwhile to consider the potential logistics challenges in connection with the construction of a larger detector.
For IceCube-Gen2, preliminary design studies have been performed 
for an optimal drilling strategy and a reduced logistics impact.  
Furthermore, strategies have been developed to ensure that such a project would not compete with other science projects for logistical support at the South Pole. The following strategies are considered: 

{\bf Drilling optimizations:}
During the initial phase and over the course of IceCube construction, 
the hot water drilling technique was refined, and detailed simulations were developed that 
accurately described the drill data \cite{Drill:Model}.  Based on 
these data, it is clear that narrower holes allow for a faster drill time. Given that and the fact that optical sensor designs 
include potentially smaller diameters, as discussed above, 
a savings of up to 40\% of fuel cost and drill time per hole appears feasible.

{\bf Mobile drill:} 
The drill is designed to be mobile in order to most efficiently drill holes over an area on the order of 10\,km$^2$. 
While for IceCube there was a seasonal drill camp, major drill subsystems for IceCube-Gen2 would be situated on large sleds that can easily be moved during a given season. The design would be optimized for less maintenance. 

{\bf Population at the South Pole station:}
An obvious concern for the construction is the required population at the South Pole. 
The South Pole station is designed for a summer population of about 160 people. 
Approximately 50 people would be required for IceCube-Gen2 construction efforts, extrapolating from IceCube construction.
A simple solution to avoid adverse impacts on other science projects would 
be to construct and run a separate summer camp where IceCube-Gen2 personnel could stay. 
Such a summer camp would be funded from the project budget and not compete with 
other program resources. 

{\bf Transport of equipment:}
For IceCube, a total of nine million pounds of cargo was transported to the South Pole by more than 
300 LC130 aircraft missions.  For IceCube-Gen2, alternate means of logistics support are 
being considered.  All major cargo would be transported by 
air-ride cargo sleds (ARCS). Such sleds are in active development already for 
high-payload traverses in Antarctica and Greenland. 
No planes would be needed to transport fuel, sensors or cables. There would be little impact on other science projects in Antarctica. Only personnel and small equipment would be required to be flown in by aircraft, 
a minor task compared to the case for IceCube where everything was transported by plane, 
even while the South Pole Telescope was being built and the South Pole station had not yet been completed. 

{\bf Drilling season:} Another variable could be the duration of the drill season. A regular summer season sets a constraint to 8 weeks of active drilling. 
Experts consider an increase of the duration by use of another type of aircraft as feasible,
allowing for a possible increase of the drill season to drill more than 30 holes.

Overall, we conclude that the logistics needs for IceCube-Gen2 can be organized to not compete with the needs of other science projects. The IceCube-Gen2 investments in logistics solutions would be of long-term benefit to 
the US Antarctic Program.

\newpage

\section{Global landscapes}
\label{sec:landscape}

Neutrino astronomy is inherently multi-messenger astronomy. As described extensively in Sections~\ref{sec:icecube} and \ref{sec:Gen2Science}, the full picture of the non-thermal universe can be obtained only through the combination of neutrino observations with measurements of the broad-band electromagnetic spectrum, gravitational waves and cosmic rays. Fortunately, IceCube-Gen2 will not stand alone but be embedded in an ever growing global landscape of multi-messenger observatories that is expected to evolve significantly over the next decades. 

In the next few years two km$^3$-scale Northern Hemisphere neutrino telescopes will complement IceCube. KM3NeT in the Mediterranean and the Gigaton Volume Detector (GVD) in Lake Baikal, are both currently under construction. Additionally, a water-based neutrino telescope P-ONE~\cite{Agostini:2020aar} is under development in the Pacific Ocean off the coast of Canada, aiming to deploy by the end of this decade. After completion, they will observe the Southern sky with a sensitivity comparable to IceCube's for the Northern Hemisphere. Taken together, KM3NeT, GVD, and IceCube will therefore have full coverage of the neutrino sky from 100 GeV to energies beyond a few PeV.

The start of IceCube-Gen2 construction is anticipated at a time when KM3NeT and GVD will likely be fully operational. Hence, the build-up of a global network of large neutrino observatories follows a staged approach. After obtaining full sky coverage with the two Northern Hemisphere detectors, IceCube-Gen2 will push for the next level in neutrino astronomy on both the intensity and energy frontiers, expanding the energy range of neutrino telescopes to EeV energies, improving source sensitivity by a factor of 5, and the observable volume for transients by a factor of at least 10.

Expected to be completed in 2033, IceCube-Gen2 will be embedded in a network of large-scale observatories surveying the sky from radio to $\gamma$-rays, detecting gravitational waves as well as the highest-energy charged cosmic rays. 
At radio wavelengths the SKA, a radio telescope with a collective aperture of 1~km$^2$ will give an unprecedented view of the radio sky. The Vera C. Rubin observatory, an 8.4~m diameter optical telescope, will perform a deep optical survey of the full sky visible from Chile. At high energies CTA, an international $\gamma$-ray observatory with both a Southern and Northern site, will give us new insights into the $\gamma$-ray sky. In parallel, AugerPrime will enhance our understanding of the ultra-high energy CR.

A new generation of gravitational wave detectors distributed over the world will complement aLIGO in creating a global network of interferometers. These will allow the detection and localization of more and more sources of gravitational waves, in particular the ones which are expected to also be observable through their neutrino emission. IceCube-Gen2, designed for more than a decade of observations, will operate in parallel to 3rd generation gravitational wave detectors like the Einstein and Cosmic Explorer telescopes.  

As the most sensitive neutrino telescope proposed, IceCube-Gen2 has a truly unique and essential role in this new multi-messenger world. Only Gen2 can provide the neutrino piece in the puzzle of understanding the most violent processes in our universe.

\newpage
\section{Glossary}
\setlength\extrarowheight{3pt}
\begin{tabularx}{0.99\textwidth}{@{}lX}
\bf AGN & Active Galactic Nuclei. The engine of AGNs is a supermassive black hole at the core of a galaxy, which is being fed material via an accretion disk \\
\bf aLIGO& advanced LIGO interferometer \newline (\url{https://www.advancedligo.mit.edu/}) \\
\bf AMANDA & Antarctic Muon And Neutrino Detector Array \\
\bf ANTARES& Astronomy with a Neutrino Telescope and Abyss environmental RESearch project \\
\bf ARA& Askaryan Radio Array  (\url{https://ara.wipac.wisc.edu}) \\
\bf ARIANNA& Antarctic Ross Ice shelf ANtenna Neutrino Array (\url{https://arianna.ps.uci.edu}) \\
\bf AugerPrime& the Pierre Auger Observatory Upgrade \newline (\url{https://www.auger.org/})\\
\bf Baikal GVD& Baikal Gigaton Volume Detector. Under construction. \\
\bf Blazar& A subclass of AGNs in which a relativistic jet points approximately along the line of sight from Earth \\
\bf BSM& Beyond the Standard Model \\
\bf CCSN(e)& Core-collapse supernova(e) \\
\bf Cosmic Explorer& Proposed future graviational wave observatory \newline (\url{https://cosmicexplorer.org/}) \\ 
\bf CR& Cosmic rays ; cosmic-ray \\
\bf CTA& Cherenkov Telescope Array \newline (\url{https://www.cta-observatory.org/})\\
\bf D-Egg& Dual optical sensors in an Ellipsoid Glass for Gen2 \\
\bf Discovery potential& Signal strength necessary to reject the null hypothesis with a given significance, e.g. 5 sigma \\
\bf DOM& Digital Optical Module \\
\bf EBL& Extragalactic background light \\
\bf Einstein Telescope& Proposed future graviational wave observatory \newline (\url{http://www.et-gw.eu/}) \\
\bf Fermi-GBM& Fermi's Gamma-ray Burst Monitor \newline (\url{https://fermi.gsfc.nasa.gov/science/instruments/gbm.html}) \\
\bf Fermi-LAT& Fermi's Large Area Telescope (\url{https://glast.sites.stanford.edu}) \\
\bf GRB& Gamma Ray Burst \\
\bf GW& Gravitational Wave \\
\bf HAWC& High-Altitude Water Cherenkov  \newline (\url{https://www.hawc-observatory.org}) \\
\bf H.E.S.S& High Energy Stereoscopic System  \newline (\url{https://www.mpi-hd.mpg.de/hfm/HESS/}) \\
\bf KM3NeT& Cubic-kilometer sized neutrino research infrastructure in the Mediterranean. 
Under construction.
 \\
\bf LBL& Long Base-Line neutrino experiment \\
\bf Long GRB& A gamma ray burst with a prompt phase, observed in the 10~keV~-- 10~MeV range,  that lasts more than 2 seconds \\
\bf  LPDA&  Log-Periodic Dipole Antenna \\
\bf  MAGIC& Major Atmospheric Gamma Imaging Cherenkov  \newline (\url{https://magic.mpp.mpg.de}) \\
\bf mDOM& Multi-PMT DOM \\
\bf NS-BH& Merger of a neutron star with a black hole. Gives rise to short GRBs and/or gravitational waves \\
\bf NS-NS& Merger of two neutron stars. Gives rise to short GRBs and/or gravitational waves \\
\bf PINGU& Precision IceCube Next Generation Upgrade \\
\bf PMT& Photomultiplier tube \\
\bf P-ONE& The Pacific Ocean Neutrino Experiment \newline  (\url{http://www.pacific-neutrino.org/}) \\
\bf RNO-G&Radio Neutrino Observatory Greenland \newline (\url{https://rno-g.org})\\
\bf SBL& Short Base-Line neutrino experiment \\ 
\bf Sensitivity& The average upper limit that would be obtained by an ensemble of experiments with the expected background and no signal \\
\bf Short GRB& A gamma ray burst with a prompt phase, observed in the 10~keV~ -- 10~MeV range,  that lasts less than 2 seconds \\
\bf SKA& Square Kilometre Array \newline (\url{https://www.skatelescope.org/}) \\
\bf SNe& supernova(e) \\
\bf TDE& Tidal Disruption Event \\
\bf UHE& Ultra-high energy (generally $>$10~PeV) \\
\bf Vera C. Rubin Observatory& formerly the Large Synoptic Survey Telescope, LSST \newline (\url{https://www.lsst.org/}) \\
\bf VERITAS& Very Energetic Radiation Imaging Telescope Array System \newline (\url{https://veritas.sao.arizona.edu})\\
\bf VIRGO& VIRGO interferometer (\url{https://www.virgo-gw.eu})\\
\bf VLBL& Very Long Base-Line neutrino experiment \\
\end{tabularx}

\newpage
\bibliographystyle{utphys}
\bibliography{gen2_loi_new}

\end{document}